\documentclass[twocolumn]{aastex631}

\usepackage{color}
\usepackage{amsmath}
\usepackage{xcolor}
\usepackage{mathrsfs}
\usepackage{natbib}
\usepackage{gensymb}
\usepackage{appendix}
\usepackage[T1]{fontenc}

\shortauthors{Bryan $\&$ Lee}
\shorttitle{GG/SE correlation}

\begin{document}

\title{Friends not Foes: Strong Correlation between Inner Super-Earths and Outer Gas Giants}

\author[0000-0002-6076-5967]{Marta L. Bryan}
\affiliation{David A. Dunlap Department of Astronomy $\&$ Astrophysics, University of Toronto, 50 St. George St., M5S 3H4 Toronto ON, Canada}
\affiliation{Department of Chemical $\&$ Physical Sciences, University of Toronto Mississauga, 3359 Mississauga Road, L5L 1C6 Mississauga ON, Canada}

\author[0000-0002-1228-9820]{Eve J.~Lee}
\affiliation{Department of Physics and Trottier Space Institute, McGill University, 3600 rue University, H3A 2T8 Montr\'eal QC, Canada}
\affiliation{Trottier Institute for Research on Exoplanets (iREx), Universit\'e de Montr\'eal, Canada}

\begin{abstract}

The connection between outer gas giants and inner super-Earths reflects their formation and evolutionary histories. Past work exploring this link has suggested a tentative positive correlation between these two populations, but these studies have been limited by small sample sizes and in some cases sample biases. Here we take a new look at this connection with a sample of 184 super-Earth systems with publicly available radial velocity data and resolved outer gas giants. We calculate the frequency of outer gas giants (GG) in super-Earth (SE) systems, dividing our sample into metal-rich ([Fe/H] $>$ 0) and metal-poor ([Fe/H]$\leq$0) hosts. We find P(GG$|$SE, [Fe/H]$>$0) = 28.0$^{+4.9}_{-4.6}\%$ and P(GG$|$SE, [Fe/H]$\leq$0) = 4.5$^{+2.6}_{-1.9}\%$. Comparing these conditional occurrence rates to field giant occurrence rates from \citet{Rosenthal2021}, we show that there is a distinct positive correlation between inner super-Earths and outer gas giants for metal-rich host stars at the 2.7$\sigma$ level, but this correlation disappears for metal-poor systems. We further find that, around metal-rich stars, the GG/SE correlation enhances slightly for systems with giants that are more distant (beyond 3 AU), more eccentric ($e > 0.2$), and/or in multi-gas giant systems. Such trends disappear around metal-poor stars with the exception of systems of multiple giants in which we observe a tentative anti-correlation. Our findings highlight the critical role metallicity (disk solid budget) plays in shaping the overall planetary architecture.

\end{abstract}

\section{Introduction}

Gas giants dominate the dynamics of the systems they are in and reflect earlier formation conditions. In our solar system, Jupiter and Saturn are believed to have played a dominant role in our system's history, from opening gaps in the disk that could shape the surface density profile and composition of planetary building blocks in the inner system, to kicking volatile rich planetesimals onto shorter period orbits \citep[e.g.,][]{Morbidelli07,Walsh11,Batygin15}. We expect extrasolar gas giants to play a significant role in the lives of small planets as well.

The connection between inner super-Earths and outer gas giants was first explored in \citet{ZhuWu18} and \citet{Bryan2019}. Both studies found that outer gas giants do not hinder the formation of super-Earths, and instead show a positive correlation between the two populations. In addition, most of the gas giant companions were found around metal-rich host stars, indicating metallicity is an important factor driving this correlation. Subsequent theoretical work explains this excess of cold Jupiters in super-Earth systems by showing that for systems with enough solids to produce a gas giant, they naturally deliver enough material to the inner system for super-Earths to grow \citep[e.g.,][]{Chachan22,Chachan23} since most of the solids are lost to radial drift rather than being coagulated into planetary cores at large orbital distances \citep[e.g.,][]{Lin18}. 

While there have been follow-up efforts that also find a positive correlation between gas giants and small planets \citep[e.g.][]{Herman19,Rosenthal2022}, some have reported a neutral or even negative correlation \citep[e.g.][]{Bonomo2023}. A resolution to this debate was presented in \citet{Zhu2023}, arguing that interpreting this correlation requires considering the host star metallicity distribution of the sample of systems. Namely, since the vast majority of the gas giants in all samples considered to date are found around metal-rich stars, having a sample of systems dominated by metal-poor stars (as was the case in \citet{Bonomo2023}) can bias the strength of the correlation. \citet{Zhu2023} showed that when only considering the metal-rich systems in \citet{ZhuWu18}, \citet{Bryan2019}, and \citet{Bonomo2023}, all three find consistent positive correlations between super-Earths and cold gas giants.

Three key takeaways from this debate are: 1) all studies to date have worked with small sample sizes, which makes interpreting the inner-outer planet correlation with statistical significance challenging; 2) properly accounting for sample biases in metallicity is a critical component of this interpretation; and 3) when sample biases are addressed, previous studies find consistent tentative positive correlations between inner super-Earths and outer gas giants.

In this paper, we explore the connection between inner super-Earths and outer gas giants with a sample of 184 systems, at least three times larger than previous samples considered, with the orbit of giant(s) resolved by radial velocity data. In Section 2, we describe how these systems were selected. Section 3 details our computation of sensitivity maps and completeness correction for each system, and Section 4 shows how we incorporate those maps into our occurrence rate calculations to compute the frequency of cold gas giants in super-Earth systems that are metal rich P(GG$|$SE, [Fe/H]$>$0) and metal poor P(GG$|$SE, [Fe/H]$\leq$0). To assess the presence and strength of a correlation, we compare our conditional probabilities to gas giant occurrence rates P(GG$|$[Fe/H]$>$0) and P(GG$|$[Fe/H]$\leq$0) from \citet{Wittenmyer2020} (hereafter denoted W20) and from \citet{Rosenthal2021} (hereafter denoted R21) in Section 5. Section 6 explores how gas giant properties shape this correlation, and Section 7 summarizes our findings, placing them in the context of formation theory.

\section{Observations}
\label{section: obs}

We select systems with the following five criteria:  the system has 1) publicly available radial velocity (RV) datasets, 2) at least one confirmed super-Earth planet (1-20 M$_{\earth}$; 1-4 R$_{\earth}$), 3) host star mass greater than 0.6 M$_{\sun}$ to exclude M-dwarfs, 4) RV datasets with baselines greater than one year and more than 20 data points, and 5) an RV semi-amplitude K$>$1m/s in the case of systems with super-Earths discovered using RVs to minimize the inclusion of false positive planets. We exclude M-dwarfs from this analysis for several reasons. Given the selection criteria above, there are 78 systems with host star masses $<$0.6 M$_{\odot}$ that we could consider. However, out of the 203 stars in W20, only 8 are M-dwarfs. While the fraction of the R21 sample that are M stars is higher (121 M-dwarfs out of 719 stars), given the relevance of the W20 sample to this study we leave the assessment of the gas giant/super-Earth correlation around M-dwarfs to subsequent work. We additionally exclude a handful of systems that do not have published metallicity measurements. This yields a total of 184 systems (see Table \ref{tab:full-sample}). 109 of these systems have one or more super-Earth initially discovered via the transit method (hereafter labeled `transit sample'), and 75 systems have one or more super-Earth initially discovered with RVs (labeled `RV sample'). As described in the following section, we initially consider these samples separately due to the significant average difference in sensitivities to distant gas giants.

\begin{figure}
\centering
\includegraphics[width=0.5\textwidth]{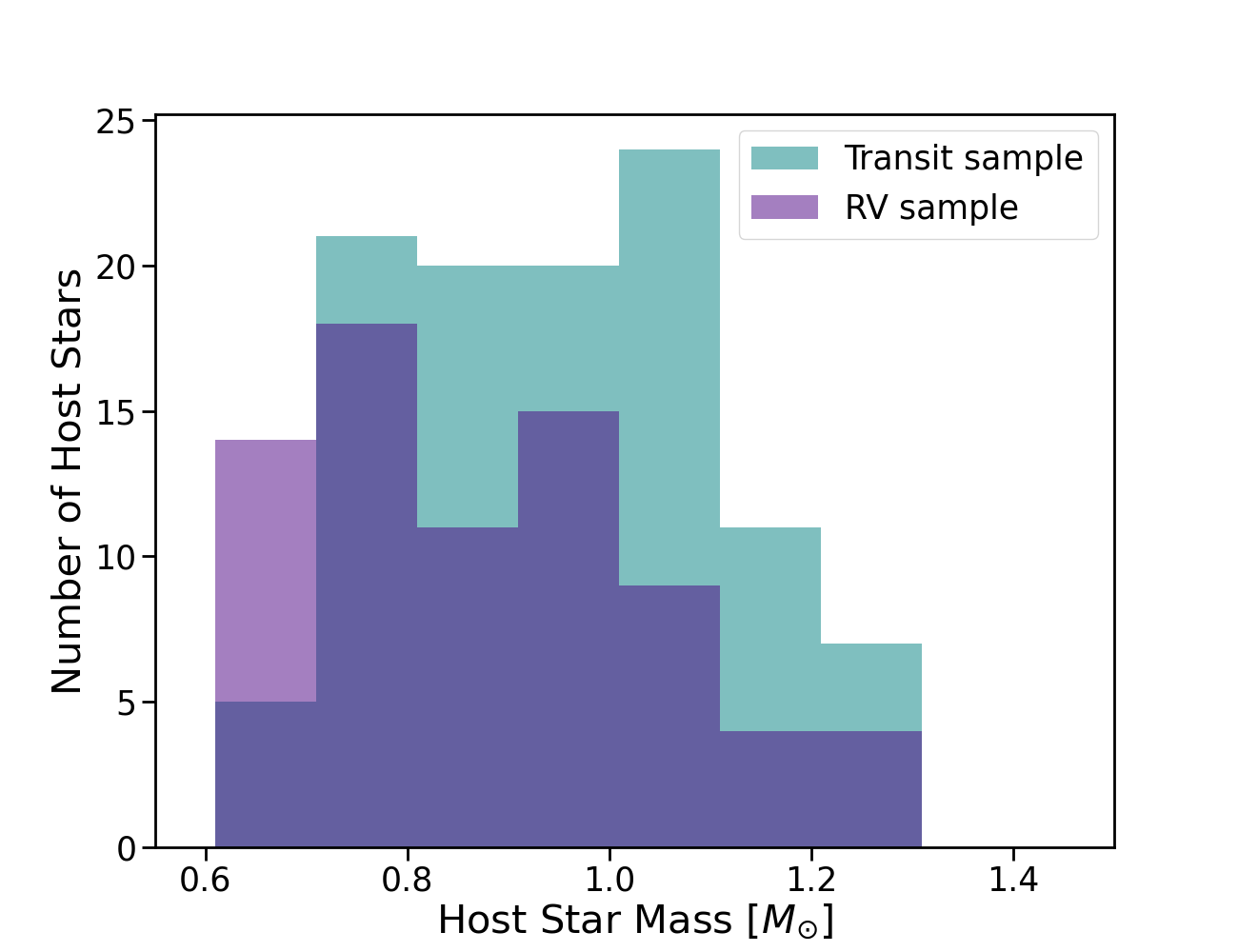}
\includegraphics[width=0.5\textwidth]{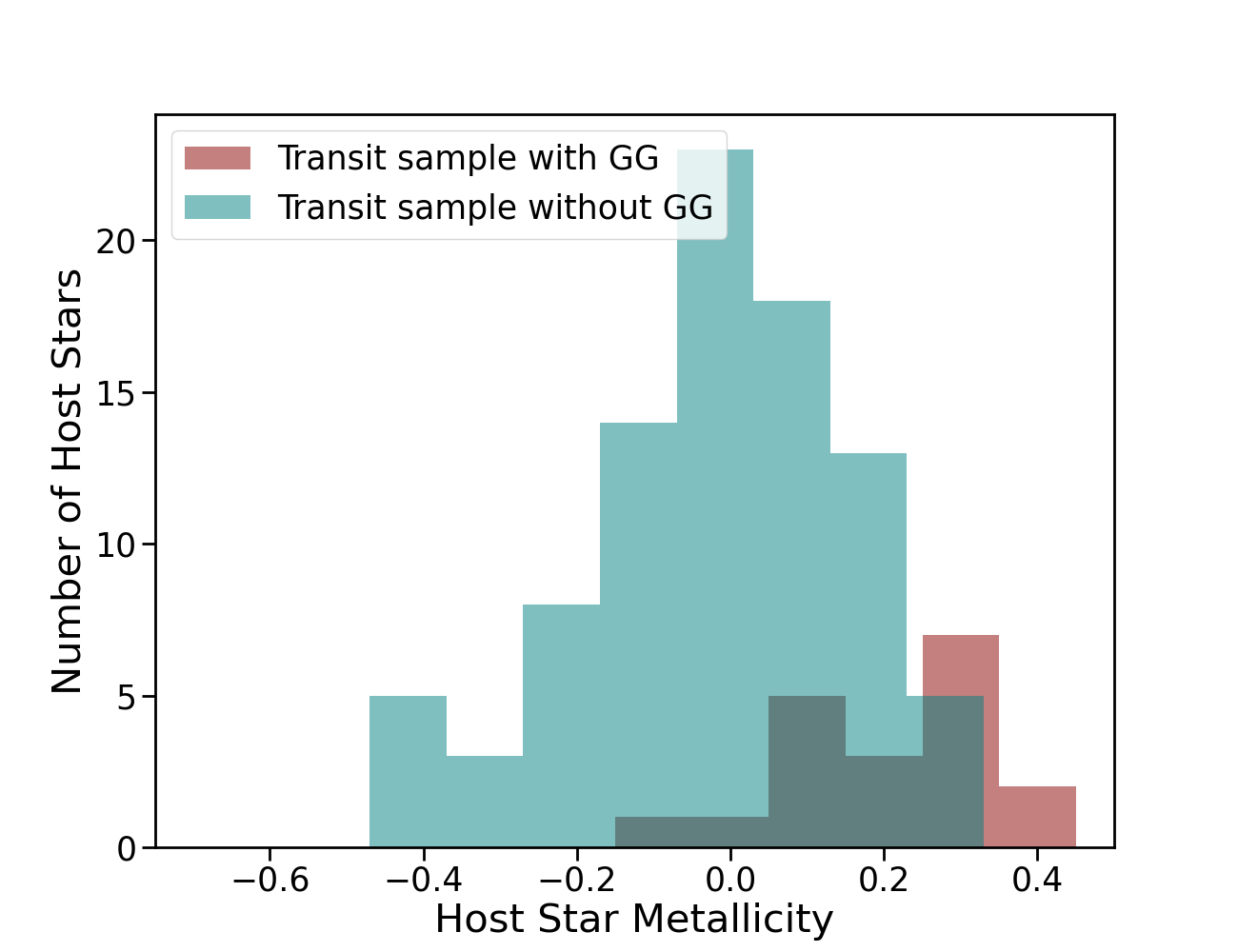}
\includegraphics[width=0.5\textwidth]{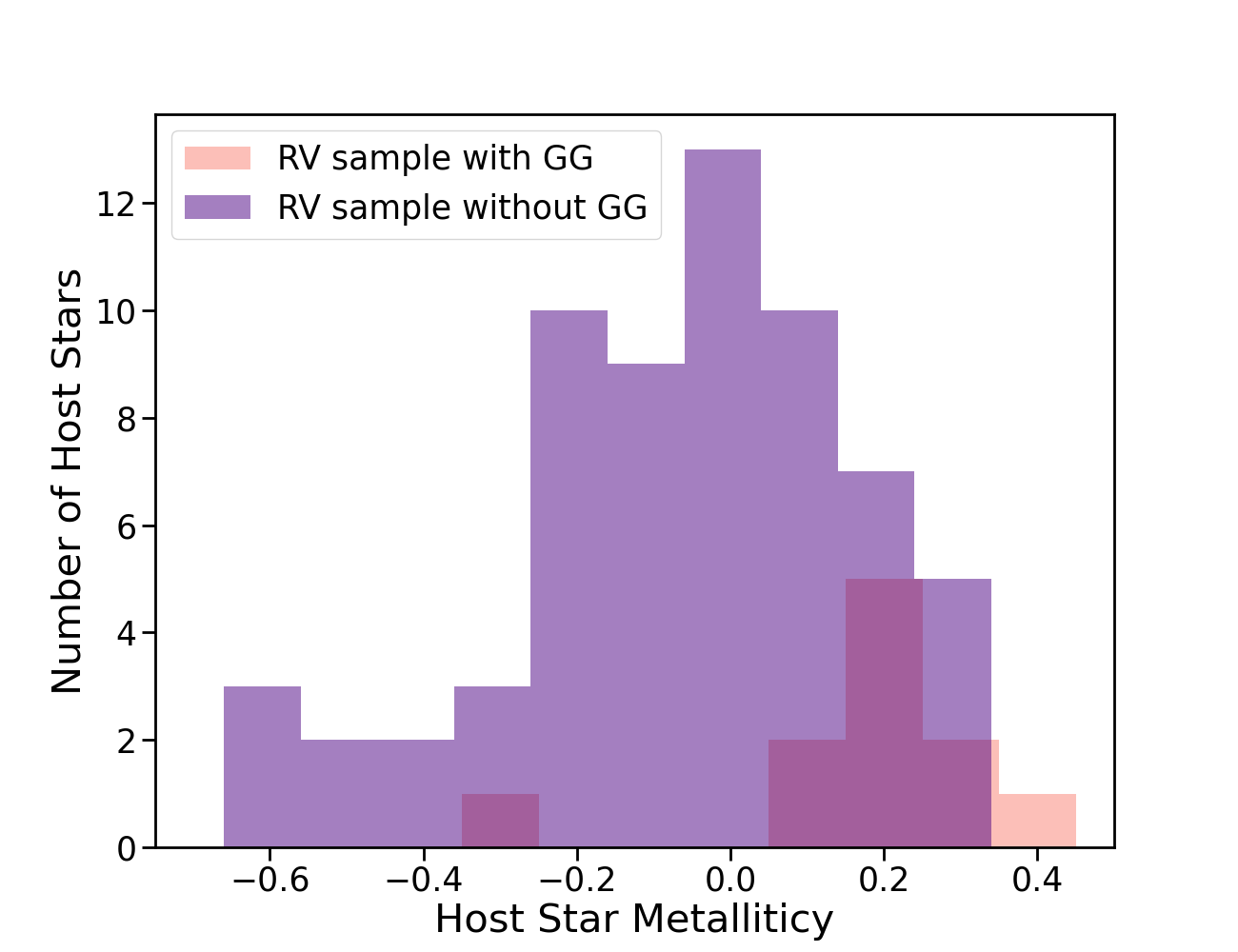}
\caption{Top: Host star mass distributions for the transit super-Earth sample (teal) and the RV super-Earth sample (purple). Middle: Host star metallicity distributions for the transit super-Earth sample without gas giants (teal), and with gas giants (maroon). Bottom: Host star metallicity distributions for the RV super-Earth sample without gas giants (purple) and with gas giants (pink). Gas giants considered are 0.5$-$20 M$_{\rm Jup}$ and $<$ 10 AU (see Table \ref{tab:GG-conf} for properties). }
\label{fig: stellar properties}
\end{figure}

Of the systems in the transit sample, 56 are metal-rich [Fe/H]$>$0, and 53 are metal-poor [Fe/H] $\leq$ 0. Similarly, the RV sample has 39 metal-rich systems and 36 metal-poor ones. We consider a system to host a gas giant (0.5$-$20 M$_{\rm Jup}$) if there is a confirmed gas giant (typically with $\gtrsim$1 measured RV orbit) in the system. Given this criteria, there are 19 systems in the transit sample hosting at least one gas giant, 17 of which are metal-rich. The RV sample has 11 systems with one or more gas giants, 10 of which are metal rich. Properties of the gas giants in this sample are shown in Table \ref{tab:GG-conf}. We show host star properties of our sample in Figure \ref{fig: stellar properties}.

\section{Sample Sensitivity Maps}
\label{sensitivity}

With a heterogeneous sample such as that considered here, individual system sensitivities to outer gas giants can vary widely due to differences in time baseline, observing cadence, number of data points, and measurement precision. For each system, we quantify sensitivity by calculating completeness as a function of mass and semi-major axis given the system's RV dataset (see Table \ref{tab:full-sample} for references).\footnote{In the case where there is more than one public RV dataset that has a time baseline $>$1 year and more than 20 data points, we select the dataset that has higher completeness.} We start with a 50$\times$50 grid in mass and semi-major axis evenly spaced in log space from 0.3$-$30 M$_{\rm Jup}$ and 0.3$-$30 AU. We inject 50 simulated planets into each grid box, each time drawing a mass and semi-major axis value from a uniform distribution across each grid box, an eccentricity from the $\beta$ distribution, an inclination $i$ from a uniform distribution in $\cos i$, and remaining orbital elements from uniform distributions. 

We calculate simulated RVs at each observational epoch for the simulated planet, and add noise by drawing from a Gaussian distribution with a width defined by the instrumental uncertainty randomly shuffled from the original dataset. To determine whether a simulated planet would be detected, we fit each simulated RV series with a one-planet orbital solution and a flat line and compare the Bayesian information criterion (BIC) values from the two fits. A simulated planet is `detected' if the one-planet model BIC is smaller than the flat line BIC by more than 10. In contrast, if the flat line is preferred or the one-planet model is preferred by a $\Delta$BIC$<$10, the simulated planet is `not detected'. After repeating this process for each injected planet, we used our `detected' vs `non-detected' results to calculate completeness across the entire mass/semi-major axis grid.  

Figure \ref{fig: average completeness} in the appendix presents average completeness maps for the RV super-Earth sample (top) and the transit super-Earth sample (bottom). Detected gas giants in each sample are overplotted. The reason why we initially separate these samples in our analysis is evident: on top of variations amongst individual systems, on average the RV sample systems have significantly higher sensitivities, driven by as much as an order of magnitude more data points and longer time baselines in comparison to the transit sample. Despite these differences, confirmed gas giants are in regions of parameter space where both transit and RV samples are on average close to, if not at, 100$\%$ completeness. This motivates our mass and semi-major axis ranges considered in our occurrence rate calculations in the following sections, namely 0.5$-$20 M$_{\rm Jup}$ and 1$-$10 AU. Individual sensitivity maps allow us to account for completeness to gas giants in this range on a system-by-system basis in these calculations, and confirm that these sensitivity differences do not bias our results.

\section{Consistent Occurrence Rates P(GG$|$SE) for Transit and RV Samples}
\label{occ rate}

To explore the connection between super-Earths and Jupiters, we start by calculating the frequency of gas giants (GG) in systems with inner super-Earths (SE), P(GG$|$SE). We define these gas giants as planets with mass 0.5$-$20 M$_{\rm Jup}$ and orbital distance 1$-$10 AU. Previous work has demonstrated the importance of host star metallicity to this GG/SE connection \citep{Zhu2023}. We split our sample into metal-rich systems [Fe/H]$>$0, and metal-poor systems [Fe/H]$\leq$0 to verify the sensitivity of this correlation to metallicity.

We determine P(GG$|$SE, [Fe/H]$>$0) and P(GG$|$SE, [Fe/H]$\leq$0) using a general binomial distribution. If all of our systems had 100$\%$ completeness across our entire integration range (0.5$-$20 M$_{\rm Jup}$ and 1$-$10 AU), the posterior distribution given number of detections $n_{\rm det}$ and total number of systems $n_{\rm tot}$ is:

\begin{equation}
    f(x;a,b) = \frac{1}{B(a,b)}x^{a-1}(1-x)^{b-1}
\end{equation}

\noindent where $B$ is the beta function, $a$ = $n_{\rm det}$+1, and $b$ = $n_{\rm tot}$-$n_{\rm det}$+1. To correct for individual system sensitivities to gas giants, we used the completeness maps we computed for each system. We sum over the 0.5$-$20 M$_{\rm Jup}$ and 1$-$10 AU range and determine the average completeness for the set of systems considered. The coefficient $b$ then becomes $b$ = $(n_{\rm eff}-n_{\rm det})+1$, where $n_{\rm eff}$ is the total number of systems modified by the sample completeness.

We initially consider the RV and transit samples separately to test whether the average differences in system sensitivity impact the resulting occurrence rates. There are 56 metal-rich transit sample systems, 14 of which host one or more gas giants 0.5$-$20 M$_{\rm Jup}$ and 1$-$10 AU, 53 metal-poor transit sample systems with 2 hosting gas giants, 39 metal-rich RV sample systems with 10 hosting gas giants, and 36 metal-poor RV sample systems with just one hosting a gas giant. We find that the frequencies of gas giants in super-Earth systems for both metal poor and metal rich systems are consistent when comparing the RV and transit samples: P(GG$|$SE, [Fe/H])$>$0) = 28.8 (+6.5 -6.0)$\%$ for the transiting sample and P(GG$|$SE, [Fe/H]$>$0) = 27.8 (+7.5 -6.8)$\%$ for the RV sample, consistent to $0.1\sigma$. Similarly, the metal poor RV and transit samples are consistent at the $0.1\sigma$ level, with P(GG$|$SE, [Fe/H])$\leq$0) = 5.5 (+3.9 -2.7)$\%$ for the transit sample and P(GG$|$SE, [Fe/H]$\leq$0) = 4.8 (+4.4 -3.8)$\%$ for the RV sample. We can therefore move forward with a combined RV$+$transit sample occurrence rate for the rest of the analyses. For this combined sample, we calculate conditional occurrence rate P(GG$|$SE, [Fe/H]$>$0) = 28.0 (+4.9 -4.6)$\%$, and a substantially lower frequency for metal poor systems, P(G$|$SE, [Fe/H]$\leq$0) = 4.5 (+2.6 -1.9)$\%$ (see Table 1).

\section{A Strong Positive Correlation Between SE/GG For Metal Rich Systems}

To assess whether there is a positive, negative, or no correlation between super-Earths and gas giants in our sample, we need to compare our conditional occurrence rate P(GG$|$SE) to the field occurrence of gas giants P(GG). For P(GG), we use the sample of systems from \citet{Rosenthal2021} (R21) and \citet{Wittenmyer2020} (W20). R21 presents the California Legacy Survey, a catalog of radial velocities for 719 stars observed over the course of 30 years. In this paper we take all systems in the R21 sample except M-star hosts ($<$0.6 M$_{\odot}$) and remaining systems with datasets $<$ 20 RV measurements (4 systems). This yields 351 metal rich systems, 43 of which host gas giants (0.5$-$20 M$_{\rm Jup}$ and 1$-$10 AU); and 243 metal-poor systems, 9 of which host gas giants (same definition as the metal-rich sample). Note again that we exclude M-dwarfs from this analysis to ensure consistent comparison. We leave the assessment of the GG/SE correlation around M-dwarfs to subsequent work. In this section we also consider the W20 sample, which comes from the Anglo-Australian Planet Search that leverages the 18-year archive of data for 203 stars. Here we use all systems in the W20 sample except ones that are M-dwarfs and ones that do not have published stellar metallicities. The W20 sample considered here has 107 metal-rich stars, 14 of which host gas giants; and 87 metal-poor stars, 5 of which have one or more gas giants. Despite the smaller sample size, we initially include the W20 P(GG) value in our analysis since previous work examining the inner-outer planet correlation \citep[e.g.,][]{Zhu2023,Bonomo2023} used the W20 sample for P(GG) when interpreting their values for P(SE|GG).

To compare our value of P(GG$|$SE) to the R21 and W20 values for P(GG), it is critical to 1) use the same definition of a gas giant, namely mass and semi-major axis ranges considered, and 2) use comparable estimates for sensitivity corrections. For the R21 sample, we calculate individual system sensitivity maps for the whole sample using our methods outlined in Section 3 using the public radial velocities. Since RVs are not public for the W20 sample, we instead compare our sensitivity estimate methods. In short, W20 carries out injection/recovery simulations where they assess whether a planet has been ``detected" based on whether a Lomb-Scargle periodogram can recover that signal with a false alarm probability (FAP) $<$ 1$\%$ (see W20 for details). To test whether our method presented in Section \ref{sensitivity} is comparable, we recalculate sensitivity maps for all of our systems using the W20 LS method. We find P(GG$|$SE, [Fe/H]$>$0) = 26.3 (+4.6 -4.4)$\%$ and P(GG$|$SE, [Fe/H]$\leq$0) = 4.2 (+2.4 -1.8)$\%$, consistent with our original values to 0.3$\sigma$ and 0.1$\sigma$ respectively. We can thus move forward with our comparisons of P(GG) and P(GG|SE).

\begin{figure}
\centering
\includegraphics[width=0.5\textwidth]{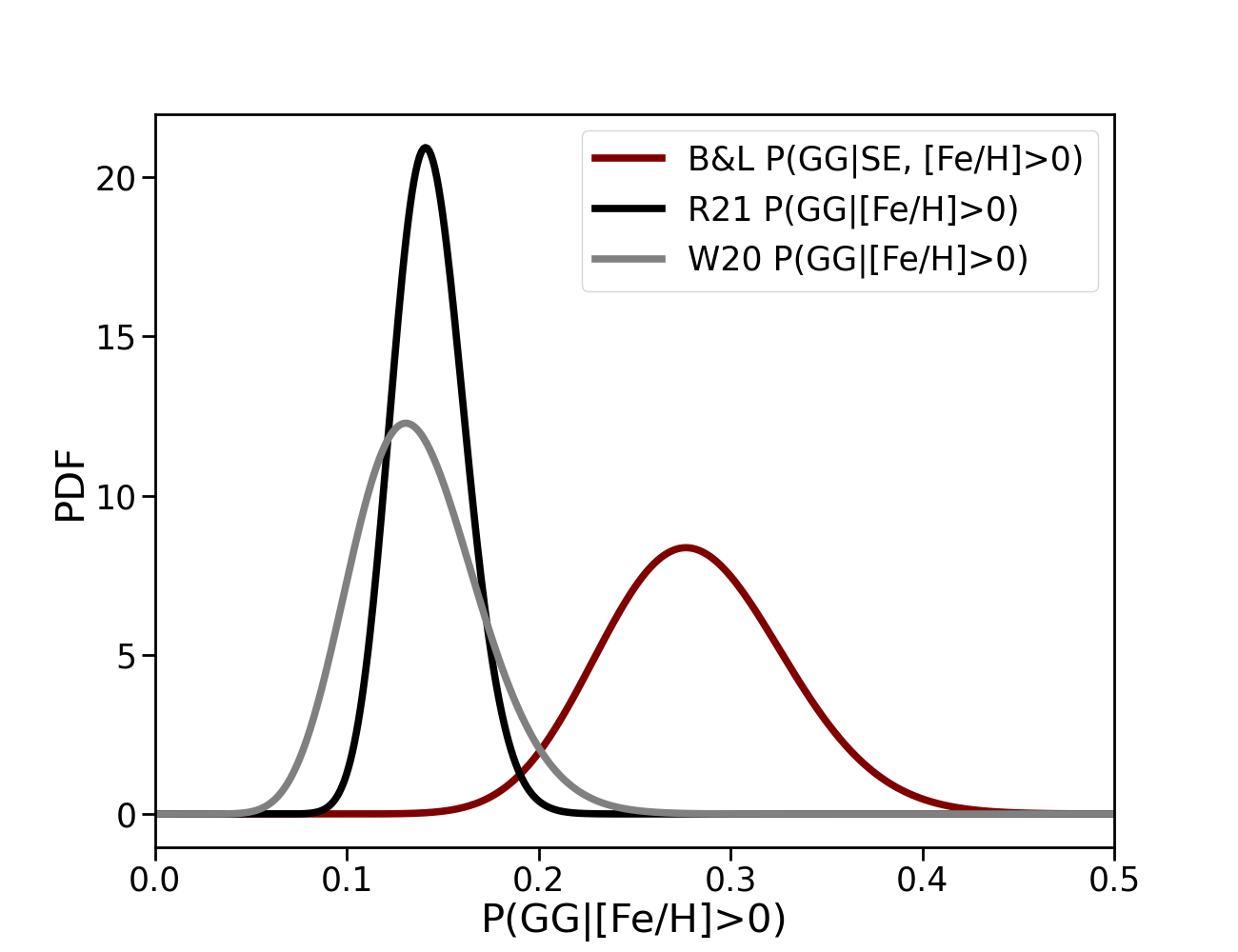}
\includegraphics[width=0.5\textwidth]{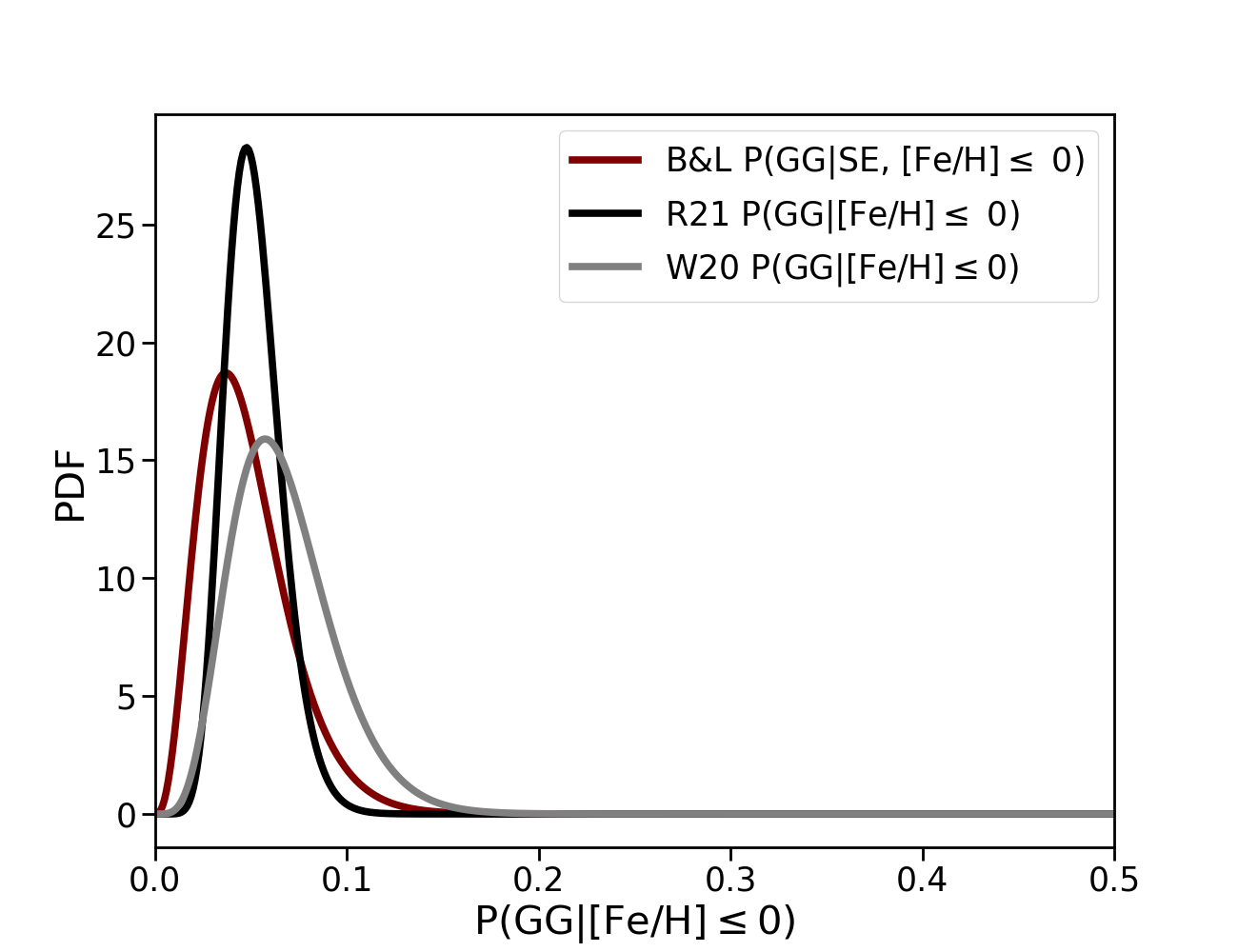}
\caption{Top: Comparison of our conditional occurrence rate P(GG|SE) (maroon) to field giant occurrence rates P(GG) from the R21 sample (black) and the W20 sample (gray) for metal-rich stars. Bottom:  The same comparison for metal poor systems. We see a positive correlation between super-Earths and gas giants for metal rich systems, namely P(GG$|$SE, [Fe/H]$>$0) $>$ P(GG$|$[Fe/H]$>$0) at the 2.7$\sigma$ level, where P(GG) is taken from the larger R21 sample. This correlation disappears when moving to metal poor systems.}
\label{fig: occ enhancement}
\end{figure}

With consistent sensitivity corrections, we calculate P(GG$|$[Fe/H]$>$0) and P(GG$|$[Fe/H]$\leq$0) for the R21 and W20 samples, using consistent definition of a gas giant (0.5$-$20 M$_{\rm Jup}$ and 1$-$10 AU). Figure \ref{fig: occ enhancement} and Table \ref{table: occ rates} show our results. For both metal rich and metal poor systems, the P(GG) frequencies from R21 and W20 are consistent to $<$1$\sigma$.  From here we solely consider the R21 P(GG) occurrence rates throughout the rest of the paper due to the greater statistical leverage stemming from a significantly larger sample size in comparison to W20. For metal-rich systems, we find a distinct positive correlation between gas giants and super-Earths at the 2.7$\sigma$ level. There are more gas giants in metal-rich super-Earth systems than you would expect just based on chance. In contrast, for metal poor systems the positive correlation disappears. These results highlight the importance of this metallicity dimension when interpreting the SE/GG correlation. If a sample is biased towards low metallicities or there is a mixture of low and high, the strength of the correlation can be artificially underestimated.

\begin{figure*}
    \begin{tabular}{cc}
    \includegraphics[width=0.5\textwidth]{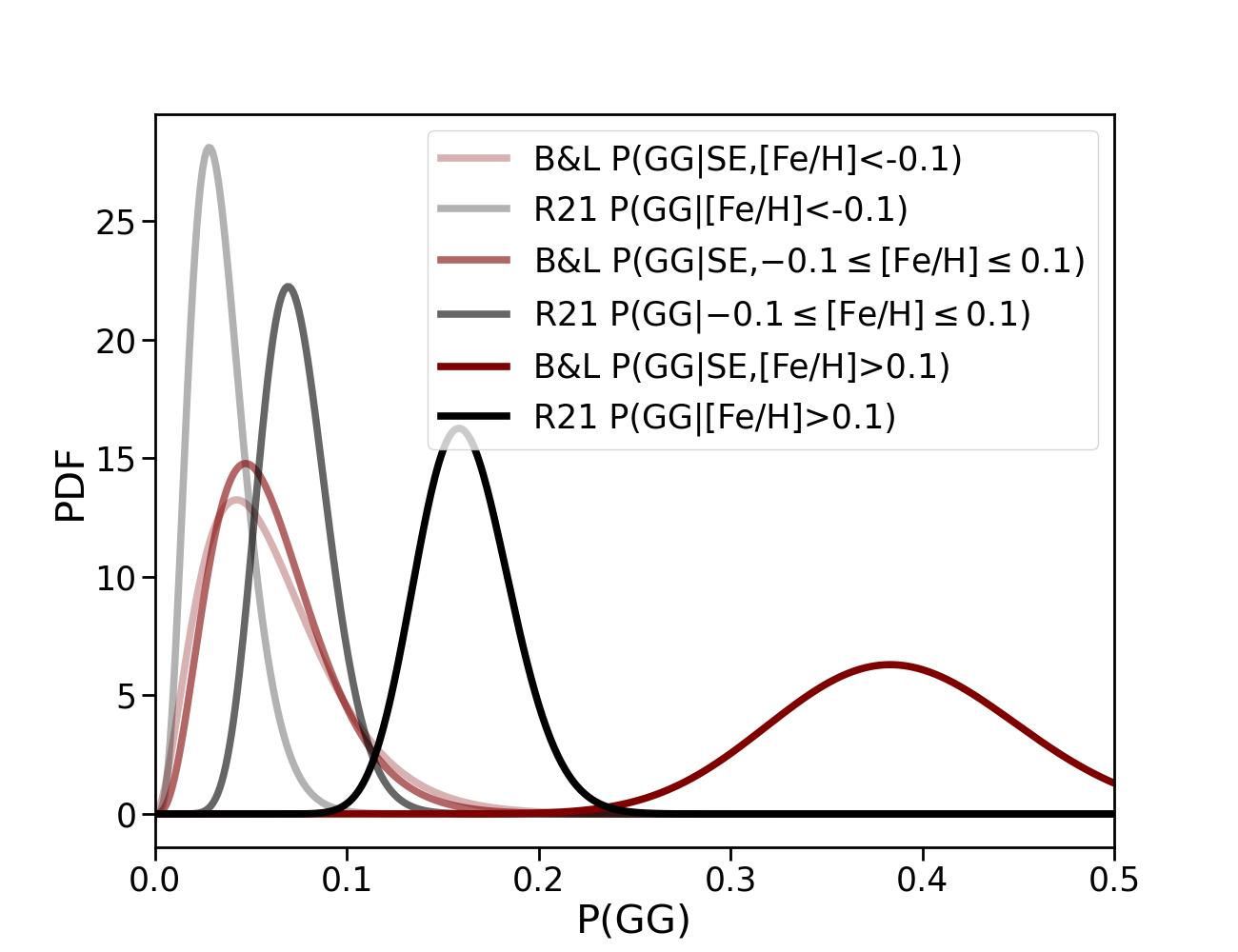} &
    \includegraphics[width=0.5\textwidth]{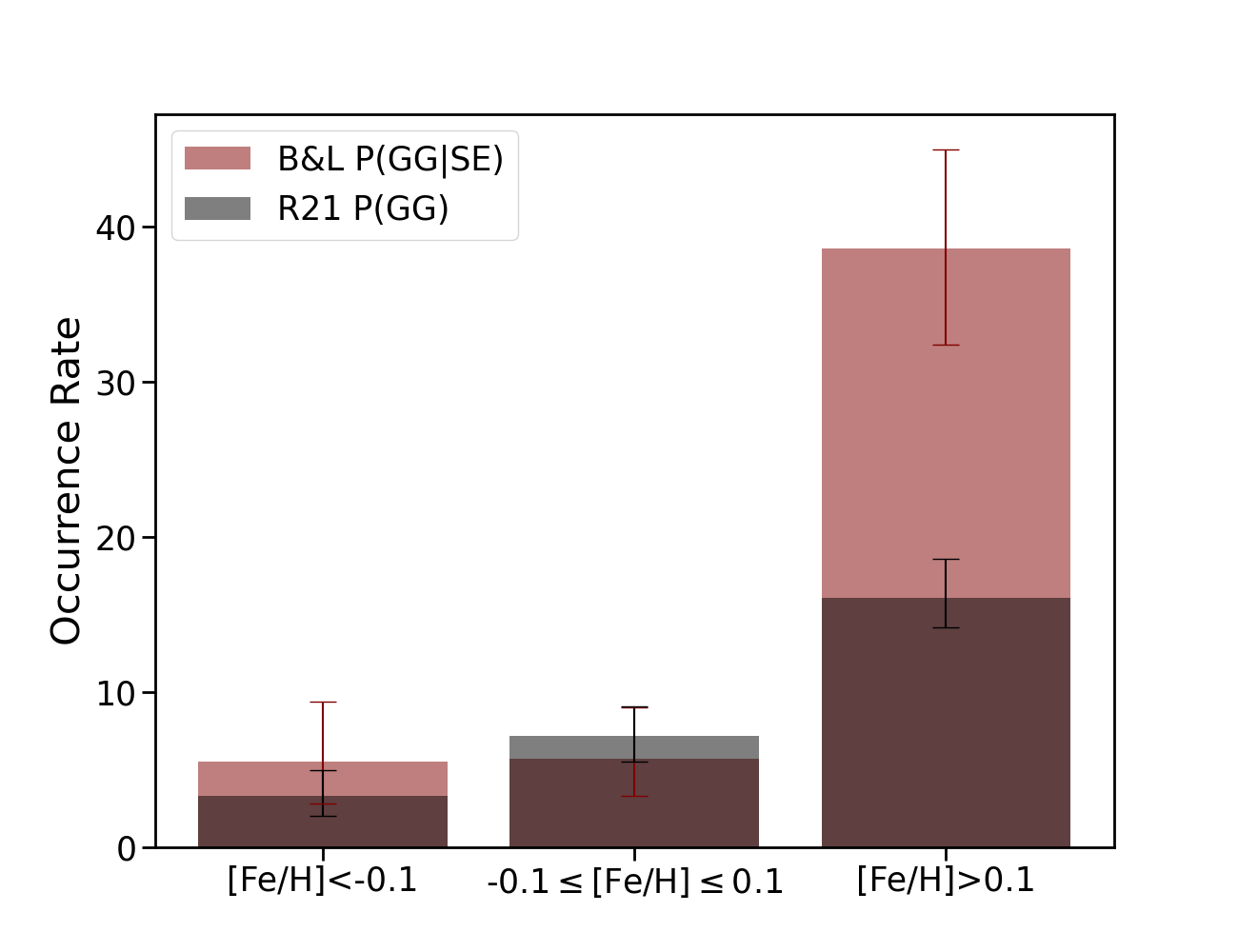}\\
    \end{tabular}

    \caption{We compare our occurrence rate P(GG|SE) to that of R21 P(GG) for three metallicity ranges: 1) [Fe/H] < -0.1; 2) -0.1 $\leq$ [Fe/H] $\leq$ 0.1 ; 3) [Fe/H] > 0.1. Left panel shows the posterior distributions, right panel shows those distributions represented as histogram bins. While the P(GG|SE) and P(GG) occurrence rates are consistent to $<$1$\sigma$ for the low and medium metallicity bins, the high metallicity bin has a significantly higher frequency of gas giants in super-Earth systems, greater than P(GG) by 3.4$\sigma$. The positive correlation gets even stronger when host star metallicities are limited to higher values.}
    \label{fig:occ_metal_res}
\end{figure*}

Given the size of our sample and that of R21, we break both samples up into three metallicity bins to explore the dependence of the correlation on metallicity further: 1) [Fe/H]$<$-0.1; 2) -0.1$\leq$[Fe/H]$\leq$0.1; 3) [Fe/H]$<$0.1. Figure \ref{fig:occ_metal_res} shows that while there is no significant correlation between super-Earths and GG around stars in the first two bins, the most metal-rich bin has an even stronger positive correlation between the two populations, at the 3.4$\sigma$ level.

\begin{deluxetable*}{lcccccc}
\tablecaption{Occurrence Rates}
\tablehead{
\colhead{Test} & \colhead{B$\&$L [Fe/H]$>$0} & \colhead{B$\&$L [Fe/H]$\leq$0} & \colhead{R21 [Fe/H]$>$0} & \colhead{R21 [Fe/H]$\leq$0} & \colhead{$\sigma$ [Fe/H]$>$0} & \colhead{$\sigma$ [Fe/H]$\leq$0}
}
\startdata
(1) Default & 28.0 (+4.9 -4.6)$\%$ & 4.5 (+2.6 -1.9)$\%$ & 14.3 (+2.0 -1.8)$\%$ & 5.0 (+1.6 -1.3)$\%$ & 2.7$\sigma$ & -0.2$\sigma$\\
(2) Close GG & 19.2 (+4.3 -3.9)$\%$ & 3.1 (+2.2 -1.5)$\%$ & 9.5 (+1.7 -1.5)$\%$ & 3.8 (+1.4 -1.1)$\%$ & 2.3$\sigma$ & -0.3$\sigma$\\
(3) Distant GG & 18.3 (+4.4 -3.9)$\%$ & 2.1 (+2.0 -1.2)$\%$ & 7.4 (+1.5 -1.3)$\%$ & 2.5 (+1.2 -0.9)$\%$ & 2.7$\sigma$ & -0.2$\sigma$\\
(4) Dynamically Hot & 20.1 (+4.5 -4.0)$\%$ &2.0 (+1.9 -1.2)$\%$ & 7.1 (+1.5 -1.3)$\%$ & 1.6 (+1.0 -0.7)$\%$ & 3.0$\sigma$ & 0.3$\sigma$\\
(5) Dynamically Cold & 12.1 (+3.7 -3.2)$\%$ & 3.2 (+2.3 -1.6)$\%$ & 7.4 (+1.5 -1.3)$\%$ & 2.9 (+1.2 -1.0)$\%$ & 1.3$\sigma$ & 0.2$\sigma$\\
(6) Single GG & 17.3 (+4.0 -3.6)$\%$ & 4.1 (+2.4 -1.7)$\%$ & 10.4 (+1.7 -1.5)$\%$ & 3.1 (+1.2 -1.0)$\%$ & 1.7$\sigma$ & 0.5$\sigma$ \\
(7) Multi GG & 8.0 (+3.0 -2.5)$\%$ & $<$2.0$\%$ & 2.2 (+0.9 -0.7)$\%$ & 0.7 (+0.7 -0.4)$\%$ & 2.2$\sigma$ & -0.3$\sigma$
\enddata
\tablecomments{(1) Default: GG 1--10AU, 0.5--20M$_{\rm Jup}$; (2) Close GG: 0.3--3AU, 0.5--20M$_{\rm Jup}$; (3) Distant GG:  3--10AU, 0.5--20M$_{\rm Jup}$; (4) Dynamically Hot: default GG with eccentricities $>$0.2; (5) Dynamically Cold: default GG with eccentricities $<$0.2; (6) Single GG: systems with a single gas giant in the default GG range; (7) Multi GG: systems with multiple gas giants in the default GG range. Median and 68$\%$ confidence intervals are quoted.}
\label{table: occ rates}

\end{deluxetable*}

\section{Effect of Gas Giant Properties on the Correlation}

We now break up our sample into gas giant orbital distance, eccentricity, and multiplicity to determine the role the outer planet plays in shaping the architecture of planetary systems.

First we consider the impact of gas giant orbital distance on the SE/GG connection. We divide our systems into two samples, those with `close' gas giants, and those with `distant' gas giants. We experiment with four definitions of `close' and `distant': 1) 0.1-3AU and 3-10AU; 2) 0.1-2AU and 2-10AU; 3) 0.3-2AU and 2-10AU; and 4) 0.3-3AU and 3-10AU. This fourth definition matches that of \citet{Rosenthal2022}, which explored the impact of gas giant location on the SE/GG correlation with a significantly smaller sample size of super-Earth systems. We divide the R21 sample in the same way. For each of these four distance combinations, we calculate the frequencies of closer and further gas giants as a function of metallicity. Note that for both our sample and R21 we treat multi-gas giant systems the same, counting systems twice if they have gas giants in both bins. For all four sets of semi-major axis ranges for `close' and `distant' gas giants, we find that the positive correlation between super-Earths and close gas giants around metal-rich stars P(GG$|$SE, [Fe/H]$>$0) $>$ P(GG$|$[Fe/H]$>$0) is slightly weaker than the positive correlation between super-Earths and distant gas giants by 0.2-0.4$\sigma$, but still significantly positive. This runs counter to the tentative findings from \citet{Rosenthal2022} that suggest close gas giants suppress super-Earth formation in their small sample of 28 super-Earth systems. We additionally find that we see no significant SE/GG correlation around metal-poor hosts regardless of gas giant semi-major axis.

How might evidence of a dynamically hot or cold history imprint on the SE/GG connection? We divide our sample into systems that have high eccentricity gas giants ($e >$ 0.2), and systems with gas giants that have low eccentricities ($e <$ 0.2). For the eccentric gas giant case around metal-rich stars, we see a slightly stronger strength of the positive correlation, P(GG$|$SE, [Fe/H]$>$0) = 20.1 (+4.5 -4.0)$\%$ $>$ P(GG$|$[Fe/H]$>$0) = 7.1 (+1.5 -1.3)$\%$ at the 3.0$\sigma$ level. However, for dynamically cold metal-rich systems this positive correlation substantially diminishes, dropping to a significance of 1.3$\sigma$ (Table \ref{table: occ rates}). True to form, metal-poor systems do not exhibit any significant correlation for either dynamically hot or cold systems.

\begin{figure*}
\begin{tabular}{cc}
\includegraphics[width=0.5\textwidth]{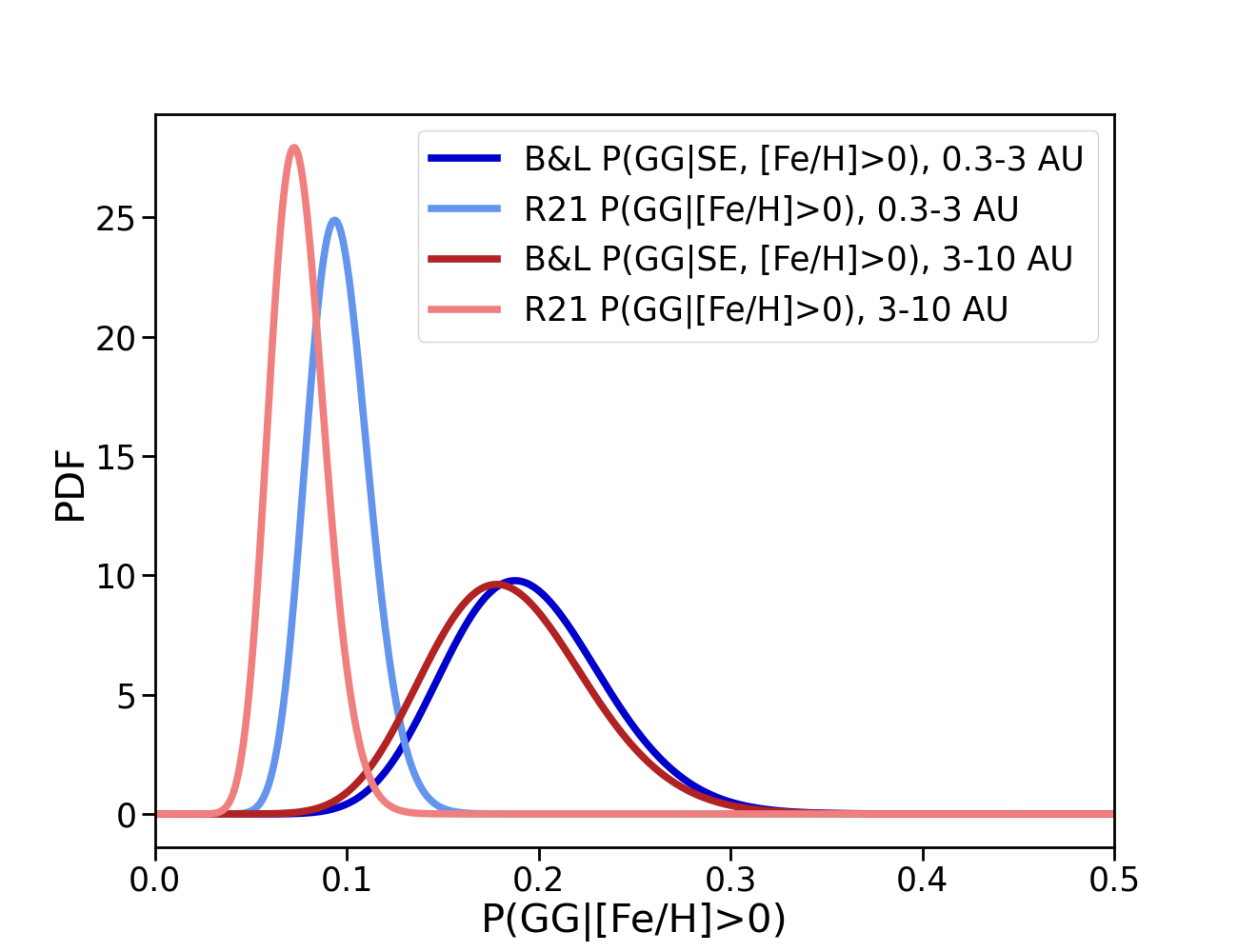} &
\includegraphics[width=0.5\textwidth]{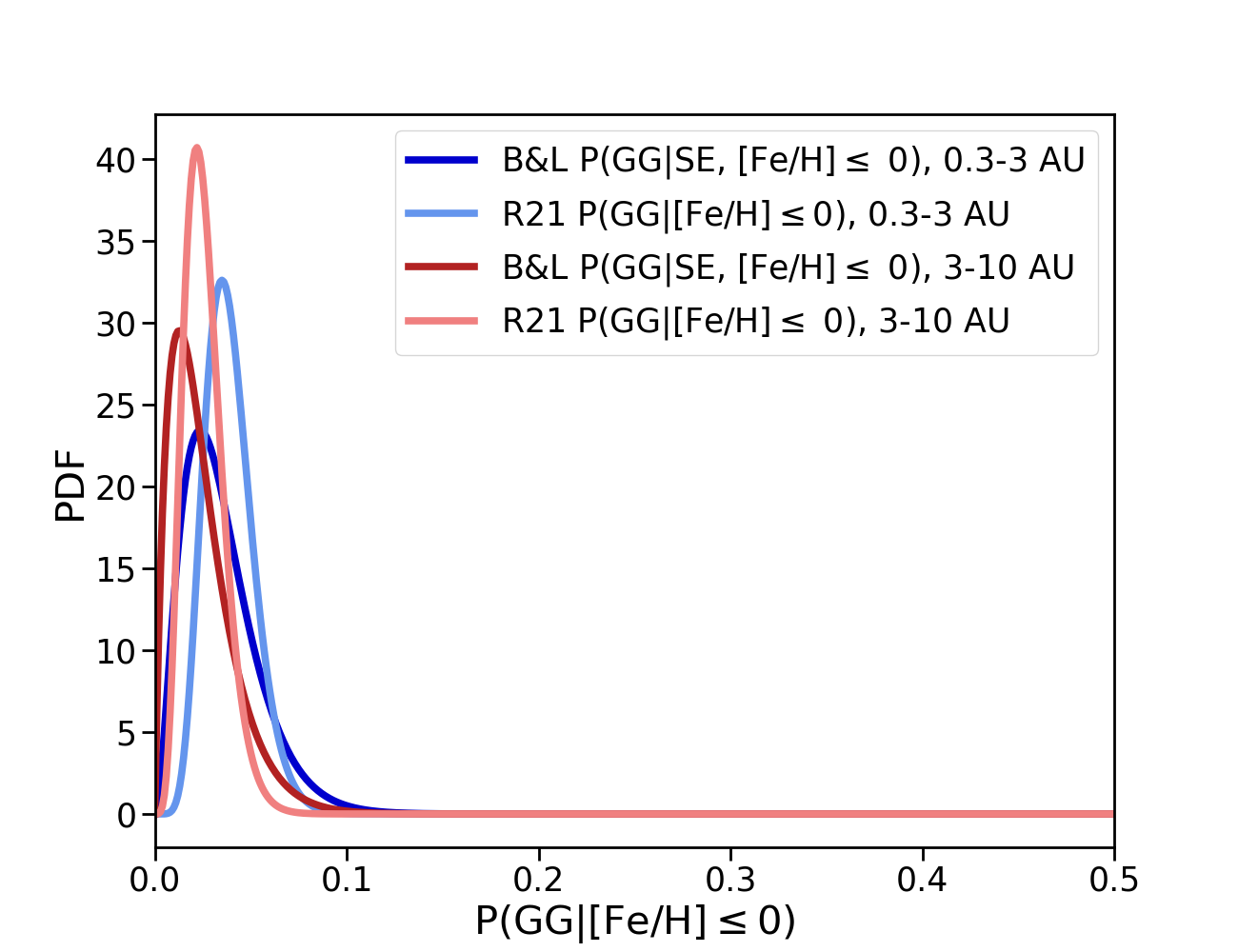} \\
\includegraphics[width=0.5\textwidth]{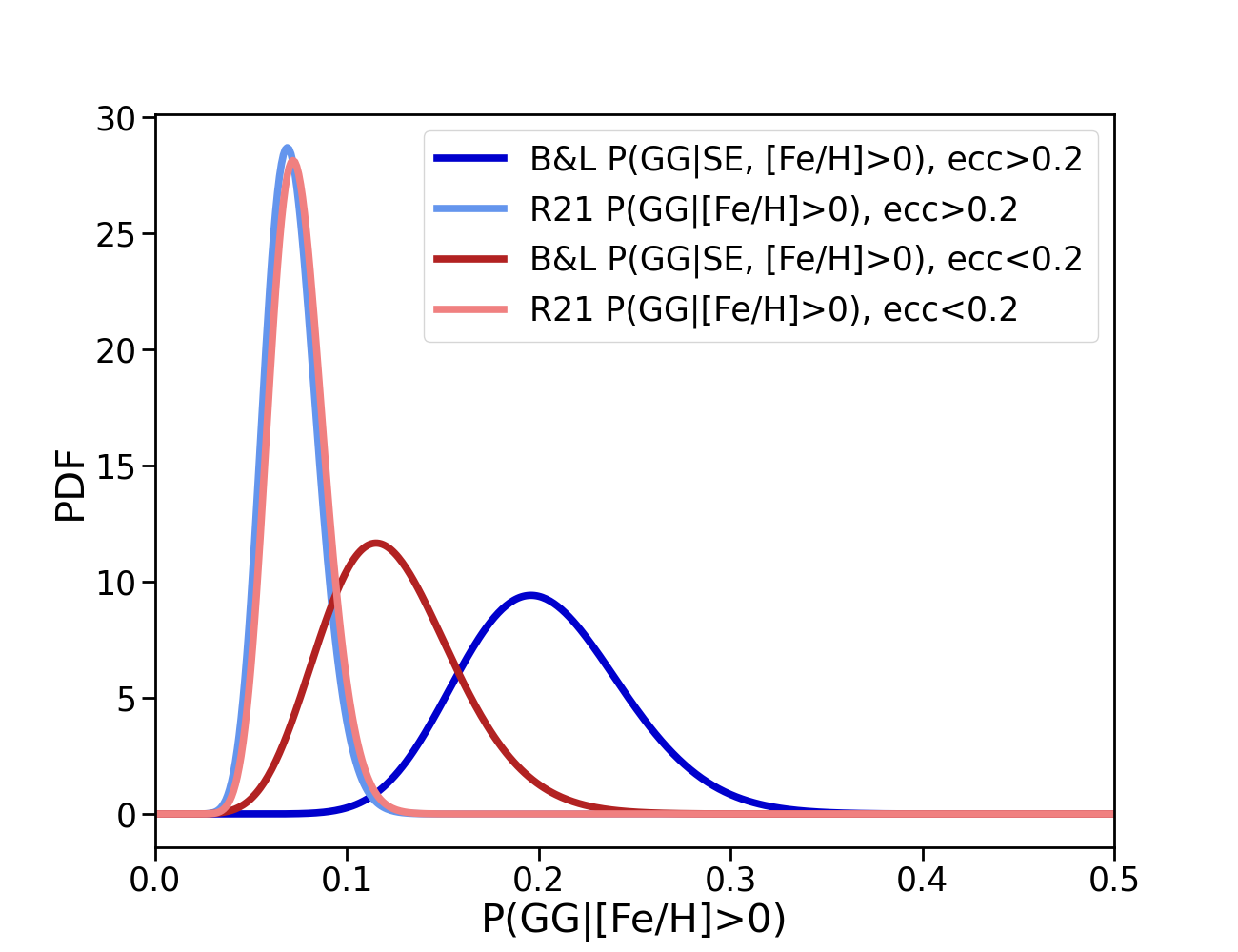} &
\includegraphics[width=0.5\textwidth]{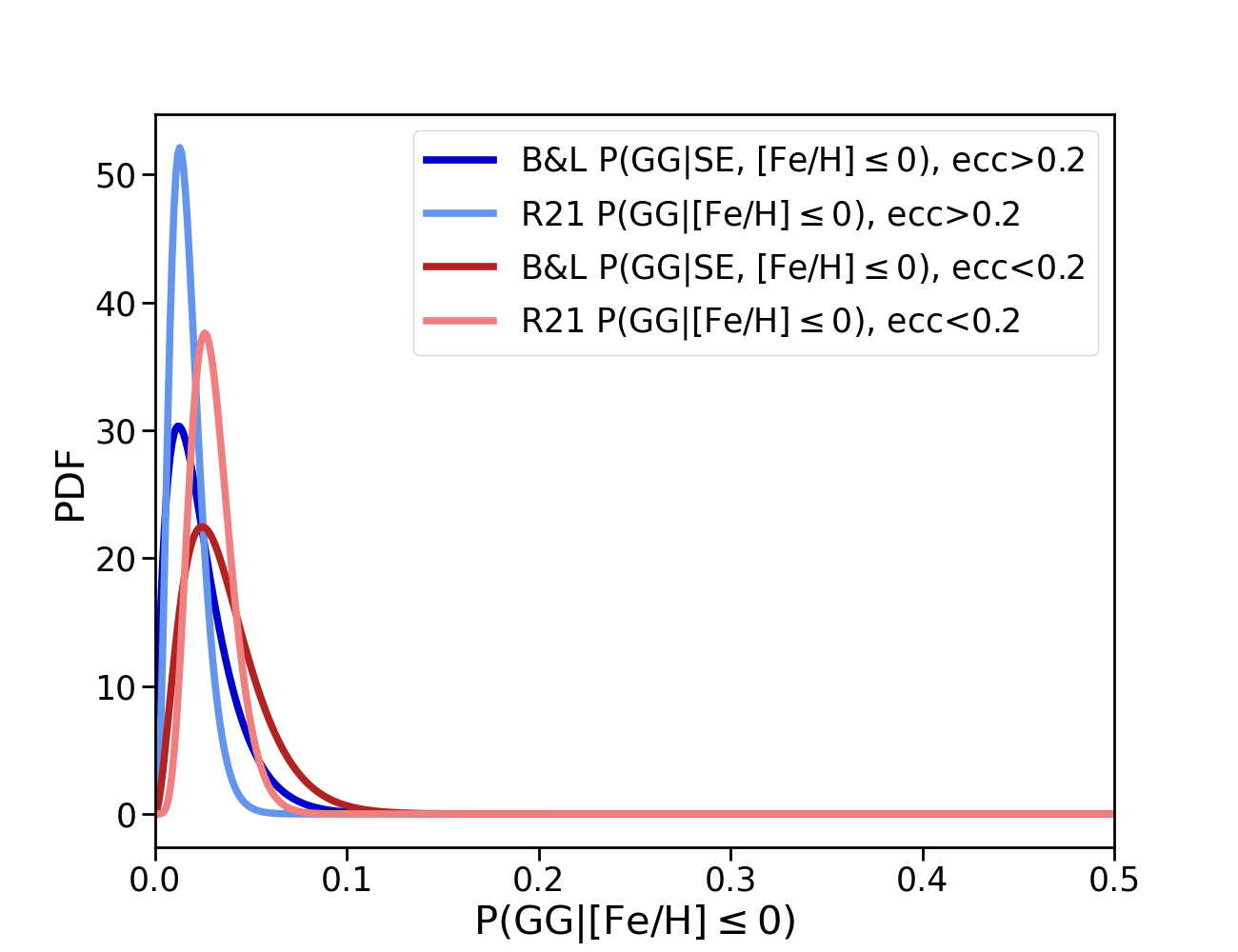} \\
\includegraphics[width=0.5\textwidth]{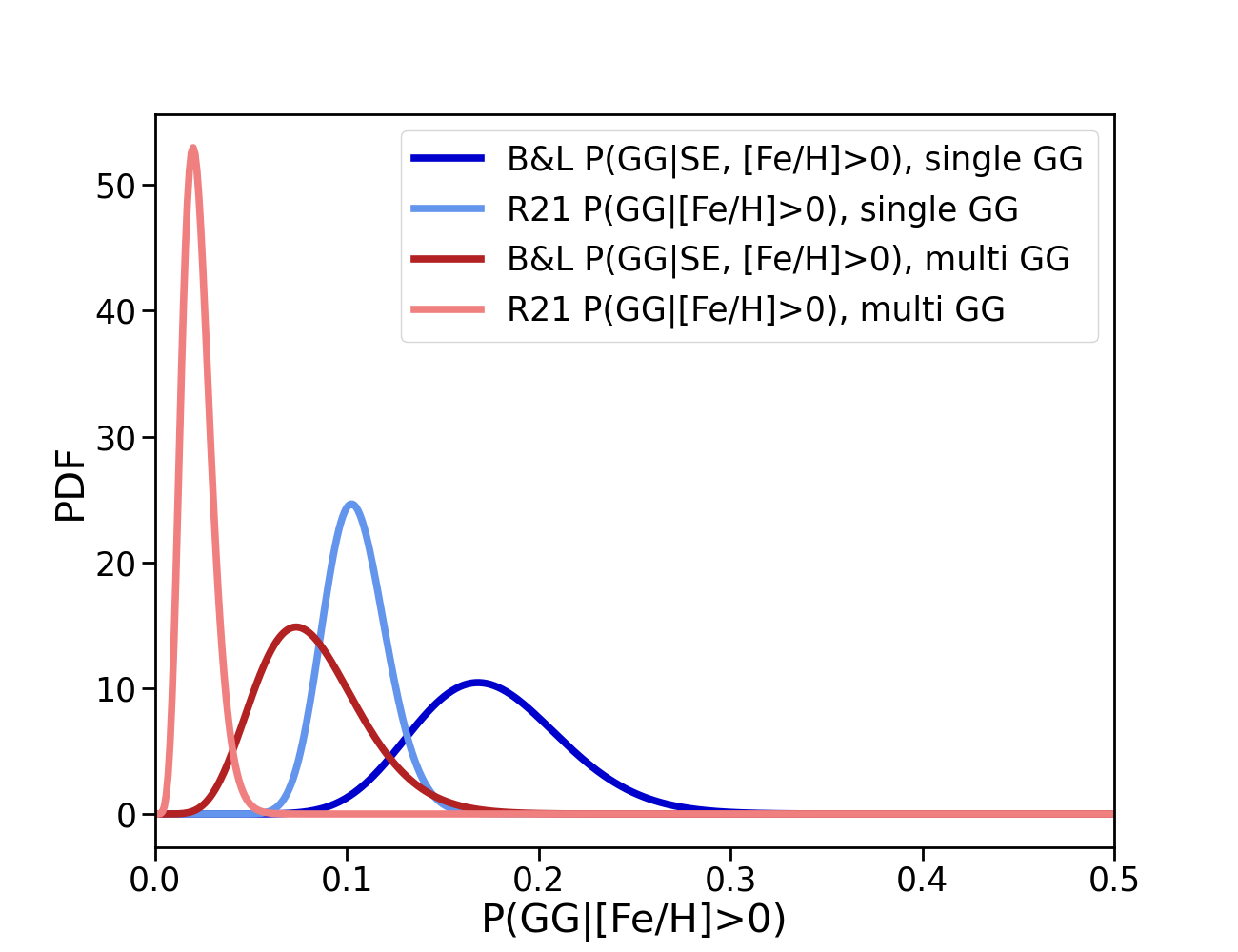} &
\includegraphics[width=0.5\textwidth]{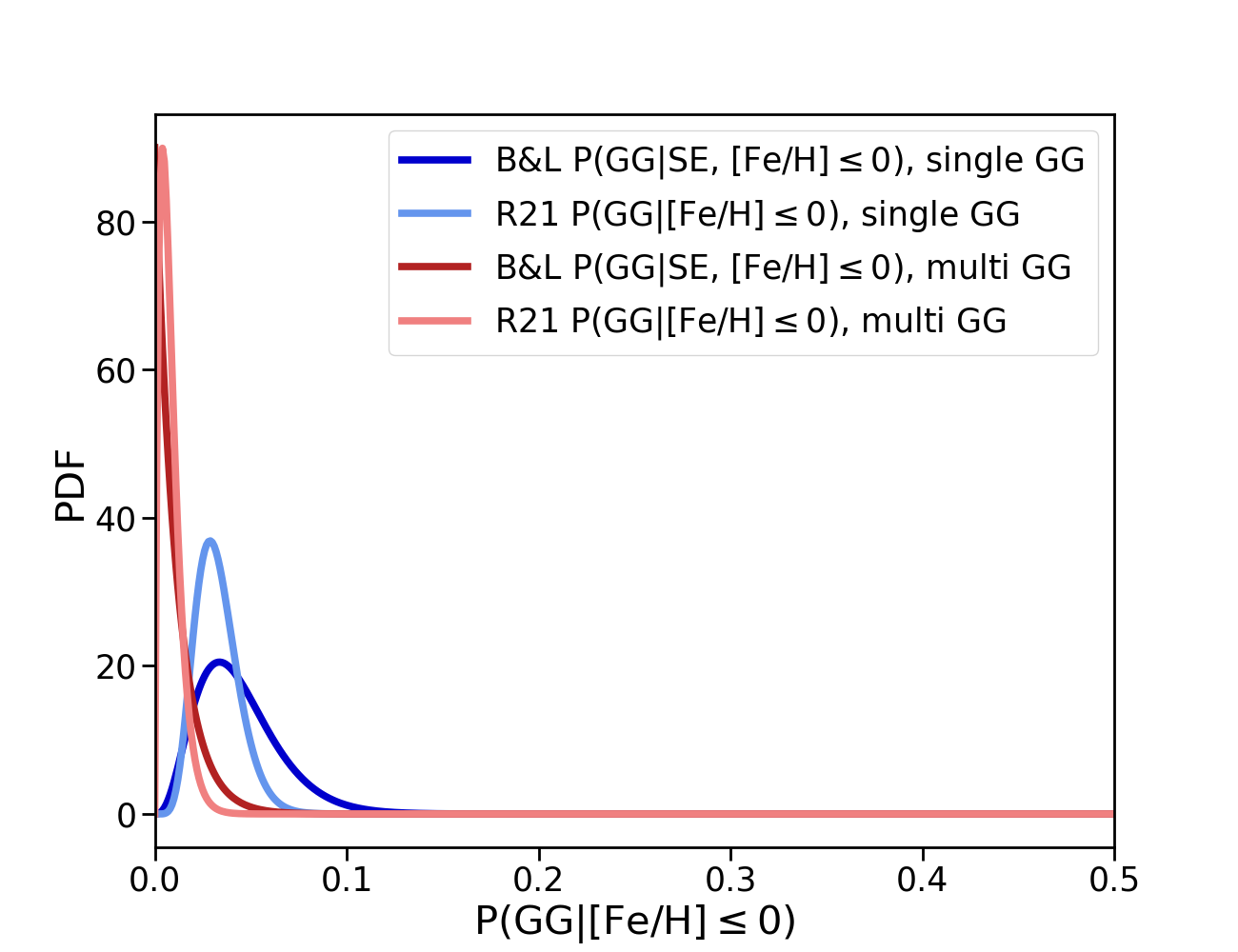} \\ 
\end{tabular}
\caption{Comparison of conditional occurrence rates P(GG|SE) (labeled B\&L) with field giant occurrence rates P(GG) (labeled R21) for metal-rich stars (left column) and metal-poor stars (right column). Top row: occurrence rates for close giants (0.3--3 AU) vs.~distant giants (3--10 AU). Middle row: occurrence rates for dynamically hot giants ($e > 0.2$) vs.~dynamically cold giants ($e < 0.2$). Bottom row: occurrence rates for systems of single giants vs.~multiple giants. In general, the GG/SE correlation enhances in systems with distant, eccentric, and/or multiple giants around metal-rich stars. Giant multiplicity impacts the level of correlation around metal-poor stars where we see a slight anti-correlation.
}
\label{fig: occ trends}
\end{figure*}

We next explore trends with gas giant multiplicity. Our sample has an average gas giant multiplicity of 1.3, while the R21 sample has an average multiplicity of 1.1. We break gas giant companions into single and multiple gas giant systems, where all gas giants considered fall in the default gas giant parameter space (1$-$10 AU, 0.5$-$20 M$_{\rm Jup}$). Table \ref{table: occ rates} and Figure \ref{fig: occ trends} show that for metal rich systems, the positive correlation gets slightly stronger when there are multiple gas giants (going from 1.7$\sigma$ for single GG to 2.2$\sigma$ for multi GG), whereas the correlation gets smaller and possibly becomes an anti-correlation for metal-poor systems with multiple gas giants (going from 0.5$\sigma$ for single GG systems to -0.3$\sigma$ for multi-GG systems.

\begin{figure}
    \centering
    \includegraphics[width=0.5\textwidth]{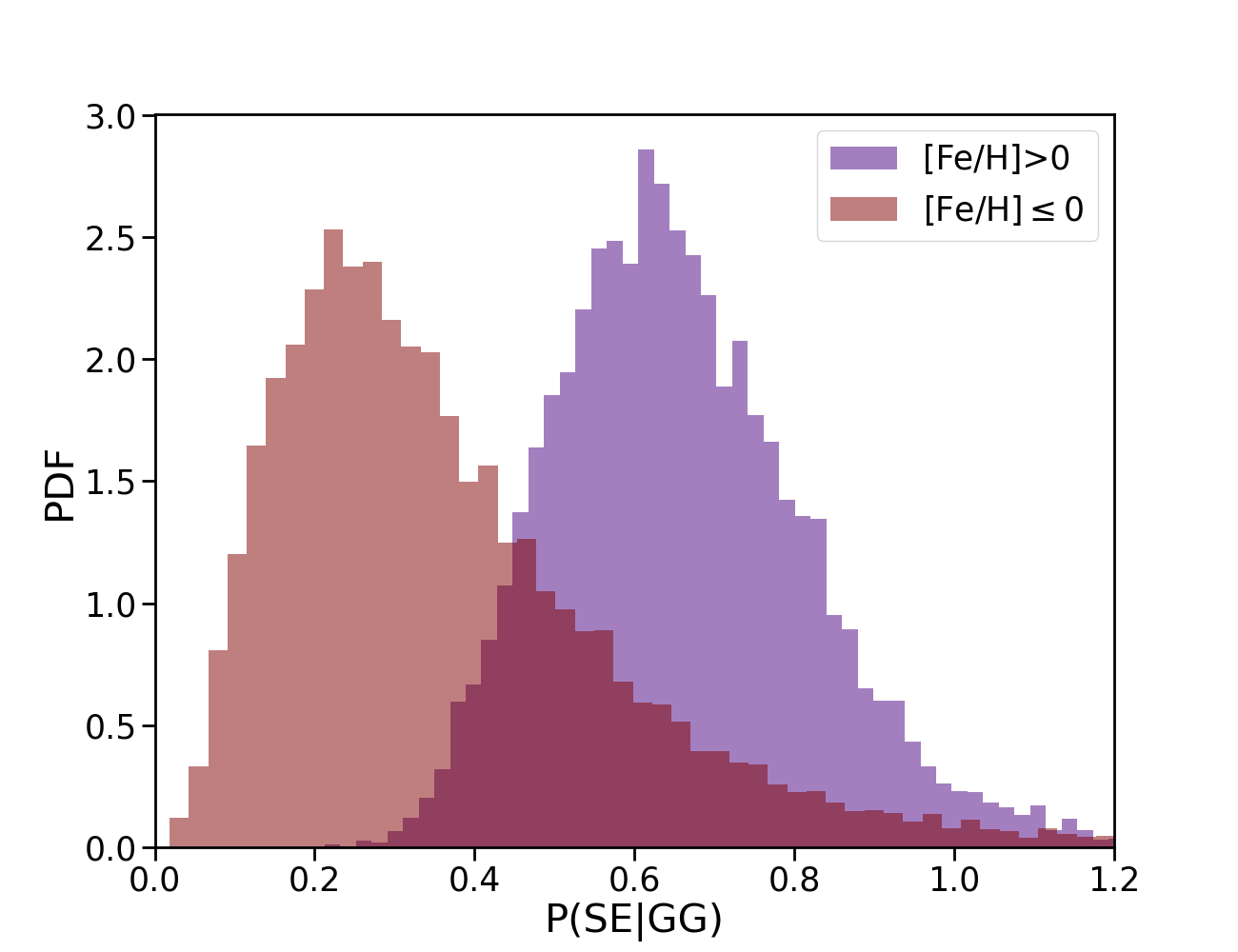}
    \caption{Comparing the inverse probability of super-Earths in systems hosting cold gas giants P(SE|GG) in metal-rich (maroon) and metal-poor (purple) systems. We find that while more than half of stars with [Fe/H]$>$0 hosting cold gas giants should also host a super-Earth, metal poor stars with gas giants may be less likely to host an inner super-Earth than other stars (P(SE)$>$P(SE$|$GG, [Fe/H]$\leq$0)).}
    \label{fig:occ_rev}
\end{figure}

Finally, we constrain the inverse conditional probability linking super-Earths and gas giants. Namely, given a cold gas giant, what is the probability of finding an inner super-Earth P(SE$|$GG)? We can estimate this using the following:

\begin{equation}
    \rm P(SE|GG) = \frac{P(GG|SE)P(SE)}{P(GG)}.
\end{equation}
We take the calculated P(GG$|$SE) and P(GG) for metal-rich and metal-poor stars as shown in Table \ref{table: occ rates}. For P(SE) we take P(SE) = 30$\pm$3$\%$ from \citet{Zhu2018b}. Around metal-rich stars, P(SE$|$GG, [Fe/H]$>$0) = 61.2 (+18.4 -12.7)$\%$, and for their metal-poor counterparts, P(SE$|$GG, [Fe/H]$\leq$0) = 21.6 (+26.3 -8.7)$\%$ (Figure \ref{fig:occ_rev}). While more than half of metal rich stars hosting cold Jupiters should also host super-Earths, metal poor stars with gas giants might be less likely to host an inner super-Earth than other stars (P(SE)$>$P(SE$|$GG, [Fe/H]$\leq$0)).

\begin{figure}
    \centering
    \includegraphics[width=0.5\textwidth]{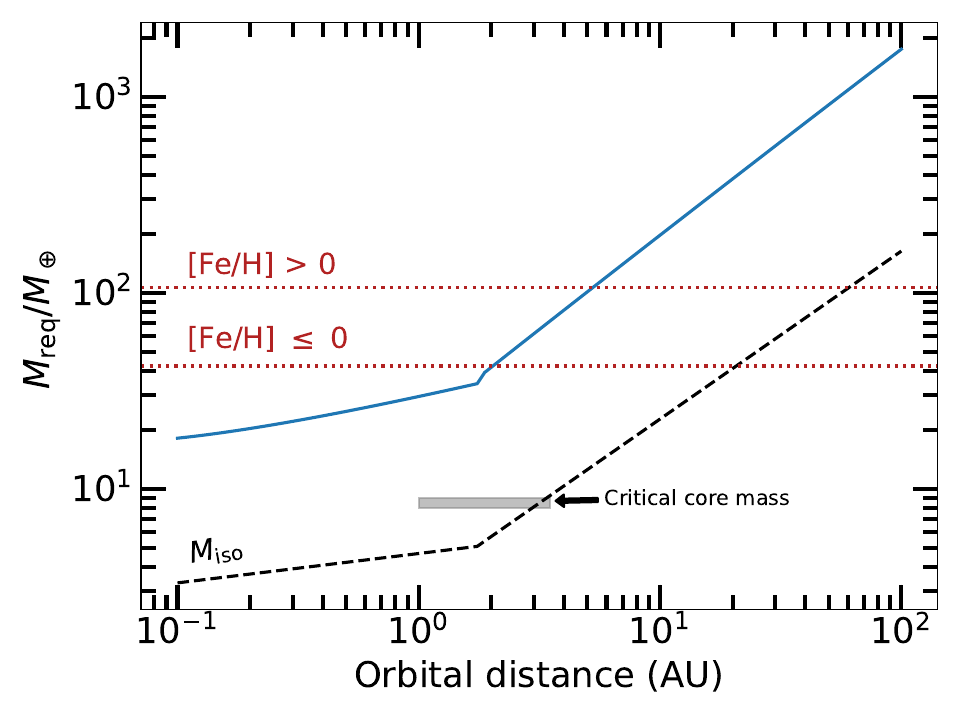}
    \caption{The required dust mass ($M_{\rm req}$; blue solid line) to create the pebble isolation mass (black dashed line) at each orbital distance following the calculation of \citet{Chachan23} adopting 1 $M_\odot$ host star, $\alpha = 10^{-3}$, $\dot{M}=10^{-8}M_\odot\,{\rm yr}^{-1}$, and the fragmentation velocity of 50 ${\rm m\,s}^{-1}$ from \citet{Kimura20}. The total disk mass for metal-poor system ([Fe/H] $\leq$ 0) and metal-rich system ([Fe/H] $>$ 0) are annotated with red dotted lines. The former value is taken as the median initial disk dust mass 42.33$M_\oplus$ for 0.6--1.4$M_\odot$ host stars \citep[see][their Figure 4]{Chachan23} while the latter value is evaluated by multiplying 42.33$M_\oplus$ by $10^{0.4}$. Planetary cores of $M_{\rm iso}$ can nucleate in regions where the total disk dust mass is larger than $M_{\rm req}$ so that the spatial extent to which disk can generate planetary cores enlarges in metal-rich systems. Under this setup, $M_{\rm iso}$ core exceeds the critical core mass (gray box; \citealt{Savignac23}) and so is expected to trigger runaway gas accretion and become a giant at $\gtrsim$3 AU.}
    \label{fig:Mreq-orbA}
\end{figure}

\section{Discussion and Conclusion}

Using the largest sample size of systems to-date with both inner super-Earths and outer gas giants, we verify the positive correlation between the two populations that were previously reported by \citet{ZhuWu18} and \citet{Bryan2019}. In particular, our results demonstrate that the correlation is particularly strong for metal-rich ([Fe/H] $>$ 0) systems with no statistically significant correlation around metal-poor ([Fe/H] $\leq$ 0) stars, verifying the argument of \citet{Zhu2023} and highlighting the importance of system metallicity in shaping planetary architecture.

By combining the theory of core coagulation by pebble accretion and the observations of dust masses in Class 0/I disks, \citet{Chachan23} showed that the positive GG/SE correlation is naturally expected because disks that are massive enough to nucleate an outer planet will have enough mass to also create the inner planets, a consequence of low planet formation efficiency at large orbital separations \citep[see also][]{Lin18, Chachan22}. Furthermore, the inner cores will coagulate before all the filtered solids drift in towards the inner edge of the protoplanetary disk \citep[see][their Figure 5, bottom panel]{Chachan23}.\footnote{The coagulation timescale being shorter than the drift time arises from the consideration of the material strength of silicate and icy grains being identical, consistent with modern lab measurements \citep[e.g.,][]{Musiolik19,Kimura20} and the ``No Snow Line'' model of \citet{Mulders21}.}

In Figure \ref{fig:Mreq-orbA}, we illustrate the decisive role system metallicity---which we trace with the total initial disk dust mass---plays in establishing inner-outer planet correlation. We reproduce the required dust mass $M_{\rm req}$ to coagulate a planetary core of pebble isolation mass $M_{\rm iso}$ at each orbital distance as outlined in \citet{Chachan23} for 1$M_\odot$ host star, turbulent $\alpha=10^{-3}$ and disk gas accretion rate $10^{-8}\,M_\odot\,{\rm yr}^{-1}$. We further adopt grain fragmentation velocity of 50 m/s following \citet[][their Figure 13; the result also holds for lower fragmentation velocities used in the original work of \citet{Chachan23}]{Kimura20}. Comparing with the critical core mass to trigger runaway gas accretion computed in \citet{Savignac23} over 1--3.5 AU (under dusty gas accretion), we find that the outer planetary cores will likely nucleate gas giants beyond $\sim$3 AU. Furthermore, and most importantly, with higher metallicity (i.e., larger solid reservoir), the spatial extent over which the disk can generate planetary cores enlarges, especially towards wider orbital separations where larger isolation mass and lower planet formation efficiency enforce larger required dust mass. This larger zone of core coagulation and the creation of giant-nucleating massive cores beyond $\sim$3 AU around higher metallicity systems therefore explains concurrently the positive correlation between inner super-Earths and outer giants around stars of [Fe/H] $>$ 0 and the slight enhancement of this correlation for giants that are farther away. The existence of giants at closer-in distances and their similarly positive correlation with the inner planets around metal-rich stars could be explained by a variation in the critical core mass owing to different opacity sources \citep[see, e.g.,][]{Piso15,Savignac23}.

In the present paper, we find a slight difference in the strength of inner-outer correlation for systems with single vs.~multiple outer giants, weakly favoring the latter. We can see in Figure \ref{fig:Mreq-orbA} that at $\sim$3 AU, to create $\sim$10$M_\oplus$ core, the total dust mass in the disk needs to be $\sim$100$M_\oplus$ and that the inner cores can be created as long as $\sim$20--30$M_\oplus$ worth of dust can be filtered through. Given the average outer giant multiplicity $\sim$1.1--1.3, the number of outer giants is likely at most 2 and so there will be more than enough dust that will drift to the inner region, consistent with the similar level of inner-outer correlation with respect to outer giant multiplicity. In low metallicity systems however, the total mass budget would be quickly depleted with the formation of multiple cores which is in line with the observed anti-correlation around [Fe/H] $\leq$ 0 stars. 

An intriguing result from our analysis is the strong inner-outer correlation for gas giants with high eccentricity ($e > 0.2$). While the observed affinity for super-Earth systems to be neighbored by eccentric outer giant(s) could be a reflection of the tendency for metal-rich systems to harbor eccentric giants \citep[see, e.g.,][for Jupiters inside 1 AU]{Dawson13}, we do not see the latter tendency in our sample, nor in R21 sample, which suggests that the metallicity-eccentricity relation is confined to warm and hot Jupiters. It may be that the more eccentric outer Jupiter aids the formation of inner super-Earths by facilitating core coagulation via orbital instabilities and merger events. Such a hypothesis will be a subject of Paper II where we will explore how the properties of the inner planets (e.g., radius, mass, multiplicity, eccentricities/inclinations, and the innermost orbital period) vary in systems with and without outer giants.

\vspace{0.5cm}
We thank Eugene Chiang, Heather Knutson, Yanqin Wu, Wei Zhu, and the referee for helpful conversations and feedback. M.L.B. acknowledges support by NSERC, the Heising-Simons Foundation, and by the Connaught New Researcher Award from the University of Toronto.  E.J.L. gratefully acknowledges support by NSERC, by FRQNT, by the Trottier Space Institute, and by the William Dawson Scholarship from McGill University.

\appendix
\counterwithin{figure}{section}
\counterwithin{table}{section}

\section{Full sample list and average sensitivities}

\startlongtable
\begin{deluxetable*}{lcccccccc}
\tabletypesize{\scriptsize}
\tablecaption{Sample of systems}
\tablewidth{0pt}
\tablehead{
\colhead{Target} & \colhead{M$_{\star}$ (M$_{\odot}$)} & \colhead{[Fe/H]} & \colhead{N$_{\rm pl}$} & \colhead{Disc. Method} & \colhead{N$_{obs}$} & \colhead{Baseline (yrs)} & \colhead{GG Companion}  & \colhead{Ref.}
}
\startdata
CoRoT-24 & 0.91$\pm$0.09 & 0.30$\pm$0.15 & 2 & Transit & 71 & 3.2 & No   & 1 \\
EPIC 220674823 & 0.95 (+0.06 -0.05) & 0.11$\pm$0.05 & 2 & Transit & 101 & 3.4 & No & 2\\
EPIC 229004835 & 0.97$\pm$0.04 & -0.12$\pm$0.05 & 1 & Transit & 126 & 2.3 & No & 2 \\
EPIC 249893012 & 1.05$\pm$0.05 & 0.20$\pm$0.05 & 3 & Transit & 99 & 1.4 & No & 3\\
GJ 143 & 0.73$\pm$0.07 & 0.00$\pm$0.06 & 2 & Transit & 112 & 15.2 & No & 4\\
GJ 9827 & 0.61$\pm$0.01 & -0.26$\pm$0.09 & 3 & Transit & 128 & 10.8 & No & 2\\
HD 137496 & 1.04$\pm$0.02 & -0.03$\pm$0.04 & 2 & Transit & 172 & 2.5 & Yes & 5\\
HD 15337 & 0.90$\pm$0.03 & 0.15$\pm$0.10 & 2 & Transit & 87 & 13.7 & No & 6\tablenotemark{1} \\
HD 183579 & 1.03$\pm$0.05 & -0.07$\pm$0.09 & 1 & Transit & 56 & 7.9 & No & 7 \\
HD 18599 & 0.86$\pm$0.02 & 0$\pm$0.10 & 1 & Transit & 103 & 4.8 & No & 8\\
HD 191939 & 0.81$\pm$0.04 & -0.15$\pm$0.06 & 6 & Transit & 182 & 1.3 & Yes & 9\tablenotemark{2} \\
HD 207897 & 0.80$\pm$0.04 & -0.04$\pm$0.04 & 1 & Transit & 122 & 17.7 & No & 11 \\
HD 213885 & 1.07$\pm$0.02 & -0.04$\pm$0.03 & 2 & Transit & 238 & 10.2 & No & 46\\
HD 23472 & 0.67$\pm$0.03 & -0.20$\pm$0.05 & 5 & Transit & 104 & 1.7 & No & 12 \\
HD 307842 & 0.91$\pm$0.10 & -0.13$\pm$0.08 & 1 & Transit & 58 & 2.1 & No & 13 \\
HD 3167 & 0.84 (+0.05 -0.04) & 0.04$\pm$0.05 & 4 & Transit & 452 & 5.3 & No & 2 \\
HD 39091 & 1.07$\pm$0.04 & 0.09$\pm$0.04 & 3 & Transit & 402 & 2.2 & Yes & 45\\
HD 5278 & 1.13$\pm$0.04 & -0.12$\pm$0.04 & 2 & Transit & 41 & 1.1 & No & 14 \\
HD 73583 & 0.73$\pm$0.02 & 0$\pm$0.09 & 2 & Transit & 118 & 3.5 & No & 15 \\
HD 80653 & 1.18$\pm$0.04 & 0.26$\pm$0.07 & 2& Transit & 208 & 2.9 & Yes & 2\\
HD 86226 & 1.02$\pm$0.06 & 0.02 (+0.06 -0.04) & 2 & Transit & 132 & 10.1 & No & 16\\
HIP 116454 & 0.76$\pm$0.03 & -0.16$\pm$0.08 & 1 & Transit & 108 & 6.5 & No & 2\\
HIP 29442 & 0.89$\pm$0.04 & 0.24$\pm$0.05 & 3 & Transit & 83 & 2.5 & No & 85\\
HIP 9618 & 1.02 (+0.04 -0.08) & -0.08$\pm$0.10\tablenotemark{3} & 2 & Transit & 36 & 10.1 & No & 17\\
K2-100 & 1.15$\pm$0.05 & 0.22$\pm$0.09 & 1 & Transit & 78 & 1.3 & No & 39 \\
K2-110 & 0.74$\pm$0.02 & -0.34$\pm$0.03 & 1 & Transit & 32 & 4.0 & No & 2\\
K2-111 & 0.84$\pm$0.02 & -0.46$\pm$0.05 & 2 & Transit & 161 & 4.4 & No & 2\\
K2-12 & 0.96 (+0.06 -0.05) & 0$\pm$0.10 & 1 & Transit & 50 & 7.0 & No & 2 \\
K2-131 & 0.80$\pm$0.03 & -0.04$\pm$0.07 & 1 &Transit & 115 & 1.3 & No & 2\\
K2-136 & 0.74$\pm$0.04 & -0.02$\pm$0.08 & 3 & Transit & 115 & 2.2 & No & 40\\
K2-141 & 0.71$\pm$0.03 & 0$\pm$0.10 & 2 & Transit & 83 & 4.1 & No & 2\\
K2-167 & 1.01 (+0.08 -0.07) & -0.40$\pm$0.06 & 1 & Transit & 82 & 6.1 & No & 2\\
K2-19 & 0.88$\pm$0.03 & 0.06$\pm$0.05 & 3 & Transit & 51 & 2.9 & No & 18\\
K2-199 & 0.71$\pm$0.02 & -0.01$\pm$0.09 & 2 & Transit & 45 & 2.9 & No & 41\\
K2-216 & 0.70$\pm$0.03 & 0$\pm$0.10 & 1 & Transit & 30 & 1.3 & No & 19\\
K2-222 & 0.99$\pm$0.07 & -0.32$\pm$0.06 & 1 & Transit & 70 & 4.4 & No & 2\\
K2-263 & 0.88$\pm$0.03 & -0.08$\pm$0.05 & 1 & Transit & 95 & 4.2 & No & 2\\
K2-265 & 0.92$\pm$0.02 & 0.08$\pm$0.02 & 1 & Transit & 144 & 1.1 & No & 20\\
K2-32 & 0.83$\pm$0.02 & -0.06$\pm$0.03 & 4 & Transit & 168 & 3.2 & No & 21\\
K2-36 & 0.79$\pm$0.01& -0.03$\pm$0.08 & 2 & Transit & 86 & 5.1 & No & 2\\
K2-38 & 1.05 (+0.07 -0.06) & 0.24$\pm$0.07 & 2 & Transit & 83 & 3.4 & No & 2\\
K2-66 & 1.11$\pm$0.04 & -0.05$\pm$0.02 & 1 & Transit & 38 & 1.3 & No & 22\\
KOI-142 & 0.99$\pm$0.02 & 0.27$\pm$0.06 & 3 & Transit & 47 & 9.0 & Yes & 92\\
KOI-351 & 1.11$\pm$0.03 & 0.08$\pm$0.03 & 8 & Transit & 34 & 11.2 & Yes & 92\\
KOI-94 & 1.28$\pm$0.05 & 0.02$\pm$0.01 & 4 & Transit & 70 & 7.3 & No & 92\\
Kepler-10 & 0.91$\pm$0.02 & -0.15$\pm$0.04 & 3 & Transit & 291 & 11.0 & No & 2\\
Kepler-100 & 1.09$\pm$0.03 & 0.07 (+0.08 -0.05) & 4 & Transit & 112 & 13.4 & No & 92\\
Kepler-101 & 1.17 (+0.07 -0.05) & 0.33$\pm$0.07 & 2 & Transit & 40 & 1.2 & No & 42\\
Kepler-102 & 0.80$\pm$0.02 & 0.11$\pm$0.04 & 5 & Transit & 147 & 10.3 & No & 2\\
Kepler-103 & 1.21 (+0.02 -0.03) & 0.16$\pm$0.04 & 2 & Transit & 60 & 4.4 & No & 2\\
Kepler-104 & 0.82$\pm$0.03 & -0.38$\pm$0.10 & 3 & Transit & 44 & 11.3 & No & 92\\
Kepler-106 & 0.96$\pm$0.03 & -0.12$\pm$0.11 & 4 & Transit & 48 & 9.8 & No & 92\\
Kepler-107 & 1.24$\pm$0.03 & 0.32$\pm$0.07 & 4 & Transit & 121 & 6.0 & No & 2\\
Kepler-109 & 1.09 (+0.09 -0.08) & -0.02$\pm$0.07 & 2 & Transit & 66 & 10.0 & No & 2\\
Kepler-11 & 0.99$\pm$0.03 & 0.07$\pm$0.10 & 6 & Transit & 31 & 11.6 & No & 92\\
Kepler-113 & 0.79$\pm$0.02 & 0.05$\pm$0.07 & 2 & Transit & 42 & 12.0 & No & 92\\
Kepler-126 & 1.11$\pm$0.03 & -0.13$\pm$0.10 & 3 & Transit & 35 & 5.3 & No & 92\\
Kepler-129 & 1.24$\pm$0.04 & 0.29$\pm$0.10 & 3 & Transit & 35 & 7.8 & Yes & 92\\
Kepler-131 & 1.08$\pm$0.02 & 0.12$\pm$0.07 & 2 & Transit & 46 & 6.9 & No & 92\\
Kepler-139 & 1.08$\pm$0.03 & 0.27$\pm$0.10 & 4 & Transit & 38 & 11.8 & Yes & 92\\
Kepler-1655 & 1.03$\pm$0.04 & -0.24$\pm$0.05 & 1 & Transit & 97 & 4.3 & No & 2\\
Kepler-1710 & 0.93$\pm$0.02 & -0.07$\pm$0.10 & 1 & Transit & 21 & 12.2 & No & 92\\
Kepler-18 & 0.98$\pm$0.02 & 0.20$\pm$0.10 & 3 & Transit & 25 & 8.8 & No & 92\\
Kepler-19 & 0.90$\pm$0.02 & -0.13$\pm$0.06 & 3 & Transit & 73 & 9.3 & No & 92\\
Kepler-20 & 0.93$\pm$0.05 & 0.07$\pm$0.08 & 6 & Transit & 161 & 10.1 & No & 2\\
Kepler-21 & 1.41 (+0.02 -0.03) & -0.03$\pm$0.10 & 1 & Transit & 98 & 5.4 & No & 2\\
Kepler-22 & 0.86 (+0.05 -0.04) & -0.26$\pm$0.07 & 1 & Transit & 71 & 10.7 & No & 2\\
Kepler-25 & 1.15$\pm$0.03 & -0.04$\pm$0.10 & 3 & Transit & 99 & 11.3 & No & 92\\
Kepler-323 & 1.02$\pm$0.07 & -0.14$\pm$0.05 & 2 & Transit & 48 & 6.1 & No & 2\\
Kepler-36 & 1.03$\pm$0.04 & -0.18$\pm$0.04 & 2 & Transit & 25 & 9.2 & No & 92\\
Kepler-37 & 0.79$\pm$0.02 & -0.36$\pm$0.05 & 4 & Transit & 108 & 12.4 & No & 92\\
Kepler-406 & 1.06$\pm$0.03 & 0.18$\pm$0.07 & 2 & Transit & 58 & 11.9 & No & 92\\
Kepler-407 & 1.10$\pm$0.03 & 0.33$\pm$0.07 & 2 & Transit & 98 & 11.0 & Yes & 92\\
Kepler-409 & 0.94$\pm$0.02 & 0.05$\pm$0.07 & 1 & Transit & 97 & 10.7 & No & 92\\
Kepler-454 & 1.03 (+0.04 -0.03) & 0.32$\pm$0.08 & 3 & Transit & 147 & 12.1 & Yes & 2\\
Kepler-48 & 0.92$\pm$0.02 & 0.17$\pm$0.07 & 4 & Transit & 59 & 12.8 & Yes & 92\\
Kepler-50 & 1.15$\pm$0.04 & -0.04$\pm$0.10 & 2 & Transit & 39 & 9.8 & No & 92\\
Kepler-507 & 1.14$\pm$0.03 & 0.16$\pm$0.10 & 1 & Transit & 49 & 10.8 & No & 92\\
Kepler-538 & 0.89 (+0.05 -0.04) & -0.09$\pm$0.07 & 1 & Transit & 111 & 9.1 & No & 2\\
Kepler-65 & 1.25$\pm$0.03 & 0.17$\pm$0.06 & 4 & Transit & 79 & 11.1 & Yes & 92\\
Kepler-68 & 1.06$\pm$0.02 & 0.11$\pm$0.06 & 4 & Transit & 225 & 12.4 & Yes & 2\\
Kepler-78 & 0.78 (+0.03 -0.05) & -0.18$\pm$0.08 & 1 & Transit & 201 & 6.3 & No & 2\\
Kepler-92 & 1.29$\pm$0.04 & 0.14$\pm$0.01 & 3 & Transit & 23 & 9.9 & No & 92\\
Kepler-93 & 0.91$\pm$0.03 & -0.18$\pm$0.10 & 1 & Transit & 153 & 12.2 & No & 2\\
Kepler-94 & 0.82$\pm$0.02 & 0.34$\pm$0.07 & 2 & Transit & 39 & 12.0 & Yes & 92\\
Kepler-95 & 1.08$\pm$0.04 & 0.30$\pm$0.10 & 1 & Transit & 36 & 7.9 & No & 92\\
Kepler-96 & 1.01$\pm$0.02 & 0.04$\pm$0.07 & 1 & Transit & 55 & 10.9 & No & 92\\
Kepler-97 & 0.90$\pm$0.03 & -0.20$\pm$0.07 & 1 & Transit & 31 & 8.0 & No & 92\\
Kepler-98 & 1.00$\pm$0.02 & 0.18$\pm$0.07 & 1 & Transit & 42 & 7.9 & No & 92\\
Kepler-99 & 0.82$\pm$0.02 & 0.18$\pm$0.07 & 1 & Transit & 45 & 7.0 & No & 92\\
TOI-1246 & 0.87$\pm$0.03 & 0.17$\pm$0.06 & 4 & Transit & 128 & 2.0 & No & 28\\
TOI-1416 & 0.80$\pm$0.04 & 0.08$\pm$0.05 & 2 & Transit & 205 & 2.2 & No & 29\\
TOI-1422 & 0.98 (+0.06 -0.07) & -0.09$\pm$0.07 & 2 & Transit & 112 & 1.6 & No & 30\\
TOI-1670 & 1.21$\pm$0.02 & 0.09$\pm$0.07 & 2 & Transit & 65 & 1.4 & Yes & 43\\
TOI-1736 & 1.08$\pm$0.04 & 0.14$\pm$0.01 & 2 & Transit & 152 & 2.6 & Yes & 81\\
TOI-1759 & 0.61$\pm$0.02 & 0.05$\pm$0.16 & 1 & Transit & 216 & 1.2 & No & 31\\
TOI-1807 & 0.76$\pm$0.03 & -0.04$\pm$0.02 & 1 & Transit & 161 & 1.3 & No & 32\\
TOI-1853 & 0.84$\pm$0.04 & 0.11$\pm$0.08 & 1 & Transit & 56 & 1.5 & No & 82\\
TOI-2134 & 0.74$\pm$0.03 & 0.12$\pm$0.02 & 2 & Transit & 221 & 2.2 & No & 83\\
TOI-2141 & 0.94$\pm$0.02 & -0.12$\pm$0.01 & 1 & Transit & 90 & 1.6 & No & 81\\
TOI-4010 & 0.88$\pm$0.03 & 0.37$\pm$0.07 & 4 & Transit & 112 & 1.5 & Yes & 33\\ 
TOI-431 & 0.78$\pm$0.07 & 0.20$\pm$0.05 & 3 & Transit & 200 & 15.8 & No & 34\\
TOI-500 & 0.74$\pm$0.02 & 0.12$\pm$0.08 & 4 & Transit & 197 & 1.0 & No & 35\\
TOI-5398 & 1.15$\pm$0.01 & 0.09$\pm$0.06 & 2 & Transit & 86 & 1.2 & No & 91\\
TOI-561 & 0.81$\pm$0.04 & -0.40$\pm$0.05 & 4 & Transit & 316 & 2.9 & No & 36\\
TOI-969 & 0.73$\pm$0.01 & 0.18$\pm$0.02 & 2 & Transit & 94 & 1.3 & Yes & 37\\
WASP-132 & 0.78$\pm$0.03 & 0.18$\pm$0.12 & 2 & Transit & 36 & 2.0 & No & 44\\
WASP-47 & 1.04$\pm$0.03 & 0.38$\pm$0.05 & 4 & Transit & 246 & 9.4 & Yes & 38\\
Wolf 503 & 0.69$\pm$0.02 & -0.47$\pm$0.08 & 1 & Transit & 115 & 3.1 & No & 2\\
55 Cnc & 0.97$\pm$0.05 & 0.38$\pm$0.06 & 5 & RV & 1584 & 30.6 & Yes & 50\\
61 Vir & 0.91$\pm$0.04 & 0.03$\pm$0.06 & 3 & RV & 954 & 28.9 & No & 50\\
BD-08 2823 & 0.74$\pm$0.07 & -0.07$\pm$0.03 & 2 & RV & 83 & 5.0 & No & 78\\
BD-11 4672 & 0.65$\pm$0.03 & -0.35$\pm$0.15 & 2 & RV & 111 & 14.2 & Yes & 68\\
DMPP-1 & 1.21$\pm$0.03 & 0.12$\pm$0.10 & 4 & RV & 148 & 2.1 & No & 48\\
DMPP-4 & 1.25$\pm$0.02 & -0.01$\pm$0.06 & 1 & RV & 71 & 4.3 & No & 88\\
GJ 160.2 & 0.69$\pm$0.10\tablenotemark{4} & -0.26$\pm$0.10\tablenotemark{3} & 1 & RV & 107 & 8.3 & No & 65\\
GJ 2056 & 0.62$\pm$0.08 & -0.08$\pm$0.10\tablenotemark{3} & 2 & RV & 51 & 9.1 & No & 59\\
GJ 3942 & 0.63$\pm$0.07 & -0.04$\pm$0.09 & 1 & RV & 145 & 3.3 & No & 58 \\
GJ 414 A & 0.69$\pm$0.01 & 0.24$\pm$0.07 & 2 & RV & 560 & 28.2 & No & 50\\
GJ 676 A & 0.62$\pm$0.06 & 0.23$\pm$0.10 & 4 & RV & 127 & 8.9 & Yes & 49\\
GJ 9404 & 0.62$\pm$0.07 & 0.16$\pm$0.10 & 1 & RV & 54 & 3.9 & No & 87\\
HD 10180 & 1.06$\pm$0.05 & 0.08$\pm$0.01 & 6 & RV & 190 & 6.7 & No & 66\\
HD 102365 & 0.85$\pm$0.10\tablenotemark{4} & -0.31$\pm$0.01 & 1 & RV & 149 & 12.5 & No & 72\\
HD 103949 & 0.77$\pm$0.04 & -0.07$\pm$0.05 & 1 & RV & 48 & 8.2 & No & 61\\
HD 107148 & 1.11$\pm$0.04 & 0.30$\pm$0.06 & 2 & RV & 114 & 20.2 & No & 50\\
HD 109271 & 1.05$\pm$0.02 & 0.10$\pm$0.05 & 2 & RV & 100 & 7.5 & No & 74\\
HD 110067 & 0.80$\pm$0.04 & -0.20$\pm$0.04 & 6 & RV & 111 & 1.9 & No & 90\\
HD 125595 & 0.76$\pm$0.02 & 0.02$\pm$0.06 & 1 & RV & 117 & 5.7 & No & 79\\
HD 125612 & 1.11$\pm$0.20 & 0.25$\pm$0.04 & 3 & RV & 72 & 12.1 & Yes & 69\\
HD 134060 & 1.09$\pm$0.10\tablenotemark{4} & 0.14$\pm$0.01 & 2 & RV & 339 & 13.2 & No & 53\\
HD 136352 & 0.87$\pm$0.04 & -0.24$\pm$0.05 & 3 & RV & 458 & 19.3 & No & 52\\
HD 13808 & 0.77$\pm$0.02 & -0.21$\pm$0.02 & 2 & RV & 246 & 11.1 & No & 64\\
HD 140901 & 0.99$\pm$0.12 & 0.09$\pm$0.01 & 2 & RV & 392 & 24.0 & Yes & 70\\
HD 141004 & 1.05$\pm$0.05 & 0.03$\pm$0.06 & 1 & RV & 1511 & 32.4 & No & 50\\
HD 1461 & 1.03$\pm$0.05& 0.18$\pm$0.06 & 2 & RV & 1040 & 23.3 & No & 50\\
HD 154088 & 0.91$\pm$0.02 & 0.28$\pm$0.03 & 1 & RV & 187 & 11.4 & No & 51\\
HD 156668 & 0.77$\pm$0.02 & 0.05$\pm$0.06 & 2 & RV & 821 & 16.8 & No & 50\\
HD 158259 & 1.08$\pm$0.10 & 0$\pm$0.12 & 5 & RV & 290 & 7.1 & No & 47\\
HD 160691 & 1.13$\pm$0.02 & 0.28$\pm$0.03 & 4 & RV & 380 & 17.1 & Yes & 80\\
HD 164595 & 0.99$\pm$0.03 & -0.04$\pm$0.08 & 1 & RV & 75 & 2.3 & No & 73\\
HD 164922 & 0.92$\pm$0.03 & 0.20$\pm$0.06 & 4 & RV & 742 & 23.1 & No & 50\\
HD 168009 & 0.99$\pm$0.04 & 0.02$\pm$0.06 & 1 & RV & 613 & 21.1 & No & 50\\
HD 175607 & 0.71$\pm$0.01 & -0.62$\pm$0.01 & 1 & RV & 110 & 9.3 & No & 62\\
HD 176986 & 0.79$\pm$0.02 & 0.03$\pm$0.05 & 2 & RV & 259 & 13.2 & No & 55\\
HD 177565 & 1.0$\pm$0.10 & 0.08$\pm$0.01 & 1 & RV & 68 & 4.6 & No & 67\\
HD 181433 & 0.78$\pm$0.10\tablenotemark{4} & 0.33$\pm$0.13 & 3 & RV & 200 & 13.8 & Yes & 56\\
HD 189567 & 0.83$\pm$0.01 & -0.24$\pm$0.01 & 2 & RV & 256 & 13.2 & No & 51\\
HD 190007 & 0.77$\pm$0.02 & 0.16$\pm$0.05 & 1 & RV & 191 & 21.3 & No & 63\\
HD 190360 & 0.99$\pm$0.04 & 0.23$\pm$0.06 & 2 & RV & 1243 & 23.5 & Yes & 50\\
HD 192310 & 0.84$\pm$0.03 & 0.08$\pm$0.06 & 2 & RV & 429 & 15.2 & No & 50\\
HD 20003 & 0.88$\pm$0.10\tablenotemark{4} & 0.04$\pm$0.02 & 2 & RV & 203 & 13.7 & No & 53\\
HD 204313 & 1.03$\pm$0.04 & 0.18$\pm$0.02 & 3 & RV & 193 & 14.2 & Yes & 75\\
HD 20781 & 0.70$\pm$0.10\tablenotemark{4}& -0.11$\pm$0.02 & 4 & RV & 237 & 13.7 & No & 53\\
HD 211970 & 0.61$\pm$0.04 & -0.50$\pm$0.10\tablenotemark{3} & 1 & RV & 52 & 8.5 & No & 61\\
HD 215497 & 0.92$\pm$0.05 & 0.23$\pm$0.07 & 2 & RV & 99 & 5.1 & No & 57 \\
HD 216520 & 0.82$\pm$0.04 & -0.16$\pm$0.10\tablenotemark{3} & 2 & RV & 804 & 18.7 & No & 63\\
HD 21693 & 0.80$\pm$0.10\tablenotemark{4} & 0$\pm$0.02 & 2 & RV & 213 & 11.3 & No & 53\\
HD 219134 & 0.79$\pm$0.03 & 0.08$\pm$0.06 & 5 & RV & 1176 & 27.1 & No & 50\\
HD 219828 & 1.23$\pm$0.10 & 0.19$\pm$0.03 & 2 & RV & 114 & 14.2 & Yes & 76\\
HD 22496 & 0.68$\pm$0.01 & -0.08$\pm$0.02 & 1 & RV & 84 & 17.4 & No & 54\\
HD 24085 & 1.22$\pm$0.07 & 0.17$\pm$0.06 & 1 & RV & 25 & 8.7 & No & 61\\
HD 26965 & 0.78$\pm$0.08 & -0.42$\pm$0.04 & 1 & RV & 1253 & 15.1 & No & 60\\
HD 31527 & 0.96$\pm$0.10\tablenotemark{4} & -0.17$\pm$0.01 & 3 & RV & 268 & 13.9 & No & 53\\
HD 34445 & 1.11$\pm$0.06 & 0.14$\pm$0.06 & 6 & RV & 222 & 21.8 & Yes & 50\\
HD 39194 & 0.67$\pm$0.04 & -0.61$\pm$0.02 & 3 & RV & 273 & 13.7 & No & 51\\
HD 39855 & 0.87$\pm$0.05 & -0.46$\pm$0.10\tablenotemark{3}& 1 & RV & 25 & 6.2 & No & 61\\
HD 40307 & 0.77$\pm$0.05 & -0.31$\pm$0.03 & 6 & RV & 226 & 10.4 & No & 49\\
HD 42618 & 0.92$\pm$0.05 & -0.09$\pm$0.06 & 1 & RV & 835 & 23.4 & No & 50\\
HD 4308 & 0.93$\pm$0.10 & -0.35$\pm$0.01 & 1 & RV & 357 & 13.9 & No & 71\\
HD 45184 & 1.01$\pm$0.05 & 0.04$\pm$0.06 & 2 & RV & 339 & 23.4 & No & 50\\
HD 51608 & 0.80$\pm$0.10\tablenotemark{4} & -0.07$\pm$0.01 & 2 & RV & 227 & 13.8 & No & 53\\
HD 64114 & 0.95$\pm$0.05 & -0.03$\pm$0.10\tablenotemark{3} & 1 & RV & 28 & 7.2 & No & 61\\
HD 69830 & 0.89$\pm$0.04 & 0.01$\pm$0.06 & 3 & RV & 845 & 19.3 & No & 50\\
HD 77338 & 0.93$\pm$0.05 & 0.28$\pm$0.04 & 1 & RV & 32 & 7.2 & No & 77\\
HD 7924 & 0.80$\pm$0.03 & -0.12$\pm$0.06 & 3 & RV & 1242 & 18.1 & No & 50\\
HD 90156 & 0.86$\pm$0.04 & -0.18$\pm$0.06 & 1 & RV & 167 & 23.2 & No & 50\\
HD 93385 & 1.04$\pm$0.01 & 0.02$\pm$0.01 & 3 & RV & 241 & 13.2 & No & 51\\
HD 96700 & 0.89$\pm$0.01 & -0.18$\pm$0.01 & 3 & RV & 254 & 14.4 & No & 51\\
HD 97658 & 0.77$\pm$0.03 & -0.23$\pm$0.06 & 1 & RV & 785 & 22.0 & No & 50\\
HD 99492 & 0.85$\pm$0.03 & 0.29$\pm$0.06 & 2 & RV & 197 & 23.1 & No & 50\\
HIP 107772 & 0.63$\pm$0.08 & -0.66$\pm$0.10\tablenotemark{3} & 1 & RV & 49 & 9.2 & No & 59\\
HIP 35173 & 0.79$\pm$0.05 & 0.13$\pm$0.10\tablenotemark{3} & 1 & RV & 39 & 8.8 & No & 61\\
HIP 38594 & 0.61$\pm$0.02 & -0.17$\pm$0.10\tablenotemark{3} & 2 & RV & 38 & 8.9 & No & 59\\
HIP 4845 & 0.62$\pm$0.04 & -0.5$\pm$0.10\tablenotemark{3}& 1 & RV & 94 & 11.1 & No & 59
\enddata
\tablecomments{References: (1) \citet{Alonso2014}, (2) \citet{Bonomo2023}, (3) \citet{Hidalgo2020}, (4) \citet{Dragomir2019}, (5) \citet{Azevedo2022}, (6) \citet{Gandolfi2019}, (7) \citet{Gan2021}, (8) \citet{Desidera2023}, (9) \citet{Lubin2022}, (10) \citet{Barros2023}, (11) \citet{Heidari2022}, (12) \citet{Barros2022}, (13) \citet{Hua2023}, (14) \citet{Sozzetti2021}, (15) \citet{Barragan2022}, (16) \citet{Teske2020}, (17) \citet{Osborn2023}, (18) \citet{Petigura2020}, (19) \citet{Persson2018}, (20) \citet{Lam2018}, (21) \citet{LilloBox2020}, (22) \citet{Sinukoff2017}, (23) \citet{Weiss2020}, (24) \citet{Weiss2013}, (25) \citet{Marcy2014}, (26) \citet{Zhang2021}, (27) \citet{Mills2019}, (28) \citet{Turtelboom2022}, (29) \citet{Deeg2023}, (30) \citet{Naponiello2022}, (31) \citet{Martioli2022}, (32) \citet{Nardiello2022}, (33) \citet{Kunimoto2023}, (34) \citet{Osborn2021}, (35) \citet{Serrano2022}, (36) \citet{Brinkman2023}, (37) \citet{LilloBox2023}, (38) \citet{Bryant2022}, (39) \citet{Barragan2019}, (40) \citet{Mayo2023}, (41) \citet{Murphy2021}, (42) \citet{Bonomo2014}, (43) \citet{Tran2022}, (44) \citet{Hellier2017}, (45) \citet{Hatzes2022}, (46) \citet{Espinoza2020}, (47) \citet{Hara2020}, (48) \citet{Staab2020}, (49) \citet{Bryan2019}, (50) \citet{Rosenthal2021}, (51) \citet{Unger2021}, (52) \citet{Kane2020}, (53)\citet{Udry2019}, (54) \citet{LilloBox2021}, (55) \citet{Suarez2018}, (56) \citet{Horner2019}, (57) \citet{LoCurto2010}, (58) \citet{Perger2017}, (59) \citet{Feng2020}, (60) \citet{Ma2018}, (61) \citet{Feng2019}, (62) \citet{Mortier2016}, (63) \citet{Burt2021}, (64) \citet{Ahrer2021}, (65) \citet{Tuomi2014}, (66) \citet{Lovis2011}, (67) \citet{Feng2017}, (68) \citet{Barbato2020}, (69) \citet{Ment2018}, (70) \citet{Feng2022}, (71) \citet{Trifonov2020}, (72) \citet{Tinney2011}, (73) \citet{Courcol2015}, (74) \citet{LoCurto2013}, (75) \citet{Diaz2016}, (76) \citet{Santos2016}, (77) \citet{Jenkins2013}, (78) \citet{Hebrard2010}, (79) \citet{Segransan2011}, (80) \citet{Benedict2022}, (81) \citet{Martioli2023}, (82) \citet{Naponiello2023}, (83) \citet{Rescigno2024}, (84) \citet{Osborne2024}, (85) \citet{Damasso2023}, (86) \citet{Dai2023}, (87) \citet{Pinamonti2022}, (88) \citet{Barnes2023}, (89) \citet{Heidari2024}, (90) \citet{Luque2023}, (91) \citet{Mantovan2024}, (92) \citet{Weiss2024} }
\tablenotetext{1}{We note that \citet{Gandolfi2019} and \citet{Dumusque2019} independently reduced the same dataset and obtained the same time baseline, comparable uncertainties, and 87/85 data points. In subsequent analyses we use the dataset from \citet{Gandolfi2019}.}
\tablenotetext{2}{\citet{Orell2023} publishes plots of additional datasets from CARMENES and HARPS-N but do not make them publicly available.}
\tablenotetext{3}{In the cases where uncertainties on stellar metallicities are not given, we assume an error of 0.1 dex.}
\tablenotetext{4}{In the cases where uncertainties on stellar mass are not given, we assume an error of 0.1 M$_{\sun}$}
\label{tab:full-sample}
\end{deluxetable*}

\begin{figure*}
\begin{tabular}{c c}
\includegraphics[width=0.5\textwidth]{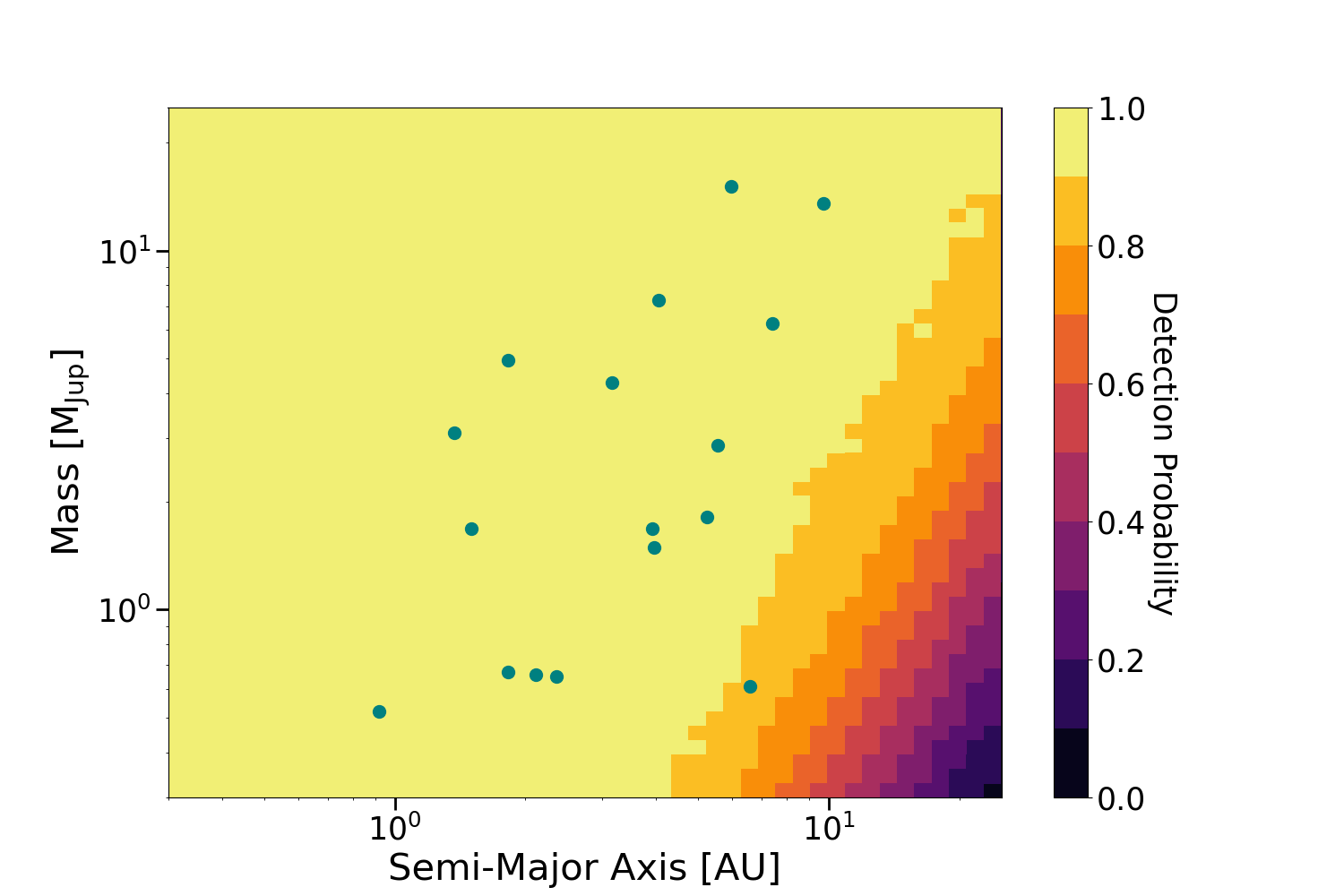}&
\includegraphics[width=0.5\textwidth]{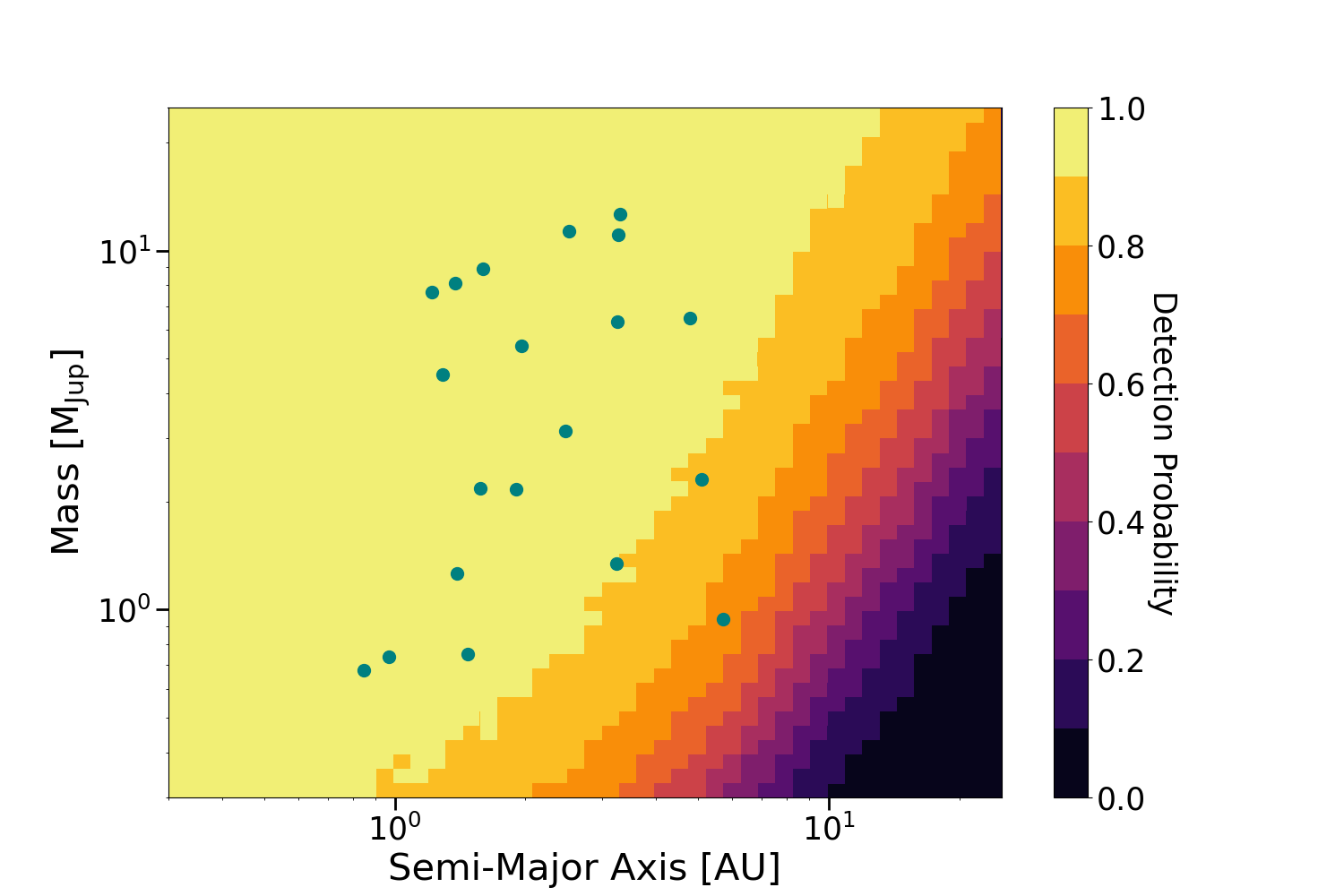}
\end{tabular}
\caption{Left: Average completeness map for the super-Earth RV sample with detected gas giants overplotted in teal. Right: Average completeness map for the super-Earth transit sample with detected gas giants overplotted in teal. Color scale represents detection probability. On average, the RV sample has higher completeness to distant gas giants typically driven by much longer time baselines and more data points. Accounting for differences in individual system sensitivity as well as these broad trends between transit and RV samples is key to calculating frequencies of gas giants in our sample of super-Earth systems.}
\label{fig: average completeness}
\end{figure*}

\startlongtable
\begin{deluxetable*}{lcccc}
\tablecaption{Confirmed Gas Giants in Our Super-Earth System Sample}
\tablehead{
\colhead{Target} & \colhead{GG Mass (M$_{\rm Jup}$)} & \colhead{GG Semi-major axis (AU)} & \colhead{M$_{\star}$ (M$_{\odot}$)} & \colhead{[Fe/H]} 
}
\startdata
HD 137496 c & 7.66$\pm$0.11 & 1.22$\pm$0.01 & 1.04$\pm$0.02 & -0.03$\pm$0.04 \\
HD 191939 f & 6.5$\pm$4.5 & 4.8$\pm$2.2 & 0.81$\pm$0.04 & -0.15$\pm$0.06 \\
HD 39091 b & 12.6$\pm$2.0 & 3.3$\pm$0.1 & 1.07$\pm$0.04 & 0.09$\pm$0.03\\
HD 80653 c & 5.41 (+0.52 -0.44) & 1.96$\pm$0.03 & 1.18$\pm$0.04 & 0.26$\pm$0.07\\
Kepler-129 d & 6.34$\pm$0.14 & 3.26 & 1.24$\pm$0.04 & 0.29$\pm$0.10\\
Kepler-139 e & 1.34$\pm$0.13 & 3.24 & 1.08$\pm$0.03 & 0.27$\pm$0.10\\
Kepler-407 c & 11.09$\pm$0.05 & 3.27 & 1.07$\pm$0.03 & 0.33$\pm$0.07\\
Kepler-454 c & 4.51$\pm$0.12 & 1.29$\pm$0.02 & 1.03 (+0.04 -0.03) & 0.32$\pm$0.08\\
Kepler-454 d & 2.31 (+0.27 -0.16) & 5.10 (+0.34 -0.19) & 1.03 (+0.04 -0.03) & 0.32$\pm$0.08\\
Kepler-48 e & 2.16$\pm$0.06 & 1.90 & 0.92$\pm$0.02 & 0.17$\pm$0.07\\
Kepler-48 f & 0.94$\pm$0.29 & 5.72 & 0.92$\pm$0.02 & 0.17$\pm$0.07\\
Kepler-65 e & 0.68$\pm$0.07 & 0.85 & 1.25$\pm$0.03 & 0.17$\pm$0.06\\
Kepler-68 d & 0.75$\pm$0.02 & 1.47$\pm$0.01 & 1.06$\pm$0.02 & 0.11$\pm$0.06\\
Kepler-94 c & 8.89$\pm$0.12 & 1.60 & 0.82$\pm$0.02 & 0.34$\pm$0.07\\ 
KOI-142 c & 0.65$\pm$0.02 & 0.15 & 0.99$\pm$0.02 & 0.27$\pm$0.06 \\
KOI-142 d & 3.15$\pm$0.17 & 2.47 & 0.99$\pm$0.02 & 0.27$\pm$0.06 \\
KOI-351 h & 0.74$\pm$0.11 & 0.97 & 1.11$\pm$0.03 & 0.08$\pm$0.03\\
TOI-1670 c & 0.63 (+0.09 -0.08) & 0.25$\pm$0.01 & 1.21$\pm$0.02 & 0.09$\pm$0.07\\
TOI-1736 c & 8.09$\pm$0.20 & 1.38$\pm$0.02 & 1.08$\pm$0.04 & 0.14$\pm$0.01\\
TOI-4010 e & 2.18 (+0.21 -0.20) & 1.57 (+0.12 -0.13) & 0.88$\pm$0.03 & 0.37$\pm$0.07\\
TOI-969 c & 11.3 (+1.1 -0.9) & 2.52 (+0.27 -0.29) & 0.73$\pm$0.01 & 0.18$\pm$0.02\\
WASP-47 c & 1.26$\pm$0.03 & 1.39$\pm$0.01 & 1.04$\pm$0.03 & 0.38$\pm$0.05\\
WASP-47 b & 1.14$\pm$0.02 & 0.05$\pm$0.01 & 1.04$\pm$0.03 & 0.38$\pm$0.05\\
55 Cnc b & 0.84$\pm$0.03 & 0.12$\pm$0.01 & 0.97$\pm$0.05 & 0.38$\pm$0.06\\
55 Cnc d & 2.86$\pm$0.25 & 5.54$\pm$0.10 & 0.97$\pm$0.05 & 0.38$\pm$0.06\\
BD-11 4672 b & 0.65 (+0.05 -0.06 ) & 2.36$\pm$0.04 & 0.65$\pm$0.03 & -0.35$\pm$0.15 \\
GJ 676 A b & 4.96$\pm$0.96 & 1.82$\pm$0.06 & 0.62$\pm$0.06 & 0.23$\pm$0.10\\
GJ 676 A c & 13.5 (+1.0 -1.1) & 9.73 (+0.63 -0.79) & 0.62$\pm$0.06 & 0.23$\pm$0.1\\
HD 125612 b & 3.1$\pm$0.4 & 1.37$\pm$0.08 & 1.11$\pm$0.20 & 0.25$\pm$0.04\\
HD 125612 d & 7.28$\pm$0.93 & 4.06$\pm$0.25 & 1.11$\pm$0.20 & 0.25$\pm$0.04\\
HD 140901 c & 6.28 (+1.34 -5.04) & 7.42 (+0.32 -0.58) & 0.99$\pm$0.12 & 0.09$\pm$0.01\\
HD 160691 b & 1.68 & 1.50 & 1.13$\pm$0.02 & 0.28$\pm$0.03 \\
HD 160691 c & 1.81 & 5.24 & 1.13$\pm$0.02 & 0.28$\pm$0.03 \\
HD 160691 e & 0.52 & 0.92 & 1.13$\pm$0.02 & 0.28$\pm$0.03 \\
HD 181433 c & 0.67$\pm$0.01 & 1.82$\pm$0.01 & 0.78$\pm$0.10& 0.33$\pm$0.13\\
HD 181433 d & 0.61$\pm$0.01 & 6.60$\pm$0.22 & 0.78$\pm$0.10 & 0.33$\pm$0.13\\
HD 190360 b & 1.49$\pm$0.04 & 3.96$\pm$0.05 & 0.99$\pm$0.04 & 0.23$\pm$0.06\\
HD 204313 b & 4.28$\pm$0.30 & 3.17$\pm$0.12 & 1.03$\pm$0.04 & 0.18$\pm$0.02\\
HD 204313 d & 1.68$\pm$0.30 & 3.93$\pm$0.14 & 1.03$\pm$0.04 & 0.18$\pm$0.02\\
HD 219828 c & 15.10$\pm$0.85 & 5.96 & 1.23$\pm$0.10 & 0.19$\pm$0.03\\
HD 34445 b & 0.66$\pm$0.04 & 2.11$\pm$0.04 & 1.11$\pm$0.06 & 0.14$\pm$0.06
\enddata
\label{tab:GG-conf}
\end{deluxetable*}

\bibliography{bibliography}{}

\begin{thebibliography}{}
\expandafter\ifx\csname natexlab\endcsname\relax\def\natexlab#1{#1}\fi
\providecommand{\url}[1]{\href{#1}{#1}}
\providecommand{\dodoi}[1]{doi:~\href{http://doi.org/#1}{\nolinkurl{#1}}}
\providecommand{\doeprint}[1]{\href{http://ascl.net/#1}{\nolinkurl{http://ascl.net/#1}}}
\providecommand{\doarXiv}[1]{\href{https://arxiv.org/abs/#1}{\nolinkurl{https://arxiv.org/abs/#1}}}

\bibitem[{{Ahrer} {et~al.}(2021){Ahrer}, {Queloz}, {Rajpaul}, {S{\'e}gransan}, {Bouchy}, {Hall}, {Handley}, {Lovis}, {Mayor}, {Mortier}, {Pepe}, {Thompson}, {Udry}, \& {Unger}}]{Ahrer2021}
{Ahrer}, E., {Queloz}, D., {Rajpaul}, V.~M., {et~al.} 2021, \mnras, 503, 1248, \dodoi{10.1093/mnras/stab373}

\bibitem[{{Akana Murphy} {et~al.}(2021){Akana Murphy}, {Kosiarek}, {Batalha}, {Gonzales}, {Isaacson}, {Petigura}, {Weiss}, {Grunblatt}, {Ciardi}, {Fulton}, {Hirsch}, {Behmard}, \& {Rosenthal}}]{Murphy2021}
{Akana Murphy}, J.~M., {Kosiarek}, M.~R., {Batalha}, N.~M., {et~al.} 2021, \aj, 162, 294, \dodoi{10.3847/1538-3881/ac2830}

\bibitem[{{Alonso} {et~al.}(2014){Alonso}, {Moutou}, {Endl}, {Almenara}, {Guenther}, {Deleuil}, {Hatzes}, {Aigrain}, {Auvergne}, {Baglin}, {Barge}, {Bonomo}, {Bord{\'e}}, {Bouchy}, {Cavarroc}, {Cabrera}, {Carpano}, {Csizmadia}, {Cochran}, {Deeg}, {D{\'\i}az}, {Dvorak}, {Erikson}, {Ferraz-Mello}, {Fridlund}, {Fruth}, {Gandolfi}, {Gillon}, {Grziwa}, {Guillot}, {H{\'e}brard}, {Jorda}, {L{\'e}ger}, {Lammer}, {Lovis}, {MacQueen}, {Mazeh}, {Ofir}, {Ollivier}, {Pasternacki}, {P{\"a}tzold}, {Queloz}, {Rauer}, {Rouan}, {Santerne}, {Schneider}, {Tadeu dos Santos}, {Tingley}, {Titz-Weider}, {Weingrill}, \& {Wuchterl}}]{Alonso2014}
{Alonso}, R., {Moutou}, C., {Endl}, M., {et~al.} 2014, \aap, 567, A112, \dodoi{10.1051/0004-6361/201118662}

\bibitem[{{Azevedo Silva} {et~al.}(2022){Azevedo Silva}, {Demangeon}, {Barros}, {Armstrong}, {Otegi}, {Bossini}, {Delgado Mena}, {Sousa}, {Adibekyan}, {Nielsen}, {Dorn}, {Lillo-Box}, {Santos}, {Hoyer}, {Stassun}, {Almenara}, {Bayliss}, {Barrado}, {Boisse}, {Brown}, {D{\'\i}az}, {Dumusque}, {Figueira}, {Hadjigeorghiou}, {Hojjatpanah}, {Mousis}, {Osborn}, {Santerne}, {Str{\o}m}, {Udry}, \& {Wheatley}}]{Azevedo2022}
{Azevedo Silva}, T., {Demangeon}, O.~D.~S., {Barros}, S.~C.~C., {et~al.} 2022, \aap, 657, A68, \dodoi{10.1051/0004-6361/202141520}

\bibitem[{{Barbato} {et~al.}(2020){Barbato}, {Pinamonti}, {Sozzetti}, {Biazzo}, {Benatti}, {Damasso}, {Desidera}, {Lanza}, {Maldonado}, {Mancini}, {Scandariato}, {Affer}, {Andreuzzi}, {Bignamini}, {Bonomo}, {Borsa}, {Carleo}, {Claudi}, {Cosentino}, {Covino}, {Fiorenzano}, {Giacobbe}, {Harutyunyan}, {Knapic}, {Leto}, {Lorenzi}, {Maggio}, {Malavolta}, {Micela}, {Molinari}, {Molinaro}, {Nascimbeni}, {Pagano}, {Pedani}, {Piotto}, {Poretti}, \& {Rainer}}]{Barbato2020}
{Barbato}, D., {Pinamonti}, M., {Sozzetti}, A., {et~al.} 2020, \aap, 641, A68, \dodoi{10.1051/0004-6361/202037954}

\bibitem[{{Barnes} {et~al.}(2023){Barnes}, {Standing}, {Haswell}, {Staab}, {Doherty}, {Waller-Bridge}, {Fossati}, {Soto}, {Anglada-Escud{\'e}}, {Llama}, {McCune}, \& {Lewis}}]{Barnes2023}
{Barnes}, J.~R., {Standing}, M.~R., {Haswell}, C.~A., {et~al.} 2023, \mnras, 524, 5196, \dodoi{10.1093/mnras/stad2109}

\bibitem[{{Barrag{\'a}n} {et~al.}(2019){Barrag{\'a}n}, {Aigrain}, {Kubyshkina}, {Gandolfi}, {Livingston}, {Fridlund}, {Fossati}, {Korth}, {Parviainen}, {Malavolta}, {Palle}, {Deeg}, {Nowak}, {Rajpaul}, {Zicher}, {Antoniciello}, {Narita}, {Albrecht}, {Bedin}, {Cabrera}, {Cochran}, {de Leon}, {Eigm{\"u}ller}, {Fukui}, {Granata}, {Grziwa}, {Guenther}, {Hatzes}, {Kusakabe}, {Latham}, {Libralato}, {Luque}, {Monta{\~n}{\'e}s-Rodr{\'\i}guez}, {Murgas}, {Nardiello}, {Pagano}, {Piotto}, {Persson}, {Redfield}, \& {Tamura}}]{Barragan2019}
{Barrag{\'a}n}, O., {Aigrain}, S., {Kubyshkina}, D., {et~al.} 2019, \mnras, 490, 698, \dodoi{10.1093/mnras/stz2569}

\bibitem[{{Barrag{\'a}n} {et~al.}(2022){Barrag{\'a}n}, {Armstrong}, {Gandolfi}, {Carleo}, {Vidotto}, {Villarreal D'Angelo}, {Oklop{\v{c}}i{\'c}}, {Isaacson}, {Oddo}, {Collins}, {Fridlund}, {Sousa}, {Persson}, {Hellier}, {Howell}, {Howard}, {Redfield}, {Eisner}, {Georgieva}, {Dragomir}, {Bayliss}, {Nielsen}, {Klein}, {Aigrain}, {Zhang}, {Teske}, {Twicken}, {Jenkins}, {Esposito}, {Van Eylen}, {Rodler}, {Adibekyan}, {Alarcon}, {Anderson}, {Akana Murphy}, {Barrado}, {Barros}, {Benneke}, {Bouchy}, {Bryant}, {Butler}, {Burt}, {Cabrera}, {Casewell}, {Chaturvedi}, {Cloutier}, {Cochran}, {Crane}, {Crossfield}, {Crouzet}, {Collins}, {Dai}, {Deeg}, {Deline}, {Demangeon}, {Dumusque}, {Figueira}, {Furlan}, {Gnilka}, {Goad}, {Goffo}, {Guti{\'e}rrez-Canales}, {Hadjigeorghiou}, {Hartman}, {Hatzes}, {Harris}, {Henderson}, {Hirano}, {Hojjatpanah}, {Hoyer}, {Kab{\'a}th}, {Korth}, {Lillo-Box}, {Luque}, {Marmier}, {Mo{\v{c}}nik}, {Muresan}, {Murgas}, {Nagel}, {Osborne}, {Osborn}, {Osborn}, {Palle}, {Raimbault}, {Ricker},
  {Rubenzahl}, {Stockdale}, {Santos}, {Scott}, {Schwarz}, {Shectman}, {Raimbault}, {Seager}, {S{\'e}gransan}, {Serrano}, {Skarka}, {Smith}, {{\v{S}}ubjak}, {Tan}, {Udry}, {Watson}, {Wheatley}, {West}, {Winn}, {Wang}, {Wolfgang}, \& {Ziegler}}]{Barragan2022}
{Barrag{\'a}n}, O., {Armstrong}, D.~J., {Gandolfi}, D., {et~al.} 2022, \mnras, 514, 1606, \dodoi{10.1093/mnras/stac638}

\bibitem[{{Barros} {et~al.}(2022){Barros}, {Demangeon}, {Alibert}, {Leleu}, {Adibekyan}, {Lovis}, {Bossini}, {Sousa}, {Hara}, {Bouchy}, {Lavie}, {Rodrigues}, {Gomes da Silva}, {Lillo-Box}, {Pepe}, {Tabernero}, {Zapatero Osorio}, {Sozzetti}, {Su{\'a}rez Mascare{\~n}o}, {Micela}, {Allende Prieto}, {Cristiani}, {Damasso}, {Di Marcantonio}, {Ehrenreich}, {Faria}, {Figueira}, {Gonz{\'a}lez Hern{\'a}ndez}, {Jenkins}, {Lo Curto}, {Martins}, {Micela}, {Nunes}, {Pall{\'e}}, {Santos}, {Rebolo}, {Seager}, {Twicken}, {Udry}, {Vanderspek}, \& {Winn}}]{Barros2022}
{Barros}, S.~C.~C., {Demangeon}, O.~D.~S., {Alibert}, Y., {et~al.} 2022, \aap, 665, A154, \dodoi{10.1051/0004-6361/202244293}

\bibitem[{{Barros} {et~al.}(2023){Barros}, {Demangeon}, {Armstrong}, {Delgado Mena}, {Acu{\~n}a}, {Fern{\'a}ndez Fern{\'a}ndez}, {Deleuil}, {Collins}, {Howell}, {Ziegler}, {Adibekyan}, {Sousa}, {Stassun}, {Grieves}, {Lillo-Box}, {Hellier}, {Wheatley}, {Brice{\~n}o}, {Collins}, {Hawthorn}, {Hoyer}, {Jenkins}, {Law}, {Mann}, {Matson}, {Mousis}, {Nielsen}, {Osborn}, {Osborn}, {Paegert}, {Papini}, {Ricker}, {Rudat}, {Santos}, {Seager}, {Stockdale}, {Str{\o}m}, {Twicken}, {Udry}, {Wang}, {Vanderspek}, \& {Winn}}]{Barros2023}
{Barros}, S.~C.~C., {Demangeon}, O.~D.~S., {Armstrong}, D.~J., {et~al.} 2023, \aap, 673, A4, \dodoi{10.1051/0004-6361/202245741}

\bibitem[{{Batygin} \& {Laughlin}(2015)}]{Batygin15}
{Batygin}, K., \& {Laughlin}, G. 2015, Proceedings of the National Academy of Science, 112, 4214, \dodoi{10.1073/pnas.1423252112}

\bibitem[{{Benedict} {et~al.}(2022){Benedict}, {McArthur}, {Nelan}, {Wittenmyer}, {Barnes}, {Smotherman}, \& {Horner}}]{Benedict2022}
{Benedict}, G.~F., {McArthur}, B.~E., {Nelan}, E.~P., {et~al.} 2022, \aj, 163, 295, \dodoi{10.3847/1538-3881/ac6ac8}

\bibitem[{{Bonomo} {et~al.}(2014){Bonomo}, {Sozzetti}, {Lovis}, {Malavolta}, {Rice}, {Buchhave}, {Sasselov}, {Cameron}, {Latham}, {Molinari}, {Pepe}, {Udry}, {Affer}, {Charbonneau}, {Cosentino}, {Dressing}, {Dumusque}, {Figueira}, {Fiorenzano}, {Gettel}, {Harutyunyan}, {Haywood}, {Horne}, {Lopez-Morales}, {Mayor}, {Micela}, {Motalebi}, {Nascimbeni}, {Phillips}, {Piotto}, {Pollacco}, {Queloz}, {S{\'e}gransan}, {Szentgyorgyi}, \& {Watson}}]{Bonomo2014}
{Bonomo}, A.~S., {Sozzetti}, A., {Lovis}, C., {et~al.} 2014, \aap, 572, A2, \dodoi{10.1051/0004-6361/201424617}

\bibitem[{{Bonomo} {et~al.}(2023){Bonomo}, {Dumusque}, {Massa}, {Mortier}, {Bongiolatti}, {Malavolta}, {Sozzetti}, {Buchhave}, {Damasso}, {Haywood}, {Morbidelli}, {Latham}, {Molinari}, {Pepe}, {Poretti}, {Udry}, {Affer}, {Boschin}, {Charbonneau}, {Cosentino}, {Cretignier}, {Ghedina}, {Lega}, {L{\'o}pez-Morales}, {Margini}, {Mart{\'\i}nez Fiorenzano}, {Mayor}, {Micela}, {Pedani}, {Pinamonti}, {Rice}, {Sasselov}, {Tronsgaard}, \& {Vanderburg}}]{Bonomo2023}
{Bonomo}, A.~S., {Dumusque}, X., {Massa}, A., {et~al.} 2023, \aap, 677, A33, \dodoi{10.1051/0004-6361/202346211}

\bibitem[{{Brinkman} {et~al.}(2023){Brinkman}, {Weiss}, {Dai}, {Huber}, {Kite}, {Valencia}, {Bean}, {Beard}, {Behmard}, {Blunt}, {Brady}, {Fulton}, {Giacalone}, {Howard}, {Isaacson}, {Kasper}, {Lubin}, {MacDougall}, {Akana Murphy}, {Plotnykov}, {Polanski}, {Rice}, {Seifahrt}, {Stef{\'a}nsson}, \& {St{\"u}rmer}}]{Brinkman2023}
{Brinkman}, C.~L., {Weiss}, L.~M., {Dai}, F., {et~al.} 2023, \aj, 165, 88, \dodoi{10.3847/1538-3881/acad83}

\bibitem[{{Bryan} {et~al.}(2019){Bryan}, {Knutson}, {Lee}, {Fulton}, {Batygin}, {Ngo}, \& {Meshkat}}]{Bryan2019}
{Bryan}, M.~L., {Knutson}, H.~A., {Lee}, E.~J., {et~al.} 2019, \aj, 157, 52, \dodoi{10.3847/1538-3881/aaf57f}

\bibitem[{{Bryant} \& {Bayliss}(2022)}]{Bryant2022}
{Bryant}, E.~M., \& {Bayliss}, D. 2022, \aj, 163, 197, \dodoi{10.3847/1538-3881/ac58ff}

\bibitem[{{Burt} {et~al.}(2021){Burt}, {Feng}, {Holden}, {Mamajek}, {Huang}, {Rosenthal}, {Wang}, {Butler}, {Vogt}, {Laughlin}, {Henry}, {Teske}, {Wang}, {Crane}, \& {Shectman}}]{Burt2021}
{Burt}, J., {Feng}, F., {Holden}, B., {et~al.} 2021, \aj, 161, 10, \dodoi{10.3847/1538-3881/abc2d0}

\bibitem[{{Chachan} \& {Lee}(2023)}]{Chachan23}
{Chachan}, Y., \& {Lee}, E.~J. 2023, \apjl, 952, L20, \dodoi{10.3847/2041-8213/ace257}

\bibitem[{{Chachan} {et~al.}(2022){Chachan}, {Dalba}, {Knutson}, {Fulton}, {Thorngren}, {Beichman}, {Ciardi}, {Howard}, \& {Van Zandt}}]{Chachan22}
{Chachan}, Y., {Dalba}, P.~A., {Knutson}, H.~A., {et~al.} 2022, \apj, 926, 62, \dodoi{10.3847/1538-4357/ac3ed6}

\bibitem[{{Courcol} {et~al.}(2015){Courcol}, {Bouchy}, {Pepe}, {Santerne}, {Delfosse}, {Arnold}, {Astudillo-Defru}, {Boisse}, {Bonfils}, {Borgniet}, {Bourrier}, {Cabrera}, {Deleuil}, {Demangeon}, {D{\'\i}az}, {Ehrenreich}, {Forveille}, {H{\'e}brard}, {Lagrange}, {Montagnier}, {Moutou}, {Rey}, {Santos}, {S{\'e}gransan}, {Udry}, \& {Wilson}}]{Courcol2015}
{Courcol}, B., {Bouchy}, F., {Pepe}, F., {et~al.} 2015, \aap, 581, A38, \dodoi{10.1051/0004-6361/201526329}

\bibitem[{{Dai} {et~al.}(2023){Dai}, {Schlaufman}, {Reggiani}, {Bouma}, {Howard}, {Chontos}, {Pidhorodetska}, {Van Zandt}, {Akana Murphy}, {Rubenzahl}, {Polanski}, {Lubin}, {Beard}, {Giacalone}, {Holcomb}, {Batalha}, {Crossfield}, {Dressing}, {Fulton}, {Huber}, {Isaacson}, {Kane}, {Petigura}, {Robertson}, {Weiss}, {Belinski}, {Boyle}, {Burke}, {Castro-Gonz{\'a}lez}, {Ciardi}, {Daylan}, {Fukui}, {Gill}, {Guerrero}, {Hellier}, {Howell}, {Lillo-Box}, {Murgas}, {Narita}, {Pall{\'e}}, {Rodriguez}, {Savel}, {Shporer}, {Stassun}, {Striegel}, {Caldwell}, {Jenkins}, {Ricker}, {Seager}, {Vanderspek}, \& {Winn}}]{Dai2023}
{Dai}, F., {Schlaufman}, K.~C., {Reggiani}, H., {et~al.} 2023, \aj, 166, 49, \dodoi{10.3847/1538-3881/acdee8}

\bibitem[{{Damasso} {et~al.}(2023){Damasso}, {Rodrigues}, {Castro-Gonz{\'a}lez}, {Lavie}, {Davoult}, {Zapatero Osorio}, {Dou}, {Sousa}, {Owen}, {Sossi}, {Adibekyan}, {Osborn}, {Leinhardt}, {Alibert}, {Lovis}, {Delgado Mena}, {Sozzetti}, {Barros}, {Bossini}, {Ziegler}, {Ciardi}, {Matthews}, {Carter}, {Lillo-Box}, {Su{\'a}rez Mascare{\~n}o}, {Cristiani}, {Pepe}, {Rebolo}, {Santos}, {Allende Prieto}, {Benatti}, {Bouchy}, {Brice{\~n}o}, {Di Marcantonio}, {D'Odorico}, {Dumusque}, {Egger}, {Ehrenreich}, {Faria}, {Figueira}, {G{\'e}nova Santos}, {Gonzales}, {Gonz{\'a}lez Hern{\'a}ndez}, {Law}, {Lo Curto}, {Mann}, {Martins}, {Mehner}, {Micela}, {Molaro}, {Nunes}, {Palle}, {Poretti}, {Schlieder}, \& {Udry}}]{Damasso2023}
{Damasso}, M., {Rodrigues}, J., {Castro-Gonz{\'a}lez}, A., {et~al.} 2023, \aap, 679, A33, \dodoi{10.1051/0004-6361/202347240}

\bibitem[{{Dawson} \& {Murray-Clay}(2013)}]{Dawson13}
{Dawson}, R.~I., \& {Murray-Clay}, R.~A. 2013, \apjl, 767, L24, \dodoi{10.1088/2041-8205/767/2/L24}

\bibitem[{{Deeg} {et~al.}(2023){Deeg}, {Georgieva}, {Nowak}, {Persson}, {Cale}, {Murgas}, {Pall{\'e}}, {Godoy-Rivera}, {Dai}, {Ciardi}, {Murphy}, {Beck}, {Burke}, {Cabrera}, {Carleo}, {Cochran}, {Collins}, {Csizmadia}, {El Mufti}, {Fridlund}, {Fukui}, {Gandolfi}, {Garc{\'\i}a}, {Guenther}, {Guerra}, {Grziwa}, {Isaacson}, {Isogai}, {Jenkins}, {K{\'a}bath}, {Korth}, {Lam}, {Latham}, {Luque}, {Lund}, {Livingston}, {Mathis}, {Mathur}, {Narita}, {Orell-Miquel}, {Osborne}, {Parviainen}, {Plavchan}, {Redfield}, {Rodriguez}, {Schwarz}, {Seager}, {Smith}, {Van Eylen}, {Van Zandt}, {Winn}, \& {Ziegler}}]{Deeg2023}
{Deeg}, H.~J., {Georgieva}, I.~Y., {Nowak}, G., {et~al.} 2023, \aap, 677, A12, \dodoi{10.1051/0004-6361/202346370}

\bibitem[{{Desidera} {et~al.}(2023){Desidera}, {Damasso}, {Gratton}, {Benatti}, {Nardiello}, {D'Orazi}, {Lanza}, {Locci}, {Marzari}, {Mesa}, {Messina}, {Pillitteri}, {Sozzetti}, {Girard}, {Maggio}, {Micela}, {Malavolta}, {Nascimbeni}, {Pinamonti}, {Squicciarini}, {Alcal{\'a}}, {Biazzo}, {Bohn}, {Bonavita}, {Brooks}, {Chauvin}, {Covino}, {Delorme}, {Hagelberg}, {Janson}, {Lagrange}, \& {Lazzoni}}]{Desidera2023}
{Desidera}, S., {Damasso}, M., {Gratton}, R., {et~al.} 2023, \aap, 675, A158, \dodoi{10.1051/0004-6361/202244611}

\bibitem[{{D{\'\i}az} {et~al.}(2016){D{\'\i}az}, {S{\'e}gransan}, {Udry}, {Lovis}, {Pepe}, {Dumusque}, {Marmier}, {Alonso}, {Benz}, {Bouchy}, {Coffinet}, {Collier Cameron}, {Deleuil}, {Figueira}, {Gillon}, {Lo Curto}, {Mayor}, {Mordasini}, {Motalebi}, {Moutou}, {Pollacco}, {Pompei}, {Queloz}, {Santos}, \& {Wyttenbach}}]{Diaz2016}
{D{\'\i}az}, R.~F., {S{\'e}gransan}, D., {Udry}, S., {et~al.} 2016, \aap, 585, A134, \dodoi{10.1051/0004-6361/201526729}

\bibitem[{{Dragomir} {et~al.}(2019){Dragomir}, {Teske}, {G{\"u}nther}, {S{\'e}gransan}, {Burt}, {Huang}, {Vanderburg}, {Matthews}, {Dumusque}, {Stassun}, {Pepper}, {Ricker}, {Vanderspek}, {Latham}, {Seager}, {Winn}, {Jenkins}, {Beatty}, {Bouchy}, {Brown}, {Butler}, {Ciardi}, {Crane}, {Eastman}, {Fossati}, {Francis}, {Fulton}, {Gaudi}, {Goeke}, {James}, {Klaus}, {Kuhn}, {Lovis}, {Lund}, {McDermott}, {Paegert}, {Pepe}, {Rodriguez}, {Sha}, {Shectman}, {Shporer}, {Siverd}, {Garcia Soto}, {Stevens}, {Twicken}, {Udry}, {Villanueva}, {Wang}, {Wohler}, {Yao}, \& {Zhan}}]{Dragomir2019}
{Dragomir}, D., {Teske}, J., {G{\"u}nther}, M.~N., {et~al.} 2019, \apjl, 875, L7, \dodoi{10.3847/2041-8213/ab12ed}

\bibitem[{{Dumusque} {et~al.}(2019){Dumusque}, {Turner}, {Dorn}, {Eastman}, {Allart}, {Adibekyan}, {Sousa}, {Santos}, {Mordasini}, {Bourrier}, {Bouchy}, {Coffinet}, {Davies}, {D{\'\i}az}, {Fausnaugh}, {Glidden}, {Guerrero}, {Henze}, {Jenkins}, {Latham}, {Lovis}, {Mayor}, {Pepe}, {Quintana}, {Ricker}, {Rowden}, {Segransan}, {Su{\'a}rez Mascare{\~n}o}, {Seager}, {Twicken}, {Udry}, {Vanderspek}, \& {Winn}}]{Dumusque2019}
{Dumusque}, X., {Turner}, O., {Dorn}, C., {et~al.} 2019, \aap, 627, A43, \dodoi{10.1051/0004-6361/201935457}

\bibitem[{{Espinoza} {et~al.}(2020){Espinoza}, {Brahm}, {Henning}, {Jord{\'a}n}, {Dorn}, {Rojas}, {Sarkis}, {Kossakowski}, {Schlecker}, {D{\'\i}az}, {Jenkins}, {Aguilera-Gomez}, {Jenkins}, {Twicken}, {Collins}, {Lissauer}, {Armstrong}, {Adibekyan}, {Barrado}, {Barros}, {Battley}, {Bayliss}, {Bouchy}, {Bryant}, {Cooke}, {Demangeon}, {Dumusque}, {Figueira}, {Giles}, {Lillo-Box}, {Lovis}, {Nielsen}, {Pepe}, {Pollacco}, {Santos}, {Sousa}, {Udry}, {Wheatley}, {Turner}, {Marmier}, {S{\'e}gransan}, {Ricker}, {Latham}, {Seager}, {Winn}, {Kielkopf}, {Hart}, {Wingham}, {Jensen}, {He{\l}miniak}, {Tokovinin}, {Brice{\~n}o}, {Ziegler}, {Law}, {Mann}, {Daylan}, {Doty}, {Guerrero}, {Boyd}, {Crossfield}, {Morris}, {Henze}, \& {Chacon}}]{Espinoza2020}
{Espinoza}, N., {Brahm}, R., {Henning}, T., {et~al.} 2020, \mnras, 491, 2982, \dodoi{10.1093/mnras/stz3150}

\bibitem[{{Feng} {et~al.}(2017){Feng}, {Tuomi}, \& {Jones}}]{Feng2017}
{Feng}, F., {Tuomi}, M., \& {Jones}, H.~R.~A. 2017, \mnras, 470, 4794, \dodoi{10.1093/mnras/stx1126}

\bibitem[{{Feng} {et~al.}(2019){Feng}, {Crane}, {Xuesong Wang}, {Teske}, {Shectman}, {D{\'\i}az}, {Thompson}, {Jones}, \& {Butler}}]{Feng2019}
{Feng}, F., {Crane}, J.~D., {Xuesong Wang}, S., {et~al.} 2019, \apjs, 242, 25, \dodoi{10.3847/1538-4365/ab1b16}

\bibitem[{{Feng} {et~al.}(2020){Feng}, {Shectman}, {Clement}, {Vogt}, {Tuomi}, {Teske}, {Burt}, {Crane}, {Holden}, {Wang}, {Thompson}, {D{\'\i}az}, \& {Butler}}]{Feng2020}
{Feng}, F., {Shectman}, S.~A., {Clement}, M.~S., {et~al.} 2020, \apjs, 250, 29, \dodoi{10.3847/1538-4365/abb139}

\bibitem[{{Feng} {et~al.}(2022){Feng}, {Butler}, {Vogt}, {Clement}, {Tinney}, {Cui}, {Aizawa}, {Jones}, {Bailey}, {Burt}, {Carter}, {Crane}, {Flammini Dotti}, {Holden}, {Ma}, {Ogihara}, {Oppenheimer}, {O'Toole}, {Shectman}, {Wittenmyer}, {Wang}, {Wright}, \& {Xuan}}]{Feng2022}
{Feng}, F., {Butler}, R.~P., {Vogt}, S.~S., {et~al.} 2022, \apjs, 262, 21, \dodoi{10.3847/1538-4365/ac7e57}

\bibitem[{{Gan} {et~al.}(2021){Gan}, {Bedell}, {Wang}, {Foreman-Mackey}, {Mel{\'e}ndez}, {Mao}, {Stassun}, {Howell}, {Ziegler}, {Wittenmyer}, {Hellier}, {Collins}, {Shporer}, {Ricker}, {Vanderspek}, {Latham}, {Seager}, {Winn}, {Jenkins}, {Addison}, {Ballard}, {Barclay}, {Bean}, {Bowler}, {Brice{\~n}o}, {Crossfield}, {Dittman}, {Horner}, {Jensen}, {Kane}, {Kielkopf}, {Kreidberg}, {Law}, {Mann}, {Mengel}, {Morgan}, {Okumura}, {Osborn}, {Paegert}, {Plavchan}, {Schwarz}, {Shiao}, {Smith}, {Spina}, {Tinney}, {Torres}, {Twicken}, {Vezie}, {Wang}, {Wright}, \& {Zhang}}]{Gan2021}
{Gan}, T., {Bedell}, M., {Wang}, S.~X., {et~al.} 2021, \mnras, 507, 2220, \dodoi{10.1093/mnras/stab2224}

\bibitem[{{Gandolfi} {et~al.}(2019){Gandolfi}, {Fossati}, {Livingston}, {Stassun}, {Grziwa}, {Barrag{\'a}n}, {Fridlund}, {Kubyshkina}, {Persson}, {Dai}, {Lam}, {Albrecht}, {Batalha}, {Beck}, {Justesen}, {Cabrera}, {Cartwright}, {Cochran}, {Csizmadia}, {Davies}, {Deeg}, {Eigm{\"u}ller}, {Endl}, {Erikson}, {Esposito}, {Garc{\'\i}a}, {Goeke}, {Gonz{\'a}lez-Cuesta}, {Guenther}, {Hatzes}, {Hidalgo}, {Hirano}, {Hjorth}, {Kabath}, {Knudstrup}, {Korth}, {Li}, {Luque}, {Mathur}, {Monta{\~n}es Rodr{\'\i}guez}, {Narita}, {Nespral}, {Niraula}, {Nowak}, {Palle}, {P{\"a}tzold}, {Prieto-Arranz}, {Rauer}, {Redfield}, {Ribas}, {Skarka}, {Smith}, {Rowden}, {Torres}, {Van Eylen}, \& {Vezie}}]{Gandolfi2019}
{Gandolfi}, D., {Fossati}, L., {Livingston}, J.~H., {et~al.} 2019, \apjl, 876, L24, \dodoi{10.3847/2041-8213/ab17d9}

\bibitem[{{Hara} {et~al.}(2020){Hara}, {Bouchy}, {Stalport}, {Boisse}, {Rodrigues}, {Delisle}, {Santerne}, {Henry}, {Arnold}, {Astudillo-Defru}, {Borgniet}, {Bonfils}, {Bourrier}, {Brugger}, {Courcol}, {Dalal}, {Deleuil}, {Delfosse}, {Demangeon}, {D{\'\i}az}, {Dumusque}, {Forveille}, {H{\'e}brard}, {Hobson}, {Kiefer}, {Lopez}, {Mignon}, {Mousis}, {Moutou}, {Pepe}, {Rey}, {Santos}, {S{\'e}gransan}, {Udry}, \& {Wilson}}]{Hara2020}
{Hara}, N.~C., {Bouchy}, F., {Stalport}, M., {et~al.} 2020, \aap, 636, L6, \dodoi{10.1051/0004-6361/201937254}

\bibitem[{{Hatzes} {et~al.}(2022){Hatzes}, {Gandolfi}, {Korth}, {Rodler}, {Sabotta}, {Esposito}, {Barrag{\'a}n}, {Van Eylen}, {Livingston}, {Serrano}, {Luque}, {Smith}, {Redfield}, {Persson}, {P{\"a}tzold}, {Palle}, {Nowak}, {Osborne}, {Narita}, {Mathur}, {Lam}, {Kab{\'a}th}, {Johnson}, {Guenther}, {Grziwa}, {Goffo}, {Fridlund}, {Endl}, {Deeg}, {Csizmadia}, {Cochran}, {Cuesta}, {Chaturvedi}, {Carleo}, {Cabrera}, {Beck}, \& {Albrecht}}]{Hatzes2022}
{Hatzes}, A.~P., {Gandolfi}, D., {Korth}, J., {et~al.} 2022, \aj, 163, 223, \dodoi{10.3847/1538-3881/ac5dcb}

\bibitem[{{H{\'e}brard} {et~al.}(2010){H{\'e}brard}, {Udry}, {Lo Curto}, {Robichon}, {Naef}, {Ehrenreich}, {Benz}, {Bouchy}, {Lecavelier Des Etangs}, {Lovis}, {Mayor}, {Moutou}, {Pepe}, {Queloz}, {Santos}, \& {S{\'e}gransan}}]{Hebrard2010}
{H{\'e}brard}, G., {Udry}, S., {Lo Curto}, G., {et~al.} 2010, \aap, 512, A46, \dodoi{10.1051/0004-6361/200913525}

\bibitem[{{Heidari} {et~al.}(2022){Heidari}, {Boisse}, {Orell-Miquel}, {H{\'e}brard}, {Acu{\~n}a}, {Hara}, {Lillo-Box}, {Eastman}, {Arnold}, {Astudillo-Defru}, {Adibekyan}, {Bieryla}, {Bonfils}, {Bouchy}, {Barclay}, {Brasseur}, {Borgniet}, {Bourrier}, {Buchhave}, {Behmard}, {Beard}, {Batalha}, {Courcol}, {Cort{\'e}s-Zuleta}, {Collins}, {Carmona}, {Crossfield}, {Chontos}, {Delfosse}, {Dalal}, {Deleuil}, {Demangeon}, {D{\'\i}az}, {Dumusque}, {Daylan}, {Dragomir}, {Delgado Mena}, {Dressing}, {Dai}, {Dalba}, {Ehrenreich}, {Forveille}, {Fulton}, {Fetherolf}, {Gaisn{\'e}}, {Giacalone}, {Riazi}, {Hoyer}, {Hobson}, {Howard}, {Huber}, {Hill}, {Hirsch}, {Isaacson}, {Jenkins}, {Kane}, {Kiefer}, {Luque}, {Latham}, {Lubin}, {Lopez}, {Mousis}, {Moutou}, {Montagnier}, {Mignon}, {Mayo}, {Mo{\v{c}}nik}, {Murphy}, {Palle}, {Pepe}, {Petigura}, {Rey}, {Ricker}, {Robertson}, {Roy}, {Rubenzahl}, {Rosenthal}, {Santerne}, {Santos}, {Sousa}, {Stassun}, {Stalport}, {Scarsdale}, {Str{\o}m}, {Seager}, {Segransan}, {Tenenbaum},
  {Tronsgaard}, {Udry}, {Vanderspek}, {Vakili}, {Winn}, \& {Weiss}}]{Heidari2022}
{Heidari}, N., {Boisse}, I., {Orell-Miquel}, J., {et~al.} 2022, \aap, 658, A176, \dodoi{10.1051/0004-6361/202141429}

\bibitem[{{Heidari} {et~al.}(2024){Heidari}, {Boisse}, {Hara}, {Wilson}, {Kiefer}, {H{\'e}brard}, {Philipot}, {Hoyer}, {Stassun}, {Henry}, {Santos}, {Acu{\~n}a}, {Almasian}, {Arnold}, {Astudillo-Defru}, {Attia}, {Bonfils}, {Bouchy}, {Bourrier}, {Collet}, {Cort{\'e}s-Zuleta}, {Carmona}, {Delfosse}, {Dalal}, {Deleuil}, {Demangeon}, {D{\'\i}az}, {Dumusque}, {Ehrenreich}, {Forveille}, {Hobson}, {Jenkins}, {Jenkins}, {Lagrange}, {Latham}, {Larue}, {Liu}, {Moutou}, {Mignon}, {Osborn}, {Pepe}, {Rapetti}, {Rodrigues}, {Santerne}, {Segransan}, {Shporer}, {Sulis}, {Torres}, {Udry}, {Vakili}, {Vanderburg}, {Venot}, {Vivien}, \& {Vines}}]{Heidari2024}
{Heidari}, N., {Boisse}, I., {Hara}, N.~C., {et~al.} 2024, \aap, 681, A55, \dodoi{10.1051/0004-6361/202347897}

\bibitem[{{Hellier} {et~al.}(2017){Hellier}, {Anderson}, {Collier Cameron}, {Delrez}, {Gillon}, {Jehin}, {Lendl}, {Maxted}, {Neveu-VanMalle}, {Pepe}, {Pollacco}, {Queloz}, {S{\'e}gransan}, {Smalley}, {Southworth}, {Triaud}, {Udry}, {Wagg}, \& {West}}]{Hellier2017}
{Hellier}, C., {Anderson}, D.~R., {Collier Cameron}, A., {et~al.} 2017, \mnras, 465, 3693, \dodoi{10.1093/mnras/stw3005}

\bibitem[{{Herman} {et~al.}(2019){Herman}, {Zhu}, \& {Wu}}]{Herman19}
{Herman}, M.~K., {Zhu}, W., \& {Wu}, Y. 2019, \aj, 157, 248, \dodoi{10.3847/1538-3881/ab1f70}

\bibitem[{{Hidalgo} {et~al.}(2020){Hidalgo}, {Pall{\'e}}, {Alonso}, {Gandolfi}, {Fridlund}, {Nowak}, {Luque}, {Hirano}, {Justesen}, {Cochran}, {Barrag{\'a}n}, {Spina}, {Rodler}, {Albrecht}, {Anderson}, {Amado}, {Bryant}, {Caballero}, {Cabrera}, {Csizmadia}, {Dai}, {De Leon}, {Deeg}, {Eigmuller}, {Endl}, {Erikson}, {Esposito}, {Figueira}, {Georgieva}, {Grziwa}, {Guenther}, {Hatzes}, {Hjorth}, {Hoeijmakers}, {Kabath}, {Korth}, {Kuzuhara}, {Lafarga}, {Lampon}, {Le{\~a}o}, {Livingston}, {Mathur}, {Monta{\~n}es-Rodriguez}, {Morales}, {Murgas}, {Nagel}, {Narita}, {Nielsen}, {Patzold}, {Persson}, {Prieto-Arranz}, {Quirrenbach}, {Rauer}, {Redfield}, {Reiners}, {Ribas}, {Smith}, {{\v{S}}ubjak}, {Van Eylen}, \& {Wilson}}]{Hidalgo2020}
{Hidalgo}, D., {Pall{\'e}}, E., {Alonso}, R., {et~al.} 2020, \aap, 636, A89, \dodoi{10.1051/0004-6361/201937080}

\bibitem[{{Horner} {et~al.}(2019){Horner}, {Wittenmyer}, {Wright}, {Hinse}, {Marshall}, {Kane}, {Clark}, {Mengel}, {Agnew}, \& {Johns}}]{Horner2019}
{Horner}, J., {Wittenmyer}, R.~A., {Wright}, D.~J., {et~al.} 2019, \aj, 158, 100, \dodoi{10.3847/1538-3881/ab2e78}

\bibitem[{{Hua} {et~al.}(2023){Hua}, {Wang}, {Teske}, {Gan}, {Shporer}, {Zhou}, {Stassun}, {Rabus}, {Howell}, {Ziegler}, {Lissauer}, {Winn}, {Jenkins}, {Ting}, {Collins}, {Mann}, {Zhu}, {Wang}, {Butler}, {Crane}, {Shectman}, {Bouma}, {Brice{\~n}o}, {Dragomir}, {Fong}, {Law}, {Medina}, {Quinn}, {Ricker}, {Schwarz}, {Seager}, {Sefako}, {Stockdale}, {Vanderspek}, \& {Villase{\~n}or}}]{Hua2023}
{Hua}, X., {Wang}, S.~X., {Teske}, J.~K., {et~al.} 2023, \aj, 166, 32, \dodoi{10.3847/1538-3881/acd751}

\bibitem[{{Jenkins} {et~al.}(2013){Jenkins}, {Jones}, {Tuomi}, {Murgas}, {Hoyer}, {Jones}, {Barnes}, {Pavlenko}, {Ivanyuk}, {Rojo}, {Jord{\'a}n}, {Day-Jones}, {Ruiz}, \& {Pinfield}}]{Jenkins2013}
{Jenkins}, J.~S., {Jones}, H.~R.~A., {Tuomi}, M., {et~al.} 2013, \apj, 766, 67, \dodoi{10.1088/0004-637X/766/2/67}

\bibitem[{{Kane} {et~al.}(2020){Kane}, {Yal{\c{c}}{\i}nkaya}, {Osborn}, {Dalba}, {Nielsen}, {Vanderburg}, {Mo{\v{c}}nik}, {Hinkel}, {Ostberg}, {Esmer}, {Udry}, {Fetherolf}, {Ba{\c{s}}t{\"u}rk}, {Ricker}, {Vanderspek}, {Latham}, {Seager}, {Winn}, {Jenkins}, {Allart}, {Bailey}, {Bean}, {Bouchy}, {Butler}, {Campante}, {Carter}, {Daylan}, {Deleuil}, {Diaz}, {Dumusque}, {Ehrenreich}, {Horner}, {Howard}, {Isaacson}, {Jones}, {Kristiansen}, {Lovis}, {Marcy}, {Marmier}, {O'Toole}, {Pepe}, {Ragozzine}, {S{\'e}gransan}, {Tinney}, {Turnbull}, {Wittenmyer}, {Wright}, \& {Wright}}]{Kane2020}
{Kane}, S.~R., {Yal{\c{c}}{\i}nkaya}, S., {Osborn}, H.~P., {et~al.} 2020, \aj, 160, 129, \dodoi{10.3847/1538-3881/aba835}

\bibitem[{{Kimura} {et~al.}(2020){Kimura}, {Wada}, {Kobayashi}, {Senshu}, {Hirai}, {Yoshida}, {Kobayashi}, {Hong}, {Arai}, {Ishibashi}, \& {Yamada}}]{Kimura20}
{Kimura}, H., {Wada}, K., {Kobayashi}, H., {et~al.} 2020, \mnras, 498, 1801, \dodoi{10.1093/mnras/staa2467}

\bibitem[{{Kunimoto} {et~al.}(2023){Kunimoto}, {Vanderburg}, {Huang}, {Davis}, {Affer}, {Cameron}, {Charbonneau}, {Cosentino}, {Damasso}, {Dumusque}, {Fiorenzano}, {Ghedina}, {Haywood}, {Lienhard}, {L{\'o}pez-Morales}, {Mayor}, {Pepe}, {Pinamonti}, {Poretti}, {Maldonado}, {Rice}, {Sozzetti}, {Wilson}, {Udry}, {Baptista}, {Barkaoui}, {Becker}, {Benni}, {Bieryla}, {Bosch-Cabot}, {Ciardi}, {Collins}, {Collins}, {Evans}, {Dupuy}, {Goliguzova}, {Guerra}, {Kraus}, {Lissauer}, {Huber}, {Murgas}, {Palle}, {Quinn}, {Safonov}, {Schwarz}, {Shporer}, {Stassun}, {Jenkins}, {Latham}, {Ricker}, {Seager}, {Vanderspek}, {Winn}, {Essack}, {Lewis}, \& {Rose}}]{Kunimoto2023}
{Kunimoto}, M., {Vanderburg}, A., {Huang}, C.~X., {et~al.} 2023, \aj, 166, 7, \dodoi{10.3847/1538-3881/acd537}

\bibitem[{{Lam} {et~al.}(2018){Lam}, {Santerne}, {Sousa}, {Vigan}, {Armstrong}, {Barros}, {Brugger}, {Adibekyan}, {Almenara}, {Delgado Mena}, {Dumusque}, {Barrado}, {Bayliss}, {Bonomo}, {Bouchy}, {Brown}, {Ciardi}, {Deleuil}, {Demangeon}, {Faedi}, {Foxell}, {Jackman}, {King}, {Kirk}, {Ligi}, {Lillo-Box}, {Lopez}, {Lovis}, {Louden}, {Nielsen}, {McCormac}, {Mousis}, {Osborn}, {Pollacco}, {Santos}, {Udry}, \& {Wheatley}}]{Lam2018}
{Lam}, K.~W.~F., {Santerne}, A., {Sousa}, S.~G., {et~al.} 2018, \aap, 620, A77, \dodoi{10.1051/0004-6361/201834073}

\bibitem[{{Lillo-Box} {et~al.}(2020){Lillo-Box}, {Lopez}, {Santerne}, {Nielsen}, {Barros}, {Deleuil}, {Acu{\~n}a}, {Mousis}, {Sousa}, {Adibekyan}, {Armstrong}, {Barrado}, {Bayliss}, {Brown}, {Demangeon}, {Dumusque}, {Figueira}, {Hojjatpanah}, {Osborn}, {Santos}, \& {Udry}}]{LilloBox2020}
{Lillo-Box}, J., {Lopez}, T.~A., {Santerne}, A., {et~al.} 2020, \aap, 640, A48, \dodoi{10.1051/0004-6361/202037896}

\bibitem[{{Lillo-Box} {et~al.}(2021){Lillo-Box}, {Faria}, {Su{\'a}rez Mascare{\~n}o}, {Figueira}, {Sousa}, {Tabernero}, {Lovis}, {Silva}, {Demangeon}, {Benatti}, {Santos}, {Mehner}, {Pepe}, {Sozzetti}, {Zapatero Osorio}, {Gonz{\'a}lez Hern{\'a}ndez}, {Micela}, {Hojjatpanah}, {Rebolo}, {Cristiani}, {Adibekyan}, {Allart}, {Allende Prieto}, {Cabral}, {Damasso}, {Di Marcantonio}, {Lo Curto}, {Martins}, {Megevand}, {Molaro}, {Nunes}, {Pall{\'e}}, {Pasquini}, {Poretti}, \& {Udry}}]{LilloBox2021}
{Lillo-Box}, J., {Faria}, J.~P., {Su{\'a}rez Mascare{\~n}o}, A., {et~al.} 2021, \aap, 654, A60, \dodoi{10.1051/0004-6361/202141714}

\bibitem[{{Lillo-Box} {et~al.}(2023){Lillo-Box}, {Gandolfi}, {Armstrong}, {Collins}, {Nielsen}, {Luque}, {Korth}, {Sousa}, {Quinn}, {Acu{\~n}a}, {Howell}, {Morello}, {Hellier}, {Giacalone}, {Hoyer}, {Stassun}, {Palle}, {Aguichine}, {Mousis}, {Adibekyan}, {Azevedo Silva}, {Barrado}, {Deleuil}, {Eastman}, {Fukui}, {Hawthorn}, {Irwin}, {Jenkins}, {Latham}, {Muresan}, {Narita}, {Persson}, {Santerne}, {Santos}, {Savel}, {Osborn}, {Teske}, {Wheatley}, {Winn}, {Barros}, {Butler}, {Caldwell}, {Charbonneau}, {Cloutier}, {Crane}, {Demangeon}, {D{\'\i}az}, {Dumusque}, {Esposito}, {Falk}, {Gill}, {Hojjatpanah}, {Kreidberg}, {Mireles}, {Osborn}, {Ricker}, {Rodriguez}, {Schwarz}, {Seager}, {Serrano Bell}, {Shectman}, {Shporer}, {Vezie}, {Wang}, \& {Zhou}}]{LilloBox2023}
{Lillo-Box}, J., {Gandolfi}, D., {Armstrong}, D.~J., {et~al.} 2023, \aap, 669, A109, \dodoi{10.1051/0004-6361/202243879}

\bibitem[{{Lin} {et~al.}(2018){Lin}, {Lee}, \& {Chiang}}]{Lin18}
{Lin}, J.~W., {Lee}, E.~J., \& {Chiang}, E. 2018, \mnras, 480, 4338, \dodoi{10.1093/mnras/sty2159}

\bibitem[{{Lo Curto} {et~al.}(2010){Lo Curto}, {Mayor}, {Benz}, {Bouchy}, {Lovis}, {Moutou}, {Naef}, {Pepe}, {Queloz}, {Santos}, {Segransan}, \& {Udry}}]{LoCurto2010}
{Lo Curto}, G., {Mayor}, M., {Benz}, W., {et~al.} 2010, \aap, 512, A48, \dodoi{10.1051/0004-6361/200913523}

\bibitem[{{Lo Curto} {et~al.}(2013){Lo Curto}, {Mayor}, {Benz}, {Bouchy}, {H{\'e}brard}, {Lovis}, {Moutou}, {Naef}, {Pepe}, {Queloz}, {Santos}, {Segransan}, \& {Udry}}]{LoCurto2013}
---. 2013, \aap, 551, A59, \dodoi{10.1051/0004-6361/201220415}

\bibitem[{{Lovis} {et~al.}(2011){Lovis}, {S{\'e}gransan}, {Mayor}, {Udry}, {Benz}, {Bertaux}, {Bouchy}, {Correia}, {Laskar}, {Lo Curto}, {Mordasini}, {Pepe}, {Queloz}, \& {Santos}}]{Lovis2011}
{Lovis}, C., {S{\'e}gransan}, D., {Mayor}, M., {et~al.} 2011, \aap, 528, A112, \dodoi{10.1051/0004-6361/201015577}

\bibitem[{{Lubin} {et~al.}(2022){Lubin}, {Van Zandt}, {Holcomb}, {Weiss}, {Petigura}, {Robertson}, {Akana Murphy}, {Scarsdale}, {Batygin}, {Polanski}, {Batalha}, {Crossfield}, {Dressing}, {Fulton}, {Howard}, {Huber}, {Isaacson}, {Kane}, {Roy}, {Beard}, {Blunt}, {Chontos}, {Dai}, {Dalba}, {Gary}, {Giacalone}, {Hill}, {Mayo}, {Mo{\v{c}}nik}, {Kosiarek}, {Rice}, {Rubenzahl}, {Latham}, {Seager}, {Winn}, \& {Gary}}]{Lubin2022}
{Lubin}, J., {Van Zandt}, J., {Holcomb}, R., {et~al.} 2022, \aj, 163, 101, \dodoi{10.3847/1538-3881/ac3d38}

\bibitem[{{Luque} {et~al.}(2023){Luque}, {Osborn}, {Leleu}, {Pall{\'e}}, {Bonfanti}, {Barrag{\'a}n}, {Wilson}, {Broeg}, {Cameron}, {Lendl}, {Maxted}, {Alibert}, {Gandolfi}, {Delisle}, {Hooton}, {Egger}, {Nowak}, {Lafarga}, {Rapetti}, {Twicken}, {Morales}, {Carleo}, {Orell-Miquel}, {Adibekyan}, {Alonso}, {Alqasim}, {Amado}, {Anderson}, {Anglada-Escud{\'e}}, {Bandy}, {B{\'a}rczy}, {Barrado Navascues}, {Barros}, {Baumjohann}, {Bayliss}, {Bean}, {Beck}, {Beck}, {Benz}, {Billot}, {Bonfils}, {Borsato}, {Boyle}, {Brandeker}, {Bryant}, {Cabrera}, {Carrazco-Gaxiola}, {Charbonneau}, {Charnoz}, {Ciardi}, {Cochran}, {Collins}, {Crossfield}, {Csizmadia}, {Cubillos}, {Dai}, {Davies}, {Deeg}, {Deleuil}, {Deline}, {Delrez}, {Demangeon}, {Demory}, {Ehrenreich}, {Erikson}, {Esparza-Borges}, {Falk}, {Fortier}, {Fossati}, {Fridlund}, {Fukui}, {Garcia-Mejia}, {Gill}, {Gillon}, {Goffo}, {G{\'o}mez Maqueo Chew}, {G{\"u}del}, {Guenther}, {G{\"u}nther}, {Hatzes}, {Helling}, {Hesse}, {Howell}, {Hoyer}, {Ikuta}, {Isaak}, {Jenkins},
  {Kagetani}, {Kiss}, {Kodama}, {Korth}, {Lam}, {Laskar}, {Latham}, {Lecavelier des Etangs}, {Leon}, {Livingston}, {Magrin}, {Matson}, {Matthews}, {Mordasini}, {Mori}, {Moyano}, {Munari}, {Murgas}, {Narita}, {Nascimbeni}, {Olofsson}, {Osborne}, {Ottensamer}, {Pagano}, {Parviainen}, {Peter}, {Piotto}, {Pollacco}, {Queloz}, {Quinn}, {Quirrenbach}, {Ragazzoni}, {Rando}, {Ratti}, {Rauer}, {Redfield}, {Ribas}, {Ricker}, {Rudat}, {Sabin}, {Salmon}, {Santos}, {Scandariato}, {Schanche}, {Schlieder}, {Seager}, {S{\'e}gransan}, {Shporer}, {Simon}, {Smith}, {Sousa}, {Stalport}, {Szab{\'o}}, {Thomas}, {Tuson}, {Udry}, {Vanderburg}, {Van Eylen}, {Van Grootel}, {Venturini}, {Walter}, {Walton}, {Watanabe}, {Winn}, \& {Zingales}}]{Luque2023}
{Luque}, R., {Osborn}, H.~P., {Leleu}, A., {et~al.} 2023, \nat, 623, 932, \dodoi{10.1038/s41586-023-06692-3}

\bibitem[{{Ma} {et~al.}(2018){Ma}, {Ge}, {Muterspaugh}, {Singer}, {Henry}, {Gonz{\'a}lez Hern{\'a}ndez}, {Sithajan}, {Jeram}, {Williamson}, {Stassun}, {Kimock}, {Varosi}, {Schofield}, {Liu}, {Powell}, {Cassette}, {Jakeman}, {Avner}, {Grieves}, {Barnes}, {Zhao}, {Gilda}, {Grantham}, {Stafford}, {Savage}, {Bland}, \& {Ealey}}]{Ma2018}
{Ma}, B., {Ge}, J., {Muterspaugh}, M., {et~al.} 2018, \mnras, 480, 2411, \dodoi{10.1093/mnras/sty1933}

\bibitem[{{Mantovan} {et~al.}(2024){Mantovan}, {Malavolta}, {Desidera}, {Zingales}, {Borsato}, {Piotto}, {Maggio}, {Locci}, {Polychroni}, {Turrini}, {Baratella}, {Biazzo}, {Nardiello}, {Stassun}, {Nascimbeni}, {Benatti}, {Anna John}, {Watkins}, {Bieryla}, {Lissauer}, {Twicken}, {Lanza}, {Winn}, {Messina}, {Montalto}, {Sozzetti}, {Boffin}, {Cheryasov}, {Strakhov}, {Murgas}, {D'Arpa}, {Barkaoui}, {Benni}, {Bignamini}, {Bonomo}, {Borsa}, {Cabona}, {Cameron}, {Claudi}, {Cochran}, {Collins}, {Damasso}, {Dong}, {Endl}, {Fukui}, {F{\H{u}}r{\'e}sz}, {Gandolfi}, {Ghedina}, {Jenkins}, {Kab{\'a}th}, {Latham}, {Lorenzi}, {Luque}, {Maldonado}, {McLeod}, {Molinaro}, {Narita}, {Nowak}, {Orell-Miquel}, {Pall{\'e}}, {Parviainen}, {Pedani}, {Quinn}, {Relles}, {Rowden}, {Scandariato}, {Schwarz}, {Seager}, {Shporer}, {Vanderburg}, \& {Wilson}}]{Mantovan2024}
{Mantovan}, G., {Malavolta}, L., {Desidera}, S., {et~al.} 2024, \aap, 682, A129, \dodoi{10.1051/0004-6361/202347472}

\bibitem[{{Marcy} {et~al.}(2014){Marcy}, {Isaacson}, {Howard}, {Rowe}, {Jenkins}, {Bryson}, {Latham}, {Howell}, {Gautier}, {Batalha}, {Rogers}, {Ciardi}, {Fischer}, {Gilliland}, {Kjeldsen}, {Christensen-Dalsgaard}, {Huber}, {Chaplin}, {Basu}, {Buchhave}, {Quinn}, {Borucki}, {Koch}, {Hunter}, {Caldwell}, {Van Cleve}, {Kolbl}, {Weiss}, {Petigura}, {Seager}, {Morton}, {Johnson}, {Ballard}, {Burke}, {Cochran}, {Endl}, {MacQueen}, {Everett}, {Lissauer}, {Ford}, {Torres}, {Fressin}, {Brown}, {Steffen}, {Charbonneau}, {Basri}, {Sasselov}, {Winn}, {Sanchis-Ojeda}, {Christiansen}, {Adams}, {Henze}, {Dupree}, {Fabrycky}, {Fortney}, {Tarter}, {Holman}, {Tenenbaum}, {Shporer}, {Lucas}, {Welsh}, {Orosz}, {Bedding}, {Campante}, {Davies}, {Elsworth}, {Handberg}, {Hekker}, {Karoff}, {Kawaler}, {Lund}, {Lundkvist}, {Metcalfe}, {Miglio}, {Silva Aguirre}, {Stello}, {White}, {Boss}, {Devore}, {Gould}, {Prsa}, {Agol}, {Barclay}, {Coughlin}, {Brugamyer}, {Mullally}, {Quintana}, {Still}, {Thompson}, {Morrison}, {Twicken},
  {D{\'e}sert}, {Carter}, {Crepp}, {H{\'e}brard}, {Santerne}, {Moutou}, {Sobeck}, {Hudgins}, {Haas}, {Robertson}, {Lillo-Box}, \& {Barrado}}]{Marcy2014}
{Marcy}, G.~W., {Isaacson}, H., {Howard}, A.~W., {et~al.} 2014, \apjs, 210, 20, \dodoi{10.1088/0067-0049/210/2/20}

\bibitem[{{Martioli} {et~al.}(2022){Martioli}, {H{\'e}brard}, {Fouqu{\'e}}, {Artigau}, {Donati}, {Cadieux}, {Bellotti}, {Lecavelier des Etangs}, {Doyon}, {do Nascimento}, {Arnold}, {Carmona}, {Cook}, {Cortes-Zuleta}, {de Almeida}, {Delfosse}, {Folsom}, {K{\"o}nig}, {Moutou}, {Ould-Elhkim}, {Petit}, {Stassun}, {Vidotto}, {Vandal}, {Benneke}, {Boisse}, {Bonfils}, {Boyd}, {Brasseur}, {Charbonneau}, {Cloutier}, {Collins}, {Cristofari}, {Crossfield}, {D{\'\i}az}, {Fausnaugh}, {Figueira}, {Forveille}, {Furlan}, {Girardin}, {Gnilka}, {Gomes da Silva}, {Gu}, {Guerra}, {Howell}, {Hussain}, {Jenkins}, {Kiefer}, {Latham}, {Matson}, {Matthews}, {Morin}, {Naves}, {Ricker}, {Seager}, {Takami}, {Twicken}, {Vanderburg}, {Vanderspek}, \& {Winn}}]{Martioli2022}
{Martioli}, E., {H{\'e}brard}, G., {Fouqu{\'e}}, P., {et~al.} 2022, \aap, 660, A86, \dodoi{10.1051/0004-6361/202142540}

\bibitem[{{Martioli} {et~al.}(2023){Martioli}, {H{\'e}brard}, {de Almeida}, {Heidari}, {Lorenzo-Oliveira}, {Kiefer}, {Almenara}, {Bieryla}, {Boisse}, {Bonfils}, {Brice{\~n}o}, {Collins}, {Cort{\'e}s-Zuleta}, {Dalal}, {Deleuil}, {Delfosse}, {Demangeon}, {Eastman}, {Forveille}, {Furlan}, {Howell}, {Hoyer}, {Jenkins}, {Latham}, {Law}, {Mann}, {Moutou}, {Santos}, {Sousa}, {Stassun}, {Stockdale}, {Torres}, {Twicken}, {Winn}, \& {Ziegler}}]{Martioli2023}
{Martioli}, E., {H{\'e}brard}, G., {de Almeida}, L., {et~al.} 2023, \aap, 680, A84, \dodoi{10.1051/0004-6361/202347744}

\bibitem[{{Mayo} {et~al.}(2023){Mayo}, {Dressing}, {Vanderburg}, {Fortenbach}, {Lienhard}, {Malavolta}, {Mortier}, {N{\'u}{\~n}ez}, {Richey-Yowell}, {Turtelboom}, {Bonomo}, {Latham}, {L{\'o}pez-Morales}, {Shkolnik}, {Sozzetti}, {Ag{\"u}eros}, {Borsato}, {Charbonneau}, {Cosentino}, {Douglas}, {Dumusque}, {Ghedina}, {Gibson}, {Granata}, {Harutyunyan}, {Haywood}, {Lacedelli}, {Lorenzi}, {Magazz{\`u}}, {Martinez Fiorenzano}, {Micela}, {Molinari}, {Montalto}, {Nardiello}, {Nascimbeni}, {Pagano}, {Piotto}, {Pino}, {Poretti}, {Scandariato}, {Udry}, \& {Buchhave}}]{Mayo2023}
{Mayo}, A.~W., {Dressing}, C.~D., {Vanderburg}, A., {et~al.} 2023, \aj, 165, 235, \dodoi{10.3847/1538-3881/acca1c}

\bibitem[{{Ment} {et~al.}(2018){Ment}, {Fischer}, {Bakos}, {Howard}, \& {Isaacson}}]{Ment2018}
{Ment}, K., {Fischer}, D.~A., {Bakos}, G., {Howard}, A.~W., \& {Isaacson}, H. 2018, \aj, 156, 213, \dodoi{10.3847/1538-3881/aae1f5}

\bibitem[{{Mills} {et~al.}(2019){Mills}, {Howard}, {Weiss}, {Steffen}, {Isaacson}, {Fulton}, {Petigura}, {Kosiarek}, {Hirsch}, \& {Boisvert}}]{Mills2019}
{Mills}, S.~M., {Howard}, A.~W., {Weiss}, L.~M., {et~al.} 2019, \aj, 157, 145, \dodoi{10.3847/1538-3881/ab0899}

\bibitem[{{Morbidelli} {et~al.}(2007){Morbidelli}, {Tsiganis}, {Crida}, {Levison}, \& {Gomes}}]{Morbidelli07}
{Morbidelli}, A., {Tsiganis}, K., {Crida}, A., {Levison}, H.~F., \& {Gomes}, R. 2007, \aj, 134, 1790, \dodoi{10.1086/521705}

\bibitem[{{Mortier} {et~al.}(2016){Mortier}, {Faria}, {Santos}, {Rajpaul}, {Figueira}, {Boisse}, {Collier Cameron}, {Dumusque}, {Lo Curto}, {Lovis}, {Mayor}, {Melo}, {Pepe}, {Queloz}, {Santerne}, {S{\'e}gransan}, {Sousa}, {Sozzetti}, \& {Udry}}]{Mortier2016}
{Mortier}, A., {Faria}, J.~P., {Santos}, N.~C., {et~al.} 2016, \aap, 585, A135, \dodoi{10.1051/0004-6361/201526905}

\bibitem[{{Mulders} {et~al.}(2021){Mulders}, {Dr{\k{a}}{\.z}kowska}, {van der Marel}, {Ciesla}, \& {Pascucci}}]{Mulders21}
{Mulders}, G.~D., {Dr{\k{a}}{\.z}kowska}, J., {van der Marel}, N., {Ciesla}, F.~J., \& {Pascucci}, I. 2021, \apjl, 920, L1, \dodoi{10.3847/2041-8213/ac2947}

\bibitem[{{Musiolik} \& {Wurm}(2019)}]{Musiolik19}
{Musiolik}, G., \& {Wurm}, G. 2019, \apj, 873, 58, \dodoi{10.3847/1538-4357/ab0428}

\bibitem[{{Naponiello} {et~al.}(2022){Naponiello}, {Mancini}, {Damasso}, {Bonomo}, {Sozzetti}, {Nardiello}, {Biazzo}, {Stognone}, {Lillo-Box}, {Lanza}, {Poretti}, {Lissauer}, {Zeng}, {Bieryla}, {H{\'e}brard}, {Basilicata}, {Benatti}, {Bignamini}, {Borsa}, {Claudi}, {Cosentino}, {Covino}, {de Gurtubai}, {Delfosse}, {Desidera}, {Dragomir}, {Eastman}, {Essack}, {Fiorenzano}, {Giacobbe}, {Harutyunyan}, {Heidari}, {Hellier}, {Jenkins}, {Knapic}, {K{\"o}nig}, {Latham}, {Magazz{\`u}}, {Maggio}, {Maldonado}, {Micela}, {Molinari}, {Molinaro}, {Morgan}, {Moutou}, {Nascimbeni}, {Pace}, {Pagano}, {Pedani}, {Piotto}, {Pinamonti}, {Quintana}, {Rainer}, {Ricker}, {Seager}, {Twicken}, {Vanderspek}, \& {Winn}}]{Naponiello2022}
{Naponiello}, L., {Mancini}, L., {Damasso}, M., {et~al.} 2022, \aap, 667, A8, \dodoi{10.1051/0004-6361/202244079}

\bibitem[{{Naponiello} {et~al.}(2023){Naponiello}, {Mancini}, {Sozzetti}, {Bonomo}, {Morbidelli}, {Dou}, {Zeng}, {Leinhardt}, {Biazzo}, {Cubillos}, {Pinamonti}, {Locci}, {Maggio}, {Damasso}, {Lanza}, {Lissauer}, {Collins}, {Carter}, {Jensen}, {Bignamini}, {Boschin}, {Bouma}, {Ciardi}, {Cosentino}, {Crossfield}, {Desidera}, {Dumusque}, {Fiorenzano}, {Fukui}, {Giacobbe}, {Gnilka}, {Ghedina}, {Guilluy}, {Harutyunyan}, {Howell}, {Jenkins}, {Lund}, {Kielkopf}, {Lester}, {Malavolta}, {Mann}, {Matson}, {Matthews}, {Nardiello}, {Narita}, {Pace}, {Pagano}, {Palle}, {Pedani}, {Seager}, {Schlieder}, {Schwarz}, {Shporer}, {Twicken}, {Winn}, {Ziegler}, \& {Zingales}}]{Naponiello2023}
{Naponiello}, L., {Mancini}, L., {Sozzetti}, A., {et~al.} 2023, \nat, 622, 255, \dodoi{10.1038/s41586-023-06499-2}

\bibitem[{{Nardiello} {et~al.}(2022){Nardiello}, {Malavolta}, {Desidera}, {Baratella}, {D'Orazi}, {Messina}, {Biazzo}, {Benatti}, {Damasso}, {Rajpaul}, {Bonomo}, {Capuzzo Dolcetta}, {Mallonn}, {Cale}, {Plavchan}, {El Mufti}, {Bignamini}, {Borsa}, {Carleo}, {Claudi}, {Covino}, {Lanza}, {Maldonado}, {Mancini}, {Micela}, {Molinari}, {Pinamonti}, {Piotto}, {Poretti}, {Scandariato}, {Sozzetti}, {Andreuzzi}, {Boschin}, {Cosentino}, {Fiorenzano}, {Harutyunyan}, {Knapic}, {Pedani}, {Affer}, {Maggio}, \& {Rainer}}]{Nardiello2022}
{Nardiello}, D., {Malavolta}, L., {Desidera}, S., {et~al.} 2022, \aap, 664, A163, \dodoi{10.1051/0004-6361/202243743}

\bibitem[{{Orell-Miquel} {et~al.}(2023){Orell-Miquel}, {Nowak}, {Murgas}, {Palle}, {Morello}, {Luque}, {Badenas-Agusti}, {Ribas}, {Lafarga}, {Espinoza}, {Morales}, {Zechmeister}, {Alqasim}, {Cochran}, {Gandolfi}, {Goffo}, {Kab{\'a}th}, {Korth}, {Lam}, {Livingston}, {Muresan}, {Persson}, \& {Van Eylen}}]{Orell2023}
{Orell-Miquel}, J., {Nowak}, G., {Murgas}, F., {et~al.} 2023, \aap, 669, A40, \dodoi{10.1051/0004-6361/202244120}

\bibitem[{{Osborn} {et~al.}(2021){Osborn}, {Armstrong}, {Cale}, {Brahm}, {Wittenmyer}, {Dai}, {Crossfield}, {Bryant}, {Adibekyan}, {Cloutier}, {Collins}, {Delgado Mena}, {Fridlund}, {Hellier}, {Howell}, {King}, {Lillo-Box}, {Otegi}, {Sousa}, {Stassun}, {Matthews}, {Ziegler}, {Ricker}, {Vanderspek}, {Latham}, {Seager}, {Winn}, {Jenkins}, {Acton}, {Addison}, {Anderson}, {Ballard}, {Barrado}, {Barros}, {Batalha}, {Bayliss}, {Barclay}, {Benneke}, {Berberian}, {Bouchy}, {Bowler}, {Brice{\~n}o}, {Burke}, {Burleigh}, {Casewell}, {Ciardi}, {Collins}, {Cooke}, {Demangeon}, {D{\'\i}az}, {Dorn}, {Dragomir}, {Dressing}, {Dumusque}, {Espinoza}, {Figueira}, {Fulton}, {Furlan}, {Gaidos}, {Geneser}, {Gill}, {Goad}, {Gonzales}, {Gorjian}, {G{\"u}nther}, {Helled}, {Henderson}, {Henning}, {Hogan}, {Hojjatpanah}, {Horner}, {Howard}, {Hoyer}, {Huber}, {Isaacson}, {Jenkins}, {Jensen}, {Jord{\'a}n}, {Kane}, {Kidwell}, {Kielkopf}, {Law}, {Lendl}, {Lund}, {Matson}, {Mann}, {McCormac}, {Mengel}, {Morales}, {Nielsen}, {Okumura},
  {Osborn}, {Petigura}, {Plavchan}, {Pollacco}, {Quintana}, {Raynard}, {Robertson}, {Rose}, {Roy}, {Reefe}, {Santerne}, {Santos}, {Sarkis}, {Schlieder}, {Schwarz}, {Scott}, {Shporer}, {Smith}, {Stibbard}, {Stockdale}, {Str{\o}m}, {Twicken}, {Tan}, {Tanner}, {Teske}, {Tilbrook}, {Tinney}, {Udry}, {Villase{\~n}or}, {Vines}, {Wang}, {Weiss}, {West}, {Wheatley}, {Wright}, {Zhang}, \& {Zohrabi}}]{Osborn2021}
{Osborn}, A., {Armstrong}, D.~J., {Cale}, B., {et~al.} 2021, \mnras, 507, 2782, \dodoi{10.1093/mnras/stab2313}

\bibitem[{{Osborn} {et~al.}(2023){Osborn}, {Nowak}, {H{\'e}brard}, {Masseron}, {Lillo-Box}, {Pall{\'e}}, {Bekkelien}, {Flor{\'e}n}, {Guterman}, {Simon}, {Adibekyan}, {Bieryla}, {Borsato}, {Brandeker}, {Ciardi}, {Collier Cameron}, {Collins}, {Egger}, {Gandolfi}, {Hooton}, {Latham}, {Lendl}, {Matthews}, {Tuson}, {Ulmer-Moll}, {Vanderburg}, {Wilson}, {Ziegler}, {Alibert}, {Alonso}, {Anglada}, {Arnold}, {Asquier}, {Barrado y Navascues}, {Baumjohann}, {Beck}, {Belinski}, {Benz}, {Biondi}, {Boisse}, {Bonfils}, {Broeg}, {Buchhave}, {B{\'a}rczy}, {Barros}, {Cabrera}, {Cardona Guillen}, {Carleo}, {Castro-Gonz{\'a}lez}, {Charnoz}, {Christiansen}, {Cortes-Zuleta}, {Csizmadia}, {Dalal}, {Davies}, {Deleuil}, {Delfosse}, {Delrez}, {Demory}, {Dunlavey}, {Ehrenreich}, {Erikson}, {Fernandes}, {Fortier}, {Forveille}, {Fossati}, {Fridlund}, {Gillon}, {Goeke}, {Goliguzova}, {Gonzales}, {G{\"u}nther}, {G{\"u}del}, {Heidari}, {Henze}, {Howell}, {Hoyer}, {Frey}, {Isaak}, {Jenkins}, {Kiefer}, {Kiss}, {Korth}, {Maxted}, {Laskar},
  {Lecavelier des Etangs}, {Lovis}, {Lund}, {Luque}, {Magrin}, {Almenara}, {Martioli}, {Mecina}, {Medina}, {Moldovan}, {Morales-Calder{\'o}n}, {Morello}, {Moutou}, {Murgas}, {Jensen}, {Nascimbeni}, {Olofsson}, {Ottensamer}, {Pagano}, {Peter}, {Piotto}, {Pollacco}, {Queloz}, {Ragazzoni}, {Rando}, {Rauer}, {Ribas}, {Ricker}, {Demangeon}, {Smith}, {Santos}, {Scandariato}, {Seager}, {Sousa}, {Steller}, {Szab{\'o}}, {S{\'e}gransan}, {Thomas}, {Udry}, {Ulmer}, {Van Grootel}, {Vanderspek}, {Walton}, \& {Winn}}]{Osborn2023}
{Osborn}, H.~P., {Nowak}, G., {H{\'e}brard}, G., {et~al.} 2023, \mnras, 523, 3069, \dodoi{10.1093/mnras/stad1319}

\bibitem[{{Osborne} {et~al.}(2024){Osborne}, {Van Eylen}, {Goffo}, {Gandolfi}, {Nowak}, {Persson}, {Livingston}, {Weeks}, {Pall{\'e}}, {Luque}, {Hellier}, {Carleo}, {Redfield}, {Hirano}, {Garbaccio Gili}, {Alarcon}, {Barrag{\'a}n}, {Casasayas-Barris}, {D{\'\i}az}, {Esposito}, {Knudstrup}, {Jenkins}, {Murgas}, {Orell-Miquel}, {Rodler}, {Serrano}, {Stangret}, {Albrecht}, {Alqasim}, {Cochran}, {Deeg}, {Fridlund}, {Hatzes}, {Korth}, \& {Lam}}]{Osborne2024}
{Osborne}, H.~L.~M., {Van Eylen}, V., {Goffo}, E., {et~al.} 2024, \mnras, 527, 11138, \dodoi{10.1093/mnras/stad3837}

\bibitem[{{Perger} {et~al.}(2017){Perger}, {Ribas}, {Damasso}, {Morales}, {Affer}, {Su{\'a}rez Mascare{\~n}o}, {Micela}, {Maldonado}, {Gonz{\'a}lez Hern{\'a}ndez}, {Rebolo}, {Scandariato}, {Leto}, {Zanmar Sanchez}, {Benatti}, {Bignamini}, {Borsa}, {Carbognani}, {Claudi}, {Desidera}, {Esposito}, {Lafarga}, {Martinez Fiorenzano}, {Herrero}, {Molinari}, {Nascimbeni}, {Pagano}, {Pedani}, {Poretti}, {Rainer}, {Rosich}, {Sozzetti}, \& {Toledo-Padr{\'o}n}}]{Perger2017}
{Perger}, M., {Ribas}, I., {Damasso}, M., {et~al.} 2017, \aap, 608, A63, \dodoi{10.1051/0004-6361/201731307}

\bibitem[{{Persson} {et~al.}(2018){Persson}, {Fridlund}, {Barrag{\'a}n}, {Dai}, {Gandolfi}, {Hatzes}, {Hirano}, {Grziwa}, {Korth}, {Prieto-Arranz}, {Fossati}, {Van Eylen}, {Justesen}, {Livingston}, {Kubyshkina}, {Deeg}, {Guenther}, {Nowak}, {Cabrera}, {Eigm{\"u}ller}, {Csizmadia}, {Smith}, {Erikson}, {Albrecht}, {Sobrino}, {Cochran}, {Endl}, {Esposito}, {Fukui}, {Heeren}, {Hidalgo}, {Hjorth}, {Kuzuhara}, {Narita}, {Nespral}, {Palle}, {P{\"a}tzold}, {Rauer}, {Rodler}, \& {Winn}}]{Persson2018}
{Persson}, C.~M., {Fridlund}, M., {Barrag{\'a}n}, O., {et~al.} 2018, \aap, 618, A33, \dodoi{10.1051/0004-6361/201832867}

\bibitem[{{Petigura} {et~al.}(2020){Petigura}, {Livingston}, {Batygin}, {Mills}, {Werner}, {Isaacson}, {Fulton}, {Howard}, {Weiss}, {Espinoza}, {Jontof-Hutter}, {Shporer}, {Bayliss}, \& {Barros}}]{Petigura2020}
{Petigura}, E.~A., {Livingston}, J., {Batygin}, K., {et~al.} 2020, \aj, 159, 2, \dodoi{10.3847/1538-3881/ab5220}

\bibitem[{{Pinamonti} {et~al.}(2022){Pinamonti}, {Sozzetti}, {Maldonado}, {Affer}, {Micela}, {Bonomo}, {Lanza}, {Perger}, {Ribas}, {Gonz{\'a}lez Hern{\'a}ndez}, {Bignamini}, {Claudi}, {Covino}, {Damasso}, {Desidera}, {Giacobbe}, {Gonz{\'a}lez-{\'A}lvarez}, {Herrero}, {Leto}, {Maggio}, {Molinari}, {Morales}, {Pagano}, {Petralia}, {Piotto}, {Poretti}, {Rebolo}, {Scandariato}, {Su{\'a}rez Mascare{\~n}o}, {Toledo-Padr{\'o}n}, \& {Zanmar S{\'a}nchez}}]{Pinamonti2022}
{Pinamonti}, M., {Sozzetti}, A., {Maldonado}, J., {et~al.} 2022, \aap, 664, A65, \dodoi{10.1051/0004-6361/202142828}

\bibitem[{{Piso} {et~al.}(2015){Piso}, {Youdin}, \& {Murray-Clay}}]{Piso15}
{Piso}, A.-M.~A., {Youdin}, A.~N., \& {Murray-Clay}, R.~A. 2015, \apj, 800, 82, \dodoi{10.1088/0004-637X/800/2/82}

\bibitem[{{Rescigno} {et~al.}(2024){Rescigno}, {H{\'e}brard}, {Vanderburg}, {Mann}, {Mortier}, {Morrell}, {Buchhave}, {Collins}, {Mann}, {Hellier}, {Haywood}, {West}, {Stalport}, {Heidari}, {Anderson}, {Huang}, {L{\'o}pez-Morales}, {Cort{\'e}s-Zuleta}, {Lewis}, {Dumusque}, {Boisse}, {Rowden}, {Collier Cameron}, {Deleuil}, {Vezie}, {Pepe}, {Delfosse}, {Charbonneau}, {Rice}, {Demangeon}, {Quinn}, {Udry}, {Forveille}, {Winn}, {Sozzetti}, {Hoyer}, {Seager}, {Wilson}, {Dalal}, {Martioli}, {Striegel}, {Boschin}, {Dragomir}, {Mart{\'\i}nez Fiorenzano}, {Cosentino}, {Ghedina}, {Malavolta}, {Affer}, {Lakeland}, {Nicholson}, {Foschino}, {W{\"u}nsche}, {Barkaoui}, {Srdoc}, {Randolph}, {Guillet}, {Conti}, {Ghachoui}, {Gillon}, {Benkhaldoun}, {Pozuelos}, {Timmermans}, {Girardin}, {Matutano}, {Bosch-Cabot}, {Mu{\~n}oz}, \& {For{\'e}s-Toribio}}]{Rescigno2024}
{Rescigno}, F., {H{\'e}brard}, G., {Vanderburg}, A., {et~al.} 2024, \mnras, 527, 5385, \dodoi{10.1093/mnras/stad3255}

\bibitem[{{Rosenthal} {et~al.}(2021){Rosenthal}, {Fulton}, {Hirsch}, {Isaacson}, {Howard}, {Dedrick}, {Sherstyuk}, {Blunt}, {Petigura}, {Knutson}, {Behmard}, {Chontos}, {Crepp}, {Crossfield}, {Dalba}, {Fischer}, {Henry}, {Kane}, {Kosiarek}, {Marcy}, {Rubenzahl}, {Weiss}, \& {Wright}}]{Rosenthal2021}
{Rosenthal}, L.~J., {Fulton}, B.~J., {Hirsch}, L.~A., {et~al.} 2021, \apjs, 255, 8, \dodoi{10.3847/1538-4365/abe23c}

\bibitem[{{Rosenthal} {et~al.}(2022){Rosenthal}, {Knutson}, {Chachan}, {Dai}, {Howard}, {Fulton}, {Chontos}, {Crepp}, {Dalba}, {Henry}, {Kane}, {Petigura}, {Weiss}, \& {Wright}}]{Rosenthal2022}
{Rosenthal}, L.~J., {Knutson}, H.~A., {Chachan}, Y., {et~al.} 2022, \apjs, 262, 1, \dodoi{10.3847/1538-4365/ac7230}

\bibitem[{{Santos} {et~al.}(2016){Santos}, {Santerne}, {Faria}, {Rey}, {Correia}, {Laskar}, {Udry}, {Adibekyan}, {Bouchy}, {Delgado-Mena}, {Melo}, {Dumusque}, {H{\'e}brard}, {Lovis}, {Mayor}, {Montalto}, {Mortier}, {Pepe}, {Figueira}, {Sahlmann}, {S{\'e}gransan}, \& {Sousa}}]{Santos2016}
{Santos}, N.~C., {Santerne}, A., {Faria}, J.~P., {et~al.} 2016, \aap, 592, A13, \dodoi{10.1051/0004-6361/201628374}

\bibitem[{{Savignac} \& {Lee}(2023)}]{Savignac23}
{Savignac}, V., \& {Lee}, E.~J. 2023, arXiv e-prints, arXiv:2310.06013, \dodoi{10.48550/arXiv.2310.06013}

\bibitem[{{S{\'e}gransan} {et~al.}(2011){S{\'e}gransan}, {Mayor}, {Udry}, {Lovis}, {Benz}, {Bouchy}, {Lo Curto}, {Mordasini}, {Moutou}, {Naef}, {Pepe}, {Queloz}, \& {Santos}}]{Segransan2011}
{S{\'e}gransan}, D., {Mayor}, M., {Udry}, S., {et~al.} 2011, \aap, 535, A54, \dodoi{10.1051/0004-6361/200913580}

\bibitem[{{Serrano} {et~al.}(2022){Serrano}, {Gandolfi}, {Mustill}, {Barrag{\'a}n}, {Korth}, {Dai}, {Redfield}, {Fridlund}, {Lam}, {D{\'\i}az}, {Grziwa}, {Collins}, {Livingston}, {Cochran}, {Hellier}, {Bellomo}, {Trifonov}, {Rodler}, {Alarcon}, {Jenkins}, {Latham}, {Ricker}, {Seager}, {Vanderspeck}, {Winn}, {Albrecht}, {Collins}, {Csizmadia}, {Daylan}, {Deeg}, {Esposito}, {Fausnaugh}, {Georgieva}, {Goffo}, {Guenther}, {Hatzes}, {Howell}, {Jensen}, {Luque}, {Mann}, {Murgas}, {Osborne}, {Palle}, {Persson}, {Rowden}, {Rudat}, {Smith}, {Twicken}, {Van Eylen}, \& {Ziegler}}]{Serrano2022}
{Serrano}, L.~M., {Gandolfi}, D., {Mustill}, A.~J., {et~al.} 2022, Nature Astronomy, 6, 736, \dodoi{10.1038/s41550-022-01641-y}

\bibitem[{{Sinukoff} {et~al.}(2017){Sinukoff}, {Howard}, {Petigura}, {Fulton}, {Crossfield}, {Isaacson}, {Gonzales}, {Crepp}, {Brewer}, {Hirsch}, {Weiss}, {Ciardi}, {Schlieder}, {Benneke}, {Christiansen}, {Dressing}, {Hansen}, {Knutson}, {Kosiarek}, {Livingston}, {Greene}, {Rogers}, \& {L{\'e}pine}}]{Sinukoff2017}
{Sinukoff}, E., {Howard}, A.~W., {Petigura}, E.~A., {et~al.} 2017, \aj, 153, 271, \dodoi{10.3847/1538-3881/aa725f}

\bibitem[{{Sozzetti} {et~al.}(2021){Sozzetti}, {Damasso}, {Bonomo}, {Alibert}, {Sousa}, {Adibekyan}, {Zapatero Osorio}, {Gonz{\'a}lez Hern{\'a}ndez}, {Barros}, {Lillo-Box}, {Stassun}, {Winn}, {Cristiani}, {Pepe}, {Rebolo}, {Santos}, {Allart}, {Barclay}, {Bouchy}, {Cabral}, {Ciardi}, {Di Marcantonio}, {D'Odorico}, {Ehrenreich}, {Fasnaugh}, {Figueira}, {Haldemann}, {Jenkins}, {Latham}, {Lavie}, {Lo Curto}, {Lovis}, {Martins}, {M{\'e}gevand}, {Mehner}, {Micela}, {Molaro}, {Nunes}, {Oshagh}, {Otegi}, {Pall{\'e}}, {Poretti}, {Ricker}, {Rodriguez}, {Seager}, {Su{\'a}rez Mascare{\~n}o}, {Twicken}, \& {Udry}}]{Sozzetti2021}
{Sozzetti}, A., {Damasso}, M., {Bonomo}, A.~S., {et~al.} 2021, \aap, 648, A75, \dodoi{10.1051/0004-6361/202040034}

\bibitem[{{Staab} {et~al.}(2020){Staab}, {Haswell}, {Barnes}, {Anglada-Escud{\'e}}, {Fossati}, {Doherty}, {Cooper}, {Jenkins}, {D{\'\i}az}, \& {Soto}}]{Staab2020}
{Staab}, D., {Haswell}, C.~A., {Barnes}, J.~R., {et~al.} 2020, Nature Astronomy, 4, 399, \dodoi{10.1038/s41550-019-0974-x}

\bibitem[{{Su{\'a}rez Mascare{\~n}o} {et~al.}(2018){Su{\'a}rez Mascare{\~n}o}, {Gonz{\'a}lez Hern{\'a}ndez}, {Rebolo}, {Velasco}, {Toledo-Padr{\'o}n}, {Udry}, {Motalebi}, {S{\'e}grasan}, {Wyttenbach}, {Mayor}, {Pepe}, {Lovis}, {Santos}, {Figueira}, \& {Esposito}}]{Suarez2018}
{Su{\'a}rez Mascare{\~n}o}, A., {Gonz{\'a}lez Hern{\'a}ndez}, J.~I., {Rebolo}, R., {et~al.} 2018, \aap, 612, A41, \dodoi{10.1051/0004-6361/201732042}

\bibitem[{{Teske} {et~al.}(2020){Teske}, {D{\'\i}az}, {Luque}, {Mo{\v{c}}nik}, {Seidel}, {Otegi}, {Feng}, {Jenkins}, {Pall{\`e}}, {S{\'e}gransan}, {Udry}, {Collins}, {Eastman}, {Ricker}, {Vanderspek}, {Latham}, {Seager}, {Winn}, {Jenkins}, {Anderson}, {Barclay}, {Bouchy}, {Burt}, {Butler}, {Caldwell}, {Collins}, {Crane}, {Dorn}, {Flowers}, {Haldemann}, {Helled}, {Hellier}, {Jensen}, {Kane}, {Law}, {Lissauer}, {Mann}, {Marmier}, {Nielsen}, {Rose}, {Shectman}, {Shporer}, {Torres}, {Wang}, {Wolfgang}, {Wong}, \& {Ziegler}}]{Teske2020}
{Teske}, J., {D{\'\i}az}, M.~R., {Luque}, R., {et~al.} 2020, \aj, 160, 96, \dodoi{10.3847/1538-3881/ab9f95}

\bibitem[{{Tinney} {et~al.}(2011){Tinney}, {Butler}, {Jones}, {Wittenmyer}, {O'Toole}, {Bailey}, \& {Carter}}]{Tinney2011}
{Tinney}, C.~G., {Butler}, R.~P., {Jones}, H. R.~A., {et~al.} 2011, \apj, 727, 103, \dodoi{10.1088/0004-637X/727/2/103}

\bibitem[{{Tran} {et~al.}(2022){Tran}, {Bowler}, {Endl}, {Cochran}, {MacQueen}, {Gandolfi}, {Persson}, {Fridlund}, {Palle}, {Nowak}, {Deeg}, {Luque}, {Livingston}, {Kab{\'a}th}, {Skarka}, {{\v{S}}ubjak}, {Howell}, {Albrecht}, {Collins}, {Esposito}, {Van Eylen}, {Grziwa}, {Goffo}, {Huang}, {Jenkins}, {Karjalainen}, {Karjalainen}, {Knudstrup}, {Korth}, {Lam}, {Latham}, {Levine}, {Osborne}, {Quinn}, {Redfield}, {Ricker}, {Seager}, {Serrano}, {Smith}, {Twicken}, \& {Winn}}]{Tran2022}
{Tran}, Q.~H., {Bowler}, B.~P., {Endl}, M., {et~al.} 2022, \aj, 163, 225, \dodoi{10.3847/1538-3881/ac5c4f}

\bibitem[{{Trifonov} {et~al.}(2020){Trifonov}, {Tal-Or}, {Zechmeister}, {Kaminski}, {Zucker}, \& {Mazeh}}]{Trifonov2020}
{Trifonov}, T., {Tal-Or}, L., {Zechmeister}, M., {et~al.} 2020, \aap, 636, A74, \dodoi{10.1051/0004-6361/201936686}

\bibitem[{{Tuomi} {et~al.}(2014){Tuomi}, {Jones}, {Barnes}, {Anglada-Escud{\'e}}, \& {Jenkins}}]{Tuomi2014}
{Tuomi}, M., {Jones}, H. R.~A., {Barnes}, J.~R., {Anglada-Escud{\'e}}, G., \& {Jenkins}, J.~S. 2014, \mnras, 441, 1545, \dodoi{10.1093/mnras/stu358}

\bibitem[{{Turtelboom} {et~al.}(2022){Turtelboom}, {Weiss}, {Dressing}, {Nowak}, {Pall{\'e}}, {Beard}, {Blunt}, {Brinkman}, {Chontos}, {Claytor}, {Dai}, {Dalba}, {Giacalone}, {Gonzales}, {Harada}, {Hill}, {Holcomb}, {Korth}, {Lubin}, {Masseron}, {MacDougall}, {Mayo}, {Mo{\v{c}}nik}, {Akana Murphy}, {Polanski}, {Rice}, {Rubenzahl}, {Scarsdale}, {Stassun}, {Tyler}, {Zandt}, {Crossfield}, {Deeg}, {Fulton}, {Gandolfi}, {Howard}, {Huber}, {Isaacson}, {Kane}, {Lam}, {Luque}, {Mart{\'\i}n}, {Morello}, {Orell-Miquel}, {Petigura}, {Robertson}, {Roy}, {Van Eylen}, {Baker}, {Belinski}, {Bieryla}, {Ciardi}, {Collins}, {Cutting}, {Della-Rose}, {Ellingsen}, {Furlan}, {Gan}, {Gnilka}, {Guerra}, {Howell}, {Jimenez}, {Latham}, {Larivi{\`e}re}, {Lester}, {Lillo-Box}, {Luker}, {Mann}, {Plavchan}, {Safonov}, {Skinner}, {Strakhov}, {Wittrock}, {Caldwell}, {Essack}, {Jenkins}, {Quintana}, {Ricker}, {Vanderspek}, {Seager}, \& {Winn}}]{Turtelboom2022}
{Turtelboom}, E.~V., {Weiss}, L.~M., {Dressing}, C.~D., {et~al.} 2022, \aj, 163, 293, \dodoi{10.3847/1538-3881/ac69e5}

\bibitem[{{Udry} {et~al.}(2019){Udry}, {Dumusque}, {Lovis}, {S{\'e}gransan}, {Diaz}, {Benz}, {Bouchy}, {Coffinet}, {Lo Curto}, {Mayor}, {Mordasini}, {Motalebi}, {Pepe}, {Queloz}, {Santos}, {Wyttenbach}, {Alonso}, {Collier Cameron}, {Deleuil}, {Figueira}, {Gillon}, {Moutou}, {Pollacco}, \& {Pompei}}]{Udry2019}
{Udry}, S., {Dumusque}, X., {Lovis}, C., {et~al.} 2019, \aap, 622, A37, \dodoi{10.1051/0004-6361/201731173}

\bibitem[{{Unger} {et~al.}(2021){Unger}, {S{\'e}gransan}, {Queloz}, {Udry}, {Lovis}, {Mordasini}, {Ahrer}, {Benz}, {Bouchy}, {Delisle}, {D{\'\i}az}, {Dumusque}, {Lo Curto}, {Marmier}, {Mayor}, {Pepe}, {Santos}, {Stalport}, {Alonso}, {Collier Cameron}, {Deleuil}, {Figueira}, {Gillon}, {Moutou}, {Pollacco}, \& {Pompei}}]{Unger2021}
{Unger}, N., {S{\'e}gransan}, D., {Queloz}, D., {et~al.} 2021, \aap, 654, A104, \dodoi{10.1051/0004-6361/202141351}

\bibitem[{{Walsh} {et~al.}(2011){Walsh}, {Morbidelli}, {Raymond}, {O'Brien}, \& {Mandell}}]{Walsh11}
{Walsh}, K.~J., {Morbidelli}, A., {Raymond}, S.~N., {O'Brien}, D.~P., \& {Mandell}, A.~M. 2011, \nat, 475, 206, \dodoi{10.1038/nature10201}

\bibitem[{{Weiss} {et~al.}(2013){Weiss}, {Marcy}, {Rowe}, {Howard}, {Isaacson}, {Fortney}, {Miller}, {Demory}, {Fischer}, {Adams}, {Dupree}, {Howell}, {Kolbl}, {Johnson}, {Horch}, {Everett}, {Fabrycky}, \& {Seager}}]{Weiss2013}
{Weiss}, L.~M., {Marcy}, G.~W., {Rowe}, J.~F., {et~al.} 2013, \apj, 768, 14, \dodoi{10.1088/0004-637X/768/1/14}

\bibitem[{{Weiss} {et~al.}(2020){Weiss}, {Fabrycky}, {Agol}, {Mills}, {Howard}, {Isaacson}, {Petigura}, {Fulton}, {Hirsch}, \& {Sinukoff}}]{Weiss2020}
{Weiss}, L.~M., {Fabrycky}, D.~C., {Agol}, E., {et~al.} 2020, \aj, 159, 242, \dodoi{10.3847/1538-3881/ab88ca}

\bibitem[{{Weiss} {et~al.}(2024){Weiss}, {Isaacson}, {Howard}, {Fulton}, {Petigura}, {Fabrycky}, {Jontof-Hutter}, {Steffen}, {Schlichting}, {Wright}, {Beard}, {Brinkman}, {Chontos}, {Giacalone}, {Hill}, {Kosiarek}, {MacDougall}, {Mo{\v{c}}nik}, {Polanski}, {Turtelboom}, {Tyler}, \& {Van Zandt}}]{Weiss2024}
{Weiss}, L.~M., {Isaacson}, H., {Howard}, A.~W., {et~al.} 2024, \apjs, 270, 8, \dodoi{10.3847/1538-4365/ad0cab}

\bibitem[{{Wittenmyer} {et~al.}(2020){Wittenmyer}, {Wang}, {Horner}, {Butler}, {Tinney}, {Carter}, {Wright}, {Jones}, {Bailey}, {O'Toole}, \& {Johns}}]{Wittenmyer2020}
{Wittenmyer}, R.~A., {Wang}, S., {Horner}, J., {et~al.} 2020, \mnras, 492, 377, \dodoi{10.1093/mnras/stz3436}

\bibitem[{{Zhang} {et~al.}(2021){Zhang}, {Weiss}, {Huber}, {Blunt}, {Chontos}, {Fulton}, {Grunblatt}, {Howard}, {Isaacson}, {Kosiarek}, {Petigura}, {Rosenthal}, \& {Rubenzahl}}]{Zhang2021}
{Zhang}, J., {Weiss}, L.~M., {Huber}, D., {et~al.} 2021, \aj, 162, 89, \dodoi{10.3847/1538-3881/ac0634}

\bibitem[{{Zhu}(2023)}]{Zhu2023}
{Zhu}, W. 2023, arXiv e-prints, arXiv:2306.16691, \dodoi{10.48550/arXiv.2306.16691}

\bibitem[{{Zhu} {et~al.}(2018){Zhu}, {Petrovich}, {Wu}, {Dong}, \& {Xie}}]{Zhu2018b}
{Zhu}, W., {Petrovich}, C., {Wu}, Y., {Dong}, S., \& {Xie}, J. 2018, \apj, 860, 101, \dodoi{10.3847/1538-4357/aac6d5}

\bibitem[{{Zhu} \& {Wu}(2018)}]{ZhuWu18}
{Zhu}, W., \& {Wu}, Y. 2018, \aj, 156, 92, \dodoi{10.3847/1538-3881/aad22a}

\end{thebibliography}

\end{document}